\documentclass[11pt,a4paper]{article}

\usepackage[pdftex]{graphics}
\usepackage{jheppub}

\usepackage{amsmath,amssymb,amsfonts}

\usepackage{enumitem}
\usepackage{multirow}

\usepackage{array,booktabs}

\usepackage{slashed}
\usepackage{mathtools}
\allowdisplaybreaks
\usepackage{tcolorbox}
\usepackage{float}
\usepackage{tikz,subcaption}
\allowdisplaybreaks
\newcommand{\be}{\begin{eqnarray}}
\newcommand{\ee}{\end{eqnarray}}
\newcommand{\eqa}{\begin{eqnarray}}
\newcommand{\eqae}{\end{eqnarray}}
\newcommand{\nn}{\nonumber}
\newcommand{\bn}{\begin{enumerate}}
\newcommand{\en}{\end{enumerate}}
\newcommand{\bl}{\begin{align}}
\newcommand{\el}{\end{align}}
\newcommand{\eq}{\begin{equation}}
\newcommand{\eqe}{\end{equation}}

\def\identity{{\rlap{1} \hskip 1.6pt \hbox{1}}}
\def\iden{\identity}

\def\IZ{\mathbb{Z}}

\def\CI{{\cal I}}

\def\CO{{\cal O}}

\def\CS{{\cal S}}

\def\a{\alpha}
\def\b{\beta}
\def\g{\gamma}

\def\e{\epsilon}
\def\ve{\varepsilon}

\def\th{\theta}

\def\k{\kappa}
\def\l{\lambda}
\def\m{\mu}
\def\n{\nu}

\def\s{\sigma}

\def\t{\tau}

\def\w{\omega}
\def\half{\frac{1}{2}}

\def\del{\nabla}

\def\identity{{\rlap{1} \hskip 1.6pt \hbox{1}}}

\newcommand{\bfig}{\begin{figure}}
\newcommand{\efig}{\end{figure}}

\def\la{{\langle}}
\def\ra{{\rangle}}

\def\bl#1\el{\begin{align} #1 \end{align}}
\def\bg#1\eg{\begin{gather} #1 \end{gather}}
\def\bld#1\eld{\begin{aligned} #1 \end{aligned}}
\def\bgd#1\egd{\begin{gathered} #1 \end{gathered}}

\newcommand{\bra}[1]{\langle{#1}|}
\newcommand{\ket}[1]{|{#1}\rangle}

\newcommand{\sbra}[1]{ [{#1} |}
\newcommand{\sket}[1]{ | {#1} ]}

\newcommand{\cSquare}{\,\rotatebox{90}{\scalebox{0.6}[0.9]{$\bowtie$}}}

\def\jmath{{j}}
\def\bl#1\el{\begin{align} #1 \end{align}}
\def\bg#1\eg{\begin{gather} #1 \end{gather}}

\def\bld#1\eld{\begin{aligned} #1 \end{aligned}}
\def\bgd#1\egd{\begin{gathered} #1 \end{gathered}}

\newcommand{\eqc}[1]{eq.(\ref{#1})}

\usetikzlibrary{arrows.meta,bending,calc,decorations.pathmorphing,decorations.markings,shapes.misc,shapes.geometric}
\pgfkeys{/tikz/.cd,
  circle color/.initial=black,
  circle color/.get=\circlecolor,
  circle color/.store in=\circlecolor,
}
\tikzset{/pgf/decoration/.cd,
    number of sines/.initial=10,
    angle step/.initial=20,
}
\newdimen\tmpdimen\pgfdeclaredecoration{complete sines}{initial}
{
    \state{initial}[
        width=+0pt,
        next state=move,
        persistent precomputation={
            \pgfmathparse{\pgfkeysvalueof{/pgf/decoration/angle step}}%
            \let\anglestep=\pgfmathresult%
            \let\currentangle=\pgfmathresult%
            \pgfmathsetlengthmacro{\pointsperanglestep}%
                {(\pgfdecoratedremainingdistance/\pgfkeysvalueof{/pgf/decoration/number of sines})/360*(\anglestep)}%
        }] {}
    \state{move}[width=+\pointsperanglestep, next state=draw]{
        \pgfpathmoveto{\pgfpointorigin}
    }
    \state{draw}[width=+\pointsperanglestep, switch if less than=1.25*\pointsperanglestep to final, 
        persistent postcomputation={
        \pgfmathparse{mod(\currentangle+\anglestep, 360)}%
        \let\currentangle=\pgfmathresult%
    }]{%
        \pgfmathsin{+\currentangle}%
        \tmpdimen=\pgfdecorationsegmentamplitude%
        \tmpdimen=\pgfmathresult\tmpdimen%
        \divide\tmpdimen by2\relax%
        \pgfpathlineto{\pgfqpoint{0pt}{\tmpdimen}}%
    }
    \state{final}{
        \ifdim\pgfdecoratedremainingdistance>0pt\relax
            \pgfpathlineto{\pgfpointdecoratedpathlast}
        \fi
   }
}

\title{Gravitational Faraday effect from on-shell amplitudes}
\author[1]{Wei-Ming Chen} 
\author[2]{Ming-Zhi Chung} 
\author[2,3]{Yu-tin Huang}
\author[4,5]{Jung-Wook Kim}

\affiliation[1]{Department of Physics, Kobe University, Kobe 657-8501, Japan}
\affiliation[2]{Department of Physics and Astronomy, National Taiwan University, Taipei 10617, Taiwan}
\affiliation[3]{Department of Physics and Center for Theoretical Physics, National Taiwan University, Taipei 10617, Taiwan}
\affiliation[4]{Centre for Theoretical Physics, Department of Physics and Astronomy,\\ Queen Mary University of London, Mile End Road, London E1 4NS, United Kingdom}
\affiliation[5]{Kavli Institute for Theoretical Physics, University of California, Santa Barbara, CA 93106, USA}

\emailAdd{tainist@gmail.com}
\emailAdd{dchung0741@gmail.com}
\emailAdd{yutin@phys.ntu.edu.tw}
\emailAdd{jung-wook.kim@qmul.ac.uk}

\abstract{Effects of massive object's spin on massive-massless $2 \to 2$ classical scattering is studied. Focus is set on the less-considered dimensionless expansion parameter $\lambda/b$, where $\lambda$ is the massless particle's wavelength and $b$ is the impact parameter. Corrections in $\lambda/b$ start to appear from $\mathcal{O}(G^2)$, with leading correction terms tied to the gravitational Faraday effect, which is a special case of the Lense-Thirring effect. We compute the eikonal phase up to $\mathcal{O}(G^2)$ and extract spin effect on the scattering angle and time delay up to 14th order in spin. The gravitational Faraday effect at linear order in spin~\cite{Ishihara:1987dv} is reproduced by $\lambda/b$ correction terms, which we compute to higher orders in spin.  We find that the equivalence principle, or universality, holds up to NLO for general spinning bodies, i.e. away from geometric optics limit.  Furthermore, in the black hole limit, we confirm the absence of particular spin structure observed~\cite{Bern:2020buy,Aoude:2020ygw,Kosmopoulos:2021zoq,Chen:2021qkk, Aoude:2022trd, Bern:2022kto, Aoude:2022thd}, along with the associated shift symmetry~\cite{Bern:2022kto}, and argue that it holds to arbitrary spin order at $\mathcal{O}(G^2)$ in the massless probe limit. 
}

\usetikzlibrary{arrows.meta,bending,calc,decorations.pathmorphing,decorations.markings,shapes.misc,shapes.geometric}
\pgfkeys{/tikz/.cd,
  circle color/.initial=black,
  circle color/.get=\circlecolor,
  circle color/.store in=\circlecolor,
}
\tikzset{/pgf/decoration/.cd,
    number of sines/.initial=10,
    angle step/.initial=20,
}
\newdimen\tmpdimen\pgfdeclaredecoration{complete sines}{initial}
{
    \state{initial}[
        width=+0pt,
        next state=move,
        persistent precomputation={
            \pgfmathparse{\pgfkeysvalueof{/pgf/decoration/angle step}}%
            \let\anglestep=\pgfmathresult%
            \let\currentangle=\pgfmathresult%
            \pgfmathsetlengthmacro{\pointsperanglestep}%
                {(\pgfdecoratedremainingdistance/\pgfkeysvalueof{/pgf/decoration/number of sines})/360*(\anglestep)}%
        }] {}
    \state{move}[width=+\pointsperanglestep, next state=draw]{
        \pgfpathmoveto{\pgfpointorigin}
    }
    \state{draw}[width=+\pointsperanglestep, switch if less than=1.25*\pointsperanglestep to final, 
        persistent postcomputation={
        \pgfmathparse{mod(\currentangle+\anglestep, 360)}%
        \let\currentangle=\pgfmathresult%
    }]{%
        \pgfmathsin{+\currentangle}%
        \tmpdimen=\pgfdecorationsegmentamplitude%
        \tmpdimen=\pgfmathresult\tmpdimen%
        \divide\tmpdimen by2\relax%
        \pgfpathlineto{\pgfqpoint{0pt}{\tmpdimen}}%
    }
    \state{final}{
        \ifdim\pgfdecoratedremainingdistance>0pt\relax
            \pgfpathlineto{\pgfpointdecoratedpathlast}
        \fi
   }
}
\begin{document}
\begin{flushright}
\vspace{10pt} \hfill{KOBE-COSMO-22-04\,,\,QMUL-PH-22-16} \vspace{20mm}
\end{flushright}

\maketitle

\newpage
\section{Introduction}
The fact that the classical limit of scattering amplitudes can correctly capture black hole dynamics, is on itself a fascinating subject irrespective of its remarkable success in applications to gravitational wave physics, which has helped brought into the spotlight since  the land mark detection by the LIGO/Virgo collaboration~\cite{LIGOScientific:2016aoc, LIGOScientific:2017vwq}. That point-like sources can faithfully reproduce effects of gravitational field in the weak field limit was appreciated long ago in the seminal work of ref.\cite{Duff:1973zz}, where it was shown that the Schwarzschild metric can be perturbatively reconstructed order by order in $G$ through Feynman diagrams. This was later bolstered by the computations of perturbative stess-energy form factors~\cite{Holstein:2004dn, Donoghue:2001qc}. Later development focused on the Post Minkowskian (PM) expansion~\cite{Bertotti:1956pxu, Kerr:1959zlt, Bertotti:1960wuq, Portilla:1979xx, Westpfahl:1979gu, Portilla:1980uz, Bel:1981be, Westpfahl:1985tsl, Ledvinka:2008tk,  Damour:2017zjx} for conservative two-body potential of binary black holes, which can be extracted from the gravitational scattering amplitude of massive scalars~\cite{Donoghue:1994dn, Bjerrum-Bohr:2002gqz, Bjerrum-Bohr:2013bxa, Neill:2013wsa}. By matching the field theory amplitude to effective field theory (EFT) description~\cite{Cheung:2018wkq}, the on-shell simplification led to rapid progress, culminating in the state of the art 4 PM conservative potential~\cite{Bern:2019nnu, Bern:2019crd, Bern:2021dqo}. As the PM expansion is exact in velocity ($v^2$), the results can be directly compared and matched to the  post-Newtonian (PN) expansion~\cite{Einstein:1938yz, Ohta:1973je, Jaranowski:1997ky, Damour:1999cr, Blanchet:2000nv, Damour:2001bu, Damour:2014jta, Jaranowski:2015lha}.

The extension to spinning black holes brings in a new angle to the correspondence. It was already shown in the early work of refs.\cite{Holstein:2008sx, Vaidya:2014kza} that the spinning Hamiltonian can be captured by the scattering of elementary spinning particle. However, the absence of elementary particle beyond spin-2 casts an intriguing conundrum for the program. This was partially overcome by the kinematic definition of minimal coupling for arbitrary spin particle~\cite{Arkani-Hamed:2017jhn} whose classical spin-limit ($\hbar\rightarrow 0$, $s\rightarrow \infty$ and $s\hbar$ held fixed) successfully reproduced the 1 PM dynamics of Kerr black hole~\cite{Maybee:2019jus, Guevara:2018wpp, Chung:2018kqs, Chung:2019duq, Guevara:2019fsj, Arkani-Hamed:2019ymq, Aoude:2020onz, Aoude:2021oqj}. For non black hole spinning bodies, in the PN EFT approach one introduces spins as extra degrees of freedom~\cite{Porto:2005ac} on the worldline for the point particle effective action, accompanied by an infinite number of spin-induced multipole operators whose strengths are parametrised by the corresponding Wilson coefficients~\cite{Porto:2008jj, Levi:2015msa}; see refs.\cite{Levi:2019kgk,Levi:2020uwu,Levi:2020lfn,Kim:2021rfj} for recent works, ref.\cite{Liu:2021zxr} in the context of PM EFT, and refs.\cite{Jakobsen:2021lvp,Jakobsen:2021zvh,Jakobsen:2022fcj} for worldline QFT. On the amplitude side, this can be incorporated by matching the Wilson coefficients to those of non-minimal couplings~\cite{Chung:2019duq, Chung:2020rrz}, or to introduce a field theory effective lagrangian~\cite{Bern:2020buy}. The latter approach has been implemented at 2 PM to obtain the spin-dependent part of the conservative hamiltonian up to fifth order in spin~\cite{Kosmopoulos:2021zoq, Bern:2022kto}.

In going beyond the quartic order in spin at 2 PM, from the on-shell approach one needs the gravitational Compton amplitude for the interaction of higher spin states which are not unique~\cite{Arkani-Hamed:2017jhn, Chung:2018kqs}. A proposal utilizing appropriately conserved off-shell current was given for spin-5/2 in ref.\cite{Chiodaroli:2021eug} (see ref.\cite{Falkowski:2020aso} for a compact expression of the ambiguity). When translated to the effective field theory Lagrangian approach in ref.\cite{Bern:2020buy}, the issue becomes the determination of black hole Wilson coefficients at 2 PM. In a nutshell, the question is \textit{can one identify the underlying principles that determine the Kerr black hole limit ?}  

A natural starting point is to examine the 1 PM dynamics and to take note of any special aspects emerging in the Kerr limit. It was observed in ref.\cite{Aoude:2020mlg} that the change in spin-entanglement entropy is minimized for minimal couplings, which was later identified as the suppression of spin-flipping sector in the eikonal phase~\cite{Chen:2021huj}. The absence of spin-flipping sectors for the 1 PM eikonal amplitude is not surprising as minimal coupling for massive spinning higher spins are defined such that it has optimal power counting in the UV, which corresponds to helicity preserving interactions.
Besides optimal UV behaviour, an interesting aspect of the 1 PM potential is the presence of shift symmetry. For general spinning bodies, the 1 PM potential depends on the spin vector through the combination $\e(q, p_2, p_1, S)$~\cite{Chung:2020rrz}, which exhibits the independence on the longitudinal part of the spin vector along the impact parameter plane, i.e. the potential is invariant under the shift symmetry proposed in ref.\cite{Bern:2022kto},
\bl\label{Shift1}
S_i^\m \to S_i^\m + \xi_i \frac{q^\m}{q^2} \,.
\el
On the other hand, preservation of this symmetry at 2 PM becomes a nontrivial constraint. Earlier, it was noted that in the spin bilinear sector the spin-spin interaction $(S_1 \cdot S_2)$ only occurs through a special combination~\cite{Holstein:2008sx,Guevara:2017csg,Damgaard:2019lfh,Bern:2020buy}
\bl
(q \cdot S_1) (q \cdot S_2) - q^2 (S_1 \cdot S_2) \,, \nn
\el
which is secretly related to shift symmetry, as will be explained. This combination can be generalised to the case where spin vectors are taken from the same body, i.e.
\bl\label{SpinComb}
(q \cdot S_i) (q \cdot S_j) - q^2 (S_i \cdot S_j) \,,
\el
where $i = j$ is also allowed. It was noted in refs.\cite{Aoude:2020ygw,Kosmopoulos:2021zoq} that for 2 PM classical amplitude, from which the conservative Hamiltonian is extracted, the black hole limit (defined by absence of tidal terms or by correct spin-quadrupole moments) is special in that the structure \eqc{SpinComb} appears, at least to \emph{quadratic or bilinear} order in spin. A related observation is that when the spin operators are expanded on the basis of Lorentz invariants   
\bl
\{ (q \cdot S_i) \,,\, (p_j \cdot S_i) \,,\, \e(q, p_1, p_2, S_i) \}\,, \label{eq:spinLorInvMassive}
\el
the black hole limit is special in that $(q \cdot S_i)$ terms all vanish, at least to \emph{quartic} order in spin~\cite{Chen:2021qkk}(appendix B).\footnote{The validity of the exponential form of the Kerr Compton amplitude used in refs.\cite{Guevara:2018wpp,Guevara:2019fsj,Bautista:2019tdr,Aoude:2020onz,Chen:2021qkk} to $\CO(S^4)$ has been confirmed by comparing to black hole perturbation theory. The authors would like to thank Justin Vines for sharing conclusions of their upcoming work~\cite{JustinNew}.} The two observations are related since the structure \eqc{SpinComb} appears when the product $\e(q, p_1, p_2, S_i) \e(q, p_1, p_2, S_j)$ is reorganised using Levi-Civita identities. Also, the latter observation is equivalent to shift symmetry \eqc{Shift1}; $\e(q, p_2, p_1, S)$ is manifestly invariant, while $(p_j \cdot S_i)$ is invariant classically. 
Recently, these equivalent conditions on the spin structures of the Hamiltonian were used with an additional assumption of favorable high energy behaviour to fix Wilson coefficients that determine spinning-spinless sectors at 2 PM~\cite{Aoude:2022trd, Bern:2022kto,Aoude:2022thd}.\footnote{The authors would like to thank Kays Haddad for clarification on the relation of the spin structure \eqc{SpinComb} to the Kerr limit.}

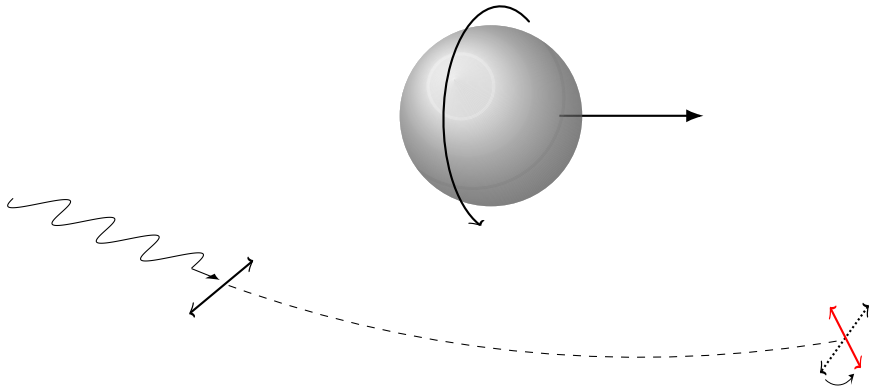
\begin{figure}
\centering
\begin{tikzpicture}[sines/.style={
        line width=0.7pt,
        line join=round, 
        draw=black, 
        decorate}
]
\draw[variable=\t,domain=-5:-2.5,samples=2000,rotate=-22,shift={(2.4,-2.77)},line cap=round]
    plot (\t,,{0.5*sin(\t*360*5/pi+5)+1.85});
 \draw[-{Latex},thick](3.1,0)--(5,0);
    \draw[-{latex[slant=0.9]},rotate=-22,shift={(0.3,-0.05)}](-1.15,-2.48)--(-0.75,-2.48);
    \draw[{to[slant=0.7]}-{to[slant=0.7]},shift={(-0.68,-4.78)},rotate=-140,thick](-1.65,-2.35)--(-0.55,-2.35);
   \draw[{to[slant=0.7]}-{to[slant=0.7]},densely dotted,shift={(8.2,-5.15)},rotate=-123,thick](-1.65,-2.3)--(-0.55,-2.35);
    \draw[color=red,{to[slant=-0.7]}-{to[slant=-0.7]},shift={(9.35,-2.6)},rotate=-60,thick](-1.4,-2.3)--(-0.5,-2.35);
   \draw[dashed,shift={(-0.75,-1.75)}] (-0.5,-0.5) arc (-110:-80.2:15.8);
  \fill[ball color = lightgray!20, opacity = 0.3] (2.2,0) circle (1.2cm);
  \draw[-{to[slant=0.8]},line cap=round,rotate=-90,shift={(-1.25,-0.3)},thick](0,3) arc (150:340:1.5 and 0.75); 
  \draw[-{stealth[slant=0.8]},line cap=round,line width=0.3pt,shift={(0,0)}](6.6,-3.5) arc (230:330:0.25 and 0.3); 
\end{tikzpicture}
\caption{A cartoon of gravitational Faraday effect. The polarisation direction (black double arrow) of the electromagnetic wave propagating along the direction of the spinning axis gets rotated (red double arrow) by the frame-dragging effect of the spinning body.
\newline
}
\label{fig:gravFaraday}
\end{figure}

In this paper, we revisit the classical limit of the gravitational Compton amplitude with spin effects included. The goal is twofold: on the one hand a rotating object will induce Lense–Thirring (or gravitomagnetic frame-dragging) effects, which can rotate the polarization plane of propagating electromagnetic or gravitational waves as in fig.\ref{fig:gravFaraday}. This effect should be visible through the classical Compton amplitude describing the scattering of photons or gravitons off a spinning object. At the same time, the role of shift symmetry or the distinctive spin-spin interaction in determining the black hole limit can be tested in this context, as the amplitude will be parameterised by the worldline Wilson coefficients for linear coupling to gravitons.

Earlier studies of gravitational Compton amplitude focused on the photon/graviton and massive scalar system, where one is interested in the corrections to the scattering angle (light bending) of null geodesics~\cite{Bjerrum-Bohr:2014zsa, Bjerrum-Bohr:2016hpa, Bai:2016ivl, Chi:2019owc} and Shapiro time delay~\cite{Camanho:2014apa, AccettulliHuber:2020oou}. An important tool is the eikonal approximation, corresponding to large center of mass energy and small scattering angle ($s\gg t$),  where the amplitude exponentiates~\cite{Cheng:1969eh, Abarbanel:1969ek, Levy:1969cr, Amati:1987wq, Amati:1987uf, Kabat:1992tb, Akhoury:2013yua, KoemansCollado:2019ggb}. Importantly, since the amplitude depends on the external helicity states, instead of the eikonal phase we have the eikonal phase matrix~\cite{Camanho:2014apa, AccettulliHuber:2020oou}
\eq
 \left(\begin{array}{cc}\delta_{{+},{+}} & \delta_{{-},{+}} \\ \delta_{{+},{-}} & \delta_{{-},{-}}\end{array}\right)=\bar{\delta}\,\mathbb{I}{+}\alpha_i\,\sigma_i\,,
\eqe 
where we use $2 \to 2$ convention for the helicity states, and the matrix is written in a manifest Hermitian form. The phases $\bar{\delta}, \alpha_i$, which are functions of impact parameter $b$ and frequency $\omega$ of the massless mode, can be perturbatively computed in powers of $G$. Previous computations focused on $\bar{\delta}=(\delta_{{+},{+}}{+}\delta_{{-},{-}})/2$, from which one extracts the scattering angle and time delay, via
\eq \label{eq:eik2obs1}
\theta=\frac{1}{\w} \frac{\partial}{ \partial b}\bar\delta(b,\w), \quad \delta t=\frac{\partial}{ \partial \w}\bar\delta(b,\w)\,.
\eqe
For scalar Compton $\alpha_3=0$ due to the lack of parity violating interactions, while $\alpha_{1,2}$ was studied for the effects of higher derivative corrections to Einstein-Hilbert term, which generate non-trivial same helicity sector at tree and one-loop~\cite{Camanho:2014apa, AccettulliHuber:2020oou}. These helicity non-preserving components modify circular polarisation states to elliptical polarised states.

We will compute the Compton eikonal matrix for general spinning bodies up to $\CO(G^2)$. The resulting scattering angle and time delay will be a function of Wilson coefficients. In contrast with the eikonal phase for massive scattering amplitudes, there is a new dimensionless parameter in the classical regime: $1/\w b =\lambda/b$ where $\lambda$ is the wavelength of the gravitational/EM wave. Thus for spinning bodies we have a double expansion in $\lambda/b$ and $a/b$, where $a = S/m$ is the spin length.  From dimensional analysis, we can see that $\lambda/b$ corrections in the classical limit only occurs at $\CO(G^2)$: 
\eqa
\CO(G^2):\quad\bar{\delta} &=& \frac{15 \pi G^2m^2\omega}{4b}\left[1{-}\frac{(\hat{k}_2\times \vec{a})\cdot\vec{b}}{3b^2}{+}\mathcal{O}\left(\frac{a^2}{b^2}\right)\right],\nonumber\\
\a := - \frac{\a_3}{|h|} &=& \frac{5 \pi  G^2m^2(\hat{k}_2\cdot \vec{a})}{4b^3}\left[1{-}(1{+}C_{\rm S^2})\frac{9(\hat{k}_2\times \vec{a})\cdot\vec{b}}{8b^2}{+}\mathcal{O}\left(\frac{a^2}{b^2}\right)\right]\,,
\eqae
where $\hat{k}_2$ is the spatial direction of the incoming massless plane wave. Here we are giving the result in terms of an $a/b$ expansion. Note that LO for $\alpha_3$ is subleading for the individual eikonal phase $\delta_{+,+}$ and $\delta_{-,-}$. We identify $\alpha$ as the rotation angle of the gravitational Faraday effect~\cite{Ishihara:1987dv, Nouri-Zonoz:1999jls, Sereno:2004jx, Brodutch:2011eh, Farooqui:2013rga, Shoom:2020zhr, Deriglazov:2021gwa, Chakraborty:2021bsb, Li:2022izh} whose name is an analogy to the magneto-optical effect bearing the same name; for linearly polarised EM waves, the polarisation direction rotates due to the ``magnetic'' background. Indeed we see that $\alpha$ is only present when there is a non-zero spin and the leading linear in spin term matches with that computed using parallel transport along null geodesics~\cite{Ishihara:1987dv}. While the leading term for $\alpha$ is universal, the subleading terms depend on the Wilson coefficients.  We provide the result up to 14th order in spin, for both photons and gravitons.

Note that the spin dependence in the computation of the classical Compton amplitude stems from the product of spinning three-point amplitudes. In the black hole limit, where one has minimal coupling, the product of two three-point amplitudes in the classical limit can be written as
\eq\label{ThreePointSum}
\left( \prod_{i = 2,3} M_{3,i} \right) \exp \left( - \sum_{i=2,3}  \frac{\epsilon^{\mu\nu\rho\sigma}(F_i)_{\mu\nu} p_{1\rho} S_\sigma}{2 (P \cdot \ve_i)}\right) 
\eqe
where $\prod_i M_{3,i}$ is the product of scalar three-point amplitudes, $(F_{i})_{\m\n} = - 2 i k_{i [\m} \ve_{i,\n]}$ is the field strength, and $\sum_i$ sums over the massless legs with $P$ representing the momentum of the intermediate massive leg ($P = - p_1 - k_2$). The expression can be derived from eq.(A.6) of ref.\cite{Chen:2021qkk} with $z_i = 1$, neglecting quantum commutator terms from the BCH formula. The expression is gauge invariant with on-shell kinematics, and a similar expression for spinning soft factor described in terms of the field strength $F_{\m\n}$ appeared in ref.\cite{Bautista:2019tdr}. A remarkable property of this form is that under the shift operation in eq.(\ref{Shift1}), it transforms as
\bl
\bld
{\rm eq}.(\ref{ThreePointSum})\bigg|_{S^\m \to S^\m + \xi \frac{q^\m}{q^2}} &= \left( \prod_{i = 2,3} M_{3,i} \right)  \exp \left( - \sum_{i=2,3}  \frac{\epsilon^{\mu\nu\rho\sigma}(F_i)_{\mu\nu} p_{1\rho} S_\sigma}{2 (P \cdot \epsilon_i)}\right)
\\ &\phantom{=asdf} \times \exp \left(- \xi  \frac{(\sum_{i{\in} {+}} k_i \cdot q){-}(\sum_{j{\in} {-}} k_j \cdot q)}{q^2}\right) 
\eld
\el
where $\sum_{i\in {+}} k_i$ sums over the momenta of positive helicity legs and $\sum_{j\in {-}}k_j$ sums over negativity helicity. Thus one immediately sees that for helicity preserving configuration, the product is shift invariant! For helicity non-conserving configuration, the shift would not be an invariance, which was already observed in ref.\cite{Aoude:2022trd}. As only the former contribute to the 2 PM classical Compton amplitude, this suggests that it is in fact shift invariant. We will demonstrate that this is indeed the case.

 Finally by considering the normalized quantity $\alpha_3/|h|$ along with $\bar\delta$, one can show that equivalence principle holds at leading order in $\lambda/b$ expansion. At subleading order  in $\lambda/b$ both of these properties begin to deviate. The deviations can be understood as the eikonal limit being the geometric optics limit, where one ignores the wave nature of electric-magnetic/gravitational waves and treat it as null rays. This is valid when the wave length is much smaller than the impact parameter $\frac{\lambda}{b}\ll1$. Thus expansion in $\frac{\lambda}{b}$ would correspond to corrections to the approximation.

This paper is organized as follows: we begin with a brief discussion of $\l/b$ corrections to the eikonal phase and their physical interpretation in sec.\ref{sec:eikphase}. Then in sec.\ref{sec:DA} we compute the $\mathcal{O}(G^1)$ eikonal phase for general spinning body, and derive physical observables. In sec.\ref{sec:2PMeik} , we present the computation and result for $\mathcal{O}(G^2)$ eikonal phase. 






\section{General discussion of $\l/b$ corrections} \label{sec:eikphase}

The kinematics of interest has four length scales; the Compton wavelength $\l_C \sim m^{-1}$ controlling the ``quantumness'' of the massive body, the wavelength $\l = \w^{-1}$ of the massless particle,\footnote{For massless particles we consider the wavenumber four-vector $\bar{k}^\m := k^\m / \hbar$ as the classical quantity.} the spin length $a \sim |\vec{S}| m^{-1}$ controlling the ``size'' of the massive body and the impact parameter ${b}$ controlling the separation. In the (classical) geometric optics regime the scale hierarchy is\footnote{The hierarchy $\l \ll a$ is only needed for the conceptual foundation of geometric optics and is irrelevant for any of the analyses in this paper. In terms of wavepackets, there is an additional scale hierarchy $\l \ll \xi \ll b$ where $\xi$ is the wavepacket size~\cite{Cristofoli:2021vyo}. JWK would like to thank Donal O'Connell for discussions on this point. 
}
\bl\label{ScaleSep}
\l_C \ll \l \ll a \ll b \,.
\el
Treating the internal nature of the massive body as quantum, the separation between ``quantum" and ``classical" effects amounts to utilizing the scale hierarchy between $\l_C$ and all other scales:  
\bl
\frac{\l_C}{b} \sim \frac{\hbar}{m b} \,,\, \frac{\l_C}{a} \sim \frac{\hbar}{|\vec{S}|} \,,\, \frac{\l_C}{\l} \sim \frac{\hbar \w}{m} \,, \label{eq:quantumexpparam}
\el
where $\hbar\w$ is the energy of the massless particle of momentum. Considering that the transfer momentum $q \sim b^{-1}$ is the Fourier dual of the impact parameter $b$, the classical limit corresponds to the following $\hbar$ scaling for the variables:
\begin{equation}\label{eq:PMexpparam}
q^{\mu} \rightarrow \hbar q^{\mu}, \quad
k^\m \rightarrow \hbar k^\m ,\quad
S^{\mu} \rightarrow 
\frac{S^{\mu}}{\hbar}, \quad
m \rightarrow m, \quad
\w \rightarrow \hbar \w\,, \quad G \rightarrow \frac{G}{\hbar} \,.
\end{equation}
Due to the separation of scale in \eqc{ScaleSep}, we have two classical expansion parameters:
\eq
\frac{a}{b} \sim \frac{|\vec{S}|}{m b},\quad \frac{\lambda}{b}=\frac{1}{b\w}\,.
\eqe  
Thus we will only compute the leading term in $\hbar \to 0$ limit, with the spin and energy expansion ${a}/{|\vec{b}|},\, (\w b)^{-1}$ being an additional expansion \textit{in the classical regime}. From now on we will suppress $\hbar$ dependence for brevity by setting $\hbar = 1$.

\begin{figure}
\begin{center}
\begin{tikzpicture}[sines/.style={
        line width=0.7pt,
        line join=round, 
        draw=black, 
        decorate, 
        decoration={complete sines, number of sines=4, amplitude=4pt}}
]
       \node[scale=1] at (-1.8,-1) {$p_1$};
       \node[,scale=1] at (-1.8,1) {$k_2$};
        \node[scale=1] at (1.8,1) {$k_3$};
        \node[scale=1] at (1.8,-1) {$p_4$};
 \draw[line width=1.5pt,line cap=round] (-2,-0.7)--(0,-0.2); 
 \draw[line width=1.5pt,line cap=round] (0,-0.2)--(2,-0.7); 
 \draw[white,postaction={sines}](-2,0.7)--(0,0.2); 
 \draw[white,postaction={sines}] (0,0.2)--(2,0.7); 
 \draw[-to,line width=0.6pt,line cap=round] (-1.5,-0.8)--(-0.7,-0.6); 
 \draw[-to,line width=0.6pt,line cap=round] (-1.5,0.8)--(-0.7,0.6); 
 \draw[to-,line width=0.6pt,line cap=round] (1.5,-0.8)--(0.7,-0.6); 
 \draw[to-,line width=0.6pt,line cap=round] (1.5,0.8)--(0.7,0.6); 
  \filldraw[fill=lightgray,draw=black,thin] (0,0) circle (0.5cm);
\end{tikzpicture}
\caption{The diagram shows the configuration of the Compton amplitudes in our consideration. The wavy line represents the massless particle and solid line is the massive spinning object. The arrows show the directions of momentum flow.}
\label{ComptonFig}
\end{center}
\end{figure}
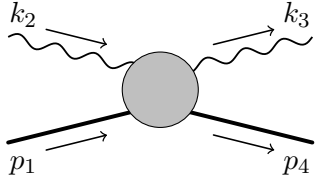

Let's consider the Compton amplitude where the incoming massless state (with $k_2$) scatters off a massive object (with momenta $p_1$) as in fig.\ref{ComptonFig}. Due to vanishing/suppressed helicity-flip amplitudes in general relativity, the outgoing massless state (with $k_3$) will be of the same helicity as the incoming state. The Compton amplitude of interest is $M(\mathbf{1} 2^{\pm h} 3^{\pm h} \mathbf{4})$, where $h=1,2$ denotes photon and graviton respectively. The eikonal approximation is an all-order resummation of $M_4$ where the amplitude is reorganised as a phase factor $i M_4 = e^{i \delta} - 1$, which has been extensively studied in the literature~\cite{Cheng:1969eh, Abarbanel:1969ek, Levy:1969cr, Amati:1987wq, Amati:1987uf, Kabat:1992tb, Akhoury:2013yua, KoemansCollado:2019ggb}. The eikonal phase $\delta (b)$ is obtained by Fourier transforming the $2 \rightarrow 2$ amplitude to impact parameter space\footnote{Different definitions for the impact parameter space are also used in the literature, which is reviewed in appendix \ref{app:eiknorm}.}
\bl
\bld
e^{i \delta_{{\pm},{\pm}} } - 1 &:= \int \frac{d^{D} q}{(2 \pi)^{D}}   \hat{\delta} (2 p_1 \cdot q) \hat{\delta} (2 k_2 \cdot q) \left[ e^{ - i b \cdot q} i M_4 (\mathbf{1} 2^{\pm h} 3^{\pm h} \mathbf{4})\right]
\\ &= \frac{i}{4m \omega} \int \frac{d^{D-2} q}{(2 \pi)^{D-2}}  \left[ e^{ i \vec{b} \cdot \vec{q}} M_4 (\mathbf{1} 2^{\pm h} 3^{\pm h} \mathbf{4})\right] \,,
\eld \label{eq:eikphasedeforig}
\el
where $\hat{\delta} (x) = 2 \pi \delta (x)$ and Mandelstam invariants are parametrised as:
\eqa
s = (p_1{+}k_2)^2 = 2m\w {+} m^2 \,,\, t = (k_3 {-} k_2)^2 = {-} |\vec{q}|^2 \,,\, u = (p_1 {-} k_3)^2 = m^2 {-}2m\w + |\vec{q}|^2 \,. \label{eq:MandelstamDef}
\eqae
This parametrises $\w = (p_1 \cdot k_2)/m$ as the frequency of the incoming massless particle in the rest frame of $p_1$. The transfer momentum from the massive to massless state is $q^\m = k_3^\m - k_2^\m$ and its spatial part in centre of momentum (COM) frame is the three-vector $\vec{q}$.
\bl
\bld
p_1^\m = (\sqrt{m^2 + k^2 } , - \vec{{k}} ) \,&,\, p_4^\m = (\sqrt{m^2 + k^2 } , - \vec{{k}} - \vec{q} ) \,,
\\ k_2^\m = ( k , \vec{{k}} ) \,&,\, k_3^\m = ( k , \vec{{k}} + \vec{q} ) \,,
\eld
\el
where the COM frame variable $k$ is given by $\w$ through the relation
\bl
(k^2 / m^2) = \frac{\sqrt{1 + 4 (\w^2 / m^2)} - 1}{2} = (\w^2 / m^2) \times [1 + \CO(\hbar^2)] \,,
\el
and can be considered as $\w$ in the classical limit.


\subsection{Dimensional analysis}
Before we begin, let's first perform a dimension analysis to determine the allowed operators that can accompany various subleading (in $\l/b$) corrections. Each element of the eikonal phase matrix, which we collectively denote by $\delta$, can be expanded as a perturbative series in the coupling $G$, the spin $S$, and the dimensionless ratio $\l / b$. 
\bl
\bgd
\delta = \sum_{L=0}^{\infty}G^{L{+}1}  \delta^{(L)} = \sum_{L=0}^{\infty}G^{L{+}1}\left( \sum_{j=0}^\infty \delta_{\text{S}^j}^{(L)}\frac{a^j}{b^j}\right) = \sum_{L=0}^{\infty} G^{L{+}1}\left(\sum_{j=0}^\infty  \frac{a^j}{b^j} \left( \sum_{i=0}^\infty \delta_{\text{S}^j}^{(L,i)}\frac{\l^i}{b^i}\right)\right)\,,
\egd \label{eq:eikphaseexp}
\el
where the superscript $(L)$ denotes the $L$-loop ($\CO(G^{L+1})$) contribution and subscript $\text{S}^n$ denotes the $\CO(S^n)$ contribution of the given $L$-loop contribution. The second subscript $i$ of the paired subscript $(L,i)$ denotes subleading $(\l/b)^i$ order contribution to the given $L$-loop contribution. 

Let's define $M_4^{(L,0,0)}$ as the terms of the Compton amplitude $M_4$ that determine the leading spin-less sector $\delta^{(L,0)}_{\text{S}^0}$ of the $L$-loop eikonal phase $\delta^{(L)}$, and $M_4^{(L,i,j)}$ as that which determines $(\l/b)^i (a/b)^j$ corrections ($\delta^{(L,i)}_{\text{S}^j}$). The classical limit of background-test mass $2 \to 2$ scattering has only one master scalar integral (the $L$-loop fan integral) at each loop order, while other remaining integrals contribute to the iteration terms required from the exponentiation in the eikonal approximation~\cite{Brandhuber:2021eyq}.  Thus as long as the eikonal approximation is valid and iteration holds, only one integral is relevant at each loop. Defining the ratio:
\bl
F_{(L,i,j)} &:= \frac{ M_4^{(L,i,j)} }{ M_4^{(L,0,0)} } \,, \label{eq:Ffactordef}
\el
becomes simply the ratios of the integral coefficient of the master scalar integral, and thus functions of Lorentz invariants (the helicity weight is cancelled in the ratio). To analyse $F_{(L,i,j)}$ we first write all possible Lorentz invariants that can be constructed from kinematics. There are Mandelstam invariants which we can combine into a dimensionless expansion parameter $\l / b$:\footnote{We ignore $\CO(1)$ numerical factors in power counting analysis.}
\bl
(s - m^2) \sim m \w \sim \frac{m}{\l} \,,\, t = - |\vec{q}|^2 \sim \frac{1}{b^2} && \Rightarrow && \left[ \frac{m \sqrt{-t}}{s - m^2} \right] \sim \frac{\l}{b} \label{eq:MandelstamLorInv}
\el
The introduction of the spin vector brings three new invariants due to the transverse condition $p \cdot S = 0$. The three Lorentz invariants can be combined into the independent dimensionless expansion parameter $a / b$ as:
\bl
\bld
(q \cdot S) &\sim \frac{m a}{b} && \Rightarrow && & \left[ \frac{q \cdot S}{m} \right] &\sim \frac{a}{b}
\\ (n \cdot S) &\sim \frac{m^2 a}{\l b} && \Rightarrow && & \left[ \frac{n \cdot S}{m (s-m^2)} \right] &\sim \frac{a}{b}
\\ (k_2 \cdot S) &\sim \frac{m a}{\l} && \Rightarrow && & \left[ \frac{\sqrt{-t}(k_2 \cdot S)}{s - m^2} \right] &\sim \frac{a}{b}\,.
\eld \label{eq:spinLorInv}
\el
The orthogonal vector $n^\m$ is defined as $n^\m = \e^{\m\a\b\g} p_{1 \a} k_{2\b} q_{\g}$. These invariants form a basis for $F_{(L,i,j)}$
\bl
F_{(L,i,j)} = \sum_{\substack{\a + \b + \g = j \\ \a,\b,\g \ge 0}} c_{i,\a,\b,\g} \left[ \frac{m \sqrt{-t}}{s - m^2} \right]^i \left[ \frac{q \cdot S}{m} \right]^\a \left[ \frac{n \cdot S}{m (s-m^2)} \right]^\b \left[ \frac{\sqrt{-t}(k_2 \cdot S)}{s - m^2} \right]^\g \,,
\el
where $c_{i,\a,\b,\g}$ are numerical coefficients. Since $F_{(L,i,j)}$ is a rational function, the powers of $\sqrt{-t}$ must add up to an even number and we have the constraint $i + \g \in 2 \IZ$. 
We list the consequences of this constraint:
\bn
\item There are no $\l / b$ corrections at $\CO(G^1)$. The scalar term $M_4^{(0,0,0)}$ contains a $t^{-1}$ pole prefactor. As any expansion in $\l / b$ must be combined to form integer powers of $-t = |\vec{q}|^2$, the corrections will cancel the $t^{-1}$ pole and result in derivatives of delta functions in impact parameter space, which do not contribute to long range effects. 

\item At $\CO(G^2)$, the scalar term $M_4^{(1,0,0)}$ contains a $1/|\vec{q}|$ prefactor from the scalar triangle integral. The non-analytic structure of the $1/|\vec{q}|$ prefactor yields long range contributions which cannot be cancelled by integer powers of $-t$, allowing arbitrary order $\l/b$ corrections. The condition $i + \g \in 2 \IZ$ forces odd powers of $\l / b$ corrections to accompany odd powers of $(k_2 \cdot S)$, and the first $\l / b$ correction to be at $(\l / b)^2$ order for spinless case ($j=0$). The corrections at this order are dependent on conventions for the eikonal phase as explained in appendix~\ref{app:eiknorm}. 
\en
As we will argue in the following, the polarisation plane rotation is an $\mathcal{O}(\l / b)$ correction from the leading eikonal approximation, thus it is first observed at one-loop order. his is consistent with the conclusion in ref.\cite{Asada:2000vn} that gravitational Faraday effect cannot be observed at $\CO(G^1)$ order.

\subsection{Subleading $\lambda/b$ corrections and rotations of polarization planes} \label{sec:NLOnAlpha}
Previous studies on eikonal phases of gravitational Compton scattering mostly ignored subleading effects in the $\l / b$ expansion. These subleading corrections are related to the wave-like nature of the massless particle since they depend on the wavelength $\l$ of the particle, but which physical properties are these terms reflecting?

To answer this question, we consider geometric optics. The propagation of light (or gravitational wave) is treated as a classical curve where the tangent vector at each point of the curve is normal to the wavefront. For an introduction, see e.g. \textsection 53 of ref.\cite{landau1975classical}. We consider the propagation of waves as a classical trajectory 
in the limit $\lambda\rightarrow 0$ or $\lambda/b\ll1$. Selecting a component of the wave (e.g. any component of $\vec{E}$ or $\vec{B}$ for a electromagnetic wave) and denoting it as $f$, we can write an ansatz\footnote{The terminology \emph{WKB approximation} is also used in the literature for this ansatz, e.g. ref.\cite{Isaacson:1968hbi}.}
\bl
f = a e^{i \psi} \,, \label{eq:fieldeikdecomp}
\el
where $a = a(x,t)$ is the amplitude of the particular component and $\psi = \psi (x,t)$ is the phase of the wave. $\psi$ is called the \emph{eikonal} and its derivative $\bar{k}_\mu = - \partial_\m \psi$ is regarded as the local wavenumber four-vector of the ray, which is normal to the wavefront and tangent to the propagation direction~\cite{landau1975classical}. Since $\psi$ changes by $2\pi$ as one transverses one $\lambda$, the limit $\l \to 0$ implies large gradient of $\psi$. Now recall that the wave equation is given by:
\bl
\frac{\partial^2 f}{\partial x_\m \partial x^\m} = \left[ \frac{\partial^2 a}{\partial x_\m \partial x^\m} e^{i\psi} + 2 i \frac{\partial a}{\partial x_\m} \frac{\partial \psi}{\partial x^\m} e^{i \psi} + \frac{\partial^2 \psi}{\partial x_\m \partial x^\m} f \right] - \frac{\partial \psi}{\partial x_\m}\frac{\partial \psi}{\partial x^\m} f = 0 \,.\label{eq:eikeqderiv}
\el
Since $\partial\psi\sim\frac{1}{\l}$ is large, we expect the first three terms in square brackets to be subleading compared to the last by a factor of $\l/b$. Amongst the three terms in the square brackets, since $\frac{\partial a}{\partial x_\m}$ is accompanied by a factor of $ \frac{\partial \psi}{\partial x^\m}$, the subleading corrections at this order will be dominated by the change in the wave amplitude denoted as $a$, which also encodes polarisation data. Thus if we can identify $\psi$ with $\delta(b)$, the $\l / b$ corrections reflect polarisation dependence. We provide an argument in appendix \ref{app:eikphase2geo} on why we can identify the eikonal $\psi$ in \eqc{eq:fieldeikdecomp} as the eikonal phase $\delta(b)$ in \eqc{eq:eikphasedeforig}. Indeed in section~\ref{sec:2PMeik}, our explicit results show that
\eq
\delta_{{+},{+}} = \delta_{{-},{-}} \times \left[ 1 + \CO \left( \l / b\right) \right] \,. \label{eq:eikphaseheldepexp}
\eqe
that is, the polarization dependence occurs at subleading order in $\l / b$.


The difference $2\a_3 \equiv (\delta_{+,+} - \delta_{-,-})$, which characterises the first subleading order in $\l/b$, has an interpretation as \emph{the rotation angle of the polarisation plane}. 
To see this, consider positive and negative helicity one-particle states propagating along the $z$ direction. The two states will acquire opposite phases under a rotation of angle $\a$ around the $z$-axes, which is simply a little group rotation.
\eq
\ket{+}'\rightarrow e^{-i \a J^{12}} \ket{+} = e^{-ih\a}\ket{+},\quad \ket{-}'\rightarrow e^{-i \a J^{12}} \ket{-} = e^{+ih\a}\ket{-} \,.
\eqe
The difference of the phase gained by positive and negative helicity photons ($|h| = 1$) is tied to the rotation angle of the polarisation plane $\a$ by
\bl
\a &= - \frac{\delta_{\g,{+},{+}} - \delta_{\g,{-},{-}}}{2} \,.
\el
There is an additional factor of 2 for gravitons ($|h| = 2$),
\bl
\a &= - \frac{\delta_{g,{+},{+}} - \delta_{g,{-},{-}}}{4} \,,
\el
which compensates for the spin of the graviton. We show that indeed the difference in eikonal phase reproduces the result of ref.\cite{Ishihara:1987dv}, where the polarisation plan rotation angle was computed by considering a parallel transport along the light ray's null geodesic. Finally, we see that the angle can be identified with the $\sigma_3$ component of the eikonal phase matrix: $\alpha=-\alpha_3/|h|$.



\section{The $\CO(G^1)$ Compton eikonal phase for spinning objects} \label{sec:DA}
We begin with the $\CO(G^1)$ Compton eikonal phase for general spinning body. This will allow us to introduce the setup which will carry over to the $\CO(G^2)$ computation in the next section. Furthermore, the $\CO(G^1)$ result will be important in checking whether the box and the cross box contributions at $\CO(G^2)$ matches with the square of the former, as required from exponentiation pattern of the eikonal amplitude. 

While helicity non-preserving amplitudes are present for the Einstein-Hilbert action~\cite{AccettulliHuber:2020oou}, they are suppressed by $\CO ( (\l/b)^4 )$ compared to their helicity preserving counterparts, both at tree and one-loop level. As we are not interested in $\l/b$ expansion to such high orders, we will solely focus on helicity preserving processes.

\subsection{The tree amplitude for eikonal phase}
Here we will only consider the helicity conserving configuration $\widetilde{M}_4 (\mathbf{1}2^{{+}h}3^{{+}h}\mathbf{4})$ of the classical amplitude. Here we use $\widetilde{\;}$ to make a distinction between the classical amplitude from ``semi"-classical amplitude. From the previous discussion we've seen that helicity dependence of the eikonal phase can only be subleading in $\l/b$, while dimensional analysis tells us that such corrections start at $\CO(G^2)$. Thus this tell us that at $\CO(G^1)$, the amplitudes $\widetilde{M}_4 (\mathbf{1}2^{{\pm}h}3^{{\pm}h}\mathbf{4})$ must be identical. Indeed this is verified through our explicit computation. 
The gravitational coupling constant is normalised as $\k = \sqrt{32 \pi G}$, where $G$ is the Newton's constant.

The eikonal phase is given by the fourier transform of the Compton amplitude in the $q^2\rightarrow 0$ limit. At $\CO(G^1)$ this limit is given by the product of two three-point amplitudes, one corresponding to graviton coupling to massive spinning and the other massless helicity states. This can be extracted from the following integral formula, motivated by the exponentiated representation for amplitudes of Kerr black holes~\cite{Guevara:2018wpp,Guevara:2019fsj,Bautista:2019tdr,Aoude:2020onz,Chen:2021qkk}\footnote{The expression is $2 \to 2$ continued from the $4 \to 0$ expression. Spinor notations follow that of ref.\cite{Chung:2018kqs}.}
\bl
\bgd
M_4^{s} = M_4^{s=0} \oint \frac{dz}{2\pi i z} \left( \sum_j C_{\text{S}^j} z^j \right) \exp \left( -i K^\m L^\n J_{\m\n} \right) \,,
\\ K^\m = -\frac{q^\m}{z} \,,\, L^\m = \frac{\bra{3} \s^\m \sket{2}}{\bra{3} p_1 \sket{2}} \,,
\egd \label{eq:tree4ptansatz}
\el
where the \emph{scalar factor} $M_4^{s=0}$ is the spin-independent part, i.e. it represents the Compton amplitude for two massive scalar and two identical helicity state generated via a graviton exchange. The above can be viewed as a semi-classical amplitude where the spin is taken to the classical limit, i.e. $s\rightarrow \infty$ and $\hbar\rightarrow 0$ with $\hbar s$ held fixed. Indeed taking $s$ finite, and $C_{\text{S}^j}=1$, this reduces to the Compton amplitude for minimally coupled spinning particle~\cite{Chen:2021qkk}. From $M_4^{s}$ we can derive the fully classical amplitude $\widetilde{M}$ by further expanding the kinematic variables with appropriate $\hbar$ counting. 

The \emph{spin factor} refers to the remaining auxiliary residue integral terms in \eqc{eq:tree4ptansatz}, where the contour is taken to be encircling the origin counter-clockwise. The spin factor is motivated to capture the full $t$-channel (graviton pole) factorization limit, which is modified from the spinless case by the Wilson coefficients $C_{\text{S}^n}$ parametrising the spin-multipole moments~\cite{Levi:2015msa,Chung:2018kqs,Chung:2019duq,Chung:2020rrz}. In particular, the $t=0$ limit can be attained by either setting $\ket{2} \propto \ket{3}$, or $|2 ] \propto |3 ]$. For the former, the scalar factor $M_4^{s=0}$ factorizes to (in all incoming convention)
\bl
-t M_4^{s=0} \to M_3^{+h,-h,+2} \times M_{3^-}^{s=0} \,,
\el
where $M_3^{+h,-h,+2}$ is the amplitude of a helicity $h$ particle
coupling to a positive helicity graviton 
and $M_{3^-}^{s=0} = - \k (p_1 \cdot \ve_{3^-})^2$ is the amplitude for a massive scalar coupling to a negative helicity graviton. The spin factor in \eqc{eq:tree4ptansatz} becomes
\bl
K^\m = - \frac{q^\m}{z} \,,\, L^\m \to \frac{\ve_{q^-}^\m}{\ve_{q^-} \cdot p_1} & && \Rightarrow && & \exp(-i K^\m L^\n J_{\m\n}) \to \exp \left( i \frac{q^\m \ve_{q^-}^\n J_{\m\n}}{(p_1 \cdot \ve_{q^-}) z}\right) \,,
\el
which reproduces the spin factor of eq.(B.1) in ref.\cite{Chen:2021qkk}.\footnote{The $\sket{2}$ spinors of $L^\m$ are interpreted as the auxiliary spinor of $\ve_{q^-}^\m$. The relative sign difference is due to outgoing convention for $q$.} The factorisation behaviour for the case $|2 ] \propto |3 ]$ can be analysed in a similar manner. Although \eqc{eq:tree4ptansatz} has spurious poles due to the factor $\bra{3} p_1 \sket{2}$ in the denominator of the exponent, the spurious poles can be resolved by terms that do not affect the $t$-channel pole which determines the $\CO(G^1)$ eikonal phase, therefore only considering the terms in \eqc{eq:tree4ptansatz} is enough to compute the $\CO(G^1)$ eikonal phase.

The scalar amplitudes for massive (uncharged) scalar interacting with massless spin-$|h|$ states through graviton exchange are given as
\bl
\frac{M_4^{s=0}}{(\k/2)^2} &= \left\{
\begin{aligned}
& - \frac{(s-u)^2}{4 t} & && h&=0
\\ & - \frac{\sbra{2}p_1\ket{3} (s-u)}{2 t} & && h&=1/2
\\ & - \frac{\sbra{2}p_1\ket{3}^2}{t} & && h&=1
\\ & \frac{\sbra{2}p_1\ket{3}^3 (s-u)}{2(s-m^2)(u-m^2)t} & && h&=3/2
\\ & \frac{\sbra{2}p_1\ket{3}^4}{(s-m^2)(u-m^2)t} & && h&=2
\end{aligned}
\right.\,. \label{eq:tree4ptscalar}
\el
Explicit evaluation of relevant spinor contractions yield
\bl
\bgd
\sbra{2} p_1 \ket{3} = \sbra{3} p_1 \ket{2} = 2m\w \sqrt{1 - \frac{|\vec{q}|^2 (1 + \frac{2\w}{m})}{4 \w^2}} = 2 m \w \times \left[1 + \CO(q^2 / \w^2 \,,\, \hbar) \right] \,,\,
\\ [23] = -\la 23 \ra = |\vec{q}| \,,
\egd \label{eq:LGfactorvalues}
\el
up to phase factors for little group scaling, thus all scalar amplitudes in \eqc{eq:tree4ptscalar} can be treated as
\bl
M_4^{s=0} \simeq \frac{32 \pi G m^2 \w^2}{\vec{q}^2} \,, \label{eq:tree4ptscalarfact}
\el
in the eikonal phase computation, regardless of the helicity. Therefore, the classical part of the eikonal phase at tree order and the observables derived from it are independent of the massless particle's helicity, which can be understood as a manifestation of the equivalence principle.

Before closing we remark that the $M_4^{s}$ in \eqc{eq:tree4ptansatz} for $h=2$ differs from the classical gravitational Compton amplitude in eq.(B.12) of ref.\cite{Chen:2021qkk}, as the latter was further constrained to have the correct massive factorization channel residues. However for our purpose, the $\CO(G^1)$ eikonal phase, both yield the same result as we demonstrate in appendix~\ref{app:KerrWilson}.

\subsection{The $\CO(G^1)$ eikonal phase and observables} 
\begin{figure}
\centering
\begin{tikzpicture}[sines/.style={
        line width=0.7pt,
        line join=round, 
        draw=black, 
        decorate}
]
  \fill[fill=black] (0,0) circle (0.5pt);
\draw[line cap=round,rotate=-47,shift={(-1.3,0.4)}](1,1) arc (150:20:0.3 and 0.15);  
  \draw[double, line width=0.2pt,-{Latex[scale=1.1]}](0,0)--(1.4,1.43);
  \draw[-{to[slant=0.5]},line cap=round,rotate=-47,shift={(-1.3,0.4)}](1,1) arc (150:310:0.3 and 0.15); 
  \draw[-{latex},line width=0.8pt ](0,0)--(-2,-0.8);
  \draw[thick,variable=\t,domain=-5:-3.05,samples=2000,rotate=-25,shift={(2.1,-0.6)},line cap=round]
    plot (\t,,{0.2*sin(\t*360*7/pi+100)+2});
    \draw[-{latex[slant=1]},thick](-1.735,0.39)--(-1.4,0.235);
    \draw[densely dotted](-1.4,0.235)--(-0.5,-0.18);
      \draw[thin] (-0.66,-0.1)--(-0.86,-0.17);
      \draw[thin,line cap=round] (-0.86,-0.17)--(-0.66,-0.26);
   \draw[-{latex},line width=0.8pt,line cap=round](0,0)--(0,2.1);
     \draw[densely dotted] (0,2.1)--(1.4,1.43);
        \draw[thin](0.17,2)--(0.17,1.77);
      \draw[thin,line cap=round] (0.17,1.77)--(0,1.85);
      \draw[thin] (0,0.2)--(-0.18,0.13);
      \draw[thin,line cap=round] (-0.18,0.13)--(-0.18,-0.08);
       \node[scale=1] at (-2.3,-0.75) {$\vec d$};
       \node[,scale=1] at (-2,1) {$\vec k_2$};
        \node[scale=1] at (0,2.4) {$\vec a_\perp$};
        \node[scale=1] at (1.7,1.7) {$\vec a$};
  \shade[ball color = gray!40, opacity = 0.4] (0,0) circle (0.5cm);
\end{tikzpicture}
\caption{Kinematic set-up of the vectors describing spin orientation. $\vec a$ is the spin-length vector and $\vec k_2$ is the momentum of the incoming massless particle. $\vec a_\perp$ is the projected component of $\vec a$ orthogonal to $\vec k_2$. The vectors $\vec d$\,, $\vec a_\perp$ and $\vec k_2$ are respectively orthogonal with $|\vec{d}| = |\vec{a}_\perp|$.}
\end{figure}
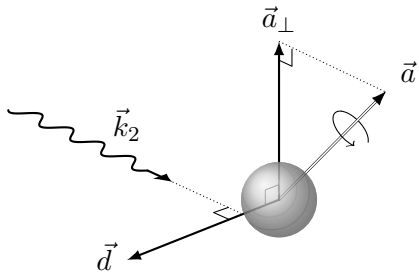

Following conventions of refs.\cite{Bern:2020buy,Kosmopoulos:2021zoq,Chen:2021qkk}, we identify the Lorentz generator $J_{\m\n}$ as the spin tensor $S_{\m\n} = - \frac{1}{m} \e_{\m\n\a\b} p_1^\a S^\b$. The exponent in the spin factor \eqc{eq:tree4ptansatz} becomes
\bl
i \frac{q^\m \bra{3} \s^\n \sket{2} S_{\m\n}}{z \bra{3} p_1 \sket{2}} = - \frac{i (n \cdot S)}{m^2 \w z} + \CO(|\vec{q}|^2, \hbar^1) \,, \label{eq:treeansatzexpred}
\el
where $n^\m = \e^{\m\a\b\g} p_{1\a} k_{2\b} q_\g$ and only the relevant term for $\CO(G^1)$ eikonal phase has been kept. Note that the apparent spurious $\bra{3} p_1 \sket{2}$ pole simply becomes $\w$ in the $q^2\rightarrow 0$ limit. Furthermore, the $\w \to 0$ limit is non-singular since $n^\m$ vanishes. Similar comment applies to our $\CO(G^2)$ analysis. The classical amplitude is then given by:
\bl
\bgd
\widetilde{M}_4|_{\mathcal{O}(G^1)} = -\frac{32 \pi G m^2 \w^2}{q^2} \oint \frac{dz}{2\pi i z} \left( \sum_j C_{\text{S}^j} z^j \right) \exp \left( - \frac{i (n \cdot S)}{m^2 \w z} \right) \,.
\egd
\el

To obtain the eikonal phase, we Fourier transform to impact parameter space, 
\bl
\bld
\delta^{(0)} (b) &= \frac{1}{4 m \w} \int \frac{d^{D-2} q}{(2\pi)^{D-2}} \left[ e^{i \vec{b} \cdot \vec{q}} \widetilde{M}_4 (q)|_{\mathcal{O}(G^1)} \right]\,,
\\ 
\eld
\el
which identifies $\del_b \Leftrightarrow i \vec{q}$. We define the following impact parameter space variables.
\bl
\bld
\vec{a} = \frac{\vec{S}}{m} \,, && \hat{k}_2 = \frac{\vec{k}_2}{\w} \,, && \vec{d} = \hat{k}_2 \times \vec{a} \,, && \vec{a}_\perp = \vec{a} - (\vec{a} \cdot \hat{k}_2) \hat{k}_2 \,.
\eld
\el
The unit vector $\hat{k}_2$ defines the direction orthogonal to the impact parameter space ($\hat{k}_2 \cdot \vec{b} = 0$), and $\vec{a}_\perp$ is the projection of the spin-length vector $\vec{a}$ onto the impact parameter plane so that $\vec{a} \cdot \vec{b} = \vec{a}_\perp \cdot \vec{b}$. Since $\vec{d} \cdot \vec{a}_\perp = 0$, the two vectors $\vec{a}_\perp$ and $\vec{d}$ span the two-dimensional impact parameter space. 
These variables are related to the spin Lorentz invariants by
\bl
\bld
- \frac{i (n \cdot S)}{m^2 \w} = \left( \hat{k}_2 \times \vec{a} \right) \cdot \left( i \vec{q} \right) &\Leftrightarrow \left( \hat{k}_2 \times \vec{a} \right) \cdot \del_{b} = \vec{d} \cdot \del_{b}
\\ - \frac{i (q \cdot S)}{m} = \vec{a} \cdot \left( i \vec{q} \right) &\Leftrightarrow \vec{a} \cdot \del_{b} = \vec{a}_\perp \cdot \del_{b}
\\ \frac{k_2 \cdot S}{m \w} &\Leftrightarrow - \hat{k}_2 \cdot \vec{a}
\eld \label{eq:spinstructuremap}
\el
The tree order eikonal phase $\delta^{(0)}$ can be obtained from the tree amplitude \eqc{eq:tree4ptansatz} by expanding \eqc{eq:eikphasedeforig} to leading coupling order,
\bl
\bld
\delta^{(0)} (b) &= \frac{1}{4 m \w} \oint \frac{dz}{2\pi i z} \left( \sum_j C_{\text{S}^j} z^j \right) \int \frac{d^{D-2} q}{(2\pi)^{D-2}} \left[ e^{i \left[ \vec{b} + ({\vec{d}}/{z}) \right] \cdot \vec{q}} \widetilde{M}^{s=0}_4 (q) \right] \,,
\eld
\el
where the order of integration has been changed, and the exponent has been simplified using \eqc{eq:treeansatzexpred}. Inserting \eqc{eq:tree4ptscalarfact} and performing the Fourier integral in $D = 4$ leads to the eikonal phase
\bl
\delta^{(0)} (b) - \delta^{(0)}_{\text{IR}} &= - \left( \frac{\k}{2} \right)^2 \frac{m \w}{2 \pi} \oint \frac{dz}{2\pi i z} \left( \sum_j C_{\text{S}^j} z^j \right) \log \left| \vec{b} + \frac{\vec{d}}{z} \right| \nn
\\ &= - 2 G m \w \left[ 2 \log b + \frac{2 \vec{d} \cdot \vec{b}}{b^2} + C_{\text{S}^2} \left( - \frac{2 (\vec{d} \cdot \vec{b})^2}{b^4} + \frac{d^2}{b^2} \right) \right. \nn
\\ &\phantom{=} \left. \phantom{asdfasdfasd}  + C_{\text{S}^3} \left( \frac{8 (\vec{d} \cdot \vec{b})^3}{3 b^6} - \frac{2 (\vec{d} \cdot \vec{b}) d^2}{b^4} \right) \right. \nn
\\ &\phantom{=} \left. \phantom{asdfasdfasd} + C_{\text{S}^4} \left( - \frac{4 (\vec{d} \cdot \vec{b})^4}{b^8} + \frac{4 (\vec{d} \cdot \vec{b})^2 d^2}{b^6} - \frac{d^4}{2 b^4} \right) + \cdots \right] \,,
\label{eq:treeeikphase}
\\ \delta^{(0)}_{\text{IR}} &= - \frac{4 G m \w}{4 - D}
\el
where we have separated the IR divergence $\delta^{(0)}_{\text{IR}}$. Setting all Wilson coefficients to Kerr value $C_{\text{S}^n} = 1$ the residue integral localises to $z=1$ and we obtain the eikonal phase eq.(5.32) of ref.\cite{Bautista:2021wfy}, reproduced below.
\bl
\delta^{(0)}_{\text{Kerr}} (b) - \delta^{(0)}_{\text{IR}}&= - 4 G m \w \log |\vec{b} + \vec{d} | \,. \nn
\el
The time delay can be computed from \eqc{eq:treeeikphase} using \eqc{eq:eik2obs1}, where the IR scale $b_0 \gg b$ was introduced as the reference so that the time delay becomes a finite quantity~\cite{Camanho:2014apa,AccettulliHuber:2020oou}
\bl
\delta t^{(0)} &= - \left( \frac{\k}{2} \right)^2 \frac{m}{2 \pi} \oint \frac{dz}{2\pi i z} \left( \sum_j C_{\text{S}^j} z^j \right) \log \left| \vec{b} + \frac{\vec{d}}{z} \right| \nn
\\ &= - 2 G m \left[ 2 \log \frac{b}{b_0} + \frac{2 \vec{d} \cdot \vec{b}}{b^2} + C_{\text{S}^2} \left( - \frac{2 (\vec{d} \cdot \vec{b})^2}{b^4} + \frac{d^2}{b^2} \right) \right. \nn
\\ &\phantom{=} \left. \phantom{asdfasdfasd}  + C_{\text{S}^3} \left( \frac{8 (\vec{d} \cdot \vec{b})^3}{3 b^6} - \frac{2 (\vec{d} \cdot \vec{b}) d^2}{b^4} \right) \right. \nn
\\ &\phantom{=} \left. \phantom{asdfasdfasd} + C_{\text{S}^4} \left( - \frac{4 (\vec{d} \cdot \vec{b})^4}{b^8} + \frac{4 (\vec{d} \cdot \vec{b})^2 d^2}{b^6} - \frac{d^4}{2 b^4} \right) + \cdots \right] \,.
\label{eq:treeetd}
\el
The conventional scattering angle is not well-defined in presence of spin, as the motion is no longer restricted to lie on the orbital plane. One way to proceed is to define the scattering angle $\th$ as a vectorial quantity $\vec{\th}$ that is directed along the direction of the impulse~\cite{Kim:2020cvf}. The impulse for the massless probe $\vec{q}$ can be computed from the eikonal via
\bl
\vec{q} &= \nabla_b \delta (b) \,,
\el
therefore the vectorial scattering angle $\vec{\th}$ is given as
\bl
\vec{\th} := \frac{\nabla_b \bar{\delta} (b,\w)}{|\vec{k}_2|}  = \frac{1}{\w} \nabla_b \bar{\delta} (b,\w) \,,
\el
since the size of the incoming momentum $\vec{k}_2$ is $\w$. The vectorial scattering angle is
\bl
\vec\th^{(0)} &= - \left( \frac{\k}{2} \right)^2 \frac{m}{2 \pi} \oint \frac{dz}{2\pi i z} \left( \sum_j C_{\text{S}^j} z^j \right) \frac{ \vec{b} + \vec{d}/z }{|\vec{b} + \vec{d}/z|^2} \nn
\\ &= - 4 G m \left[ \frac{\vec{b}}{b^2} + \left( \frac{\vec{d}}{b^2} - \frac{2 (\vec{d} \cdot \vec{b}) }{b^4} \vec{b} \right) - C_{\text{S}^2} \left( \frac{2 (\vec{d} \cdot \vec{b})^2}{b^4} \vec{d} + \frac{b^2 d^2 - 4 (\vec{d} \cdot \vec{b})^2}{b^6} \vec{b} \right) \right. \nn
\\ &\phantom{=} \left. \phantom{asdfasdf} + C_{\text{S}^3} \left( \frac{4(\vec{d} \cdot \vec{b})^2 - b^2 d^2}{b^6} \vec{d} + \frac{4 (\vec{d} \cdot \vec{b}) (b^2 d^2 - 2 (\vec{d} \cdot \vec{b})^2)}{b^8} \vec{b} \right) \right. \nn
\\ &\phantom{=} \left. \phantom{asdfasdf} + C_{\text{S}^4} \left( \frac{4 (\vec{d} \cdot \vec{b}) (b^2 d^2 - 2 (\vec{d} \cdot \vec{b})^2)}{b^8} \vec{d} \right. \right. \nn
\\ &\phantom{=} \left. \left. \phantom{asdfasdfasdfasdfasdf} + \frac{b^4 d^4 - 12 b^2 d^2 (\vec{d} \cdot \vec{b})^2 + 16 (\vec{d} \cdot \vec{b})^4}{b^{10}} \vec{b} \right) + \cdots \right] \,.
\label{eq:TreeScAng}
\el
Calculation of $\CO(G^2)$ scattering angle requires a modified equation suggested in ref.\cite{Bern:2020buy} as eq.(7.15), but since the scattering angle is not the main focus of this paper the calculations will be omitted.

\section{The $\CO(G^2)$ eikonal phase and observables} \label{sec:2PMeik}
\begin{figure}
\centering
\begin{tikzpicture}[line width=1. pt, scale=1.75,
sines/.style={
        line width=0.7pt,
        line join=round, 
        draw=black, 
        decorate, 
        decoration={complete sines, number of sines=4, amplitude=4pt}},
        sines2/.style={
        line width=1pt,
        line join=round, 
        draw=black, 
        decorate, 
        decoration={complete sines, number of sines=7, amplitude=4pt}},
        sines3/.style={
        line width=0.7pt,
        line join=round, 
        draw=black, 
        decorate, 
        decoration={complete sines, number of sines=7, amplitude=4pt}}
]
\def\AA{3};
\def\AAA{6};
\def\AAAA{9};
%
%
\draw[black, postaction={decorate},line width=2pt] (-0.5+\AA,0) -- (1.5+\AA,0);
\draw[white,postaction={sines2}](-0.5+\AA,1) -- (1.5+\AA,1);
\draw[white,postaction={sines,double}] (0+\AA,0) -- (0.5+\AA,1);
\draw[white,postaction={sines,double}] (1+\AA,0) -- (0.5+\AA,1);
\filldraw[color=black, fill=gray!70,thin](0+\AA,0) circle[radius=0.1] ;
\filldraw[color=black, fill=gray!70,thin](1+\AA,0) circle[radius=0.1] ;
\node[scale=1] at (-0.7+\AA,0) {$1$};
\node[scale=1] at (-0.7+\AA,1) {$2$};
\node[scale=1] at (1.7+\AA,1) {$3$};
\node[scale=1] at (1.7+\AA,0) {$4$};
\node[scale=1] at (-0.1+\AA,0.35) {$\ell$};
\draw[-stealth,color=gray,line width=0.8pt] (0.05+\AA,0.4)--(-0.1+\AA,0.15);
\node[ellipse,fill=gray!70,  thin,draw = black,
    minimum width = 0.7cm, 
    minimum height = 0.5cm] (e) at (0.5+\AA,1) {};
\draw[white,line width=3pt] (0.3+\AA,0.4)--(0.1+\AA,0.5);
\draw[white,line width=3pt](0.7+\AA,0.4)--(0.9+\AA,0.5);
\draw[dotted,color=red]  (0.3+\AA,0.4)--(0.1+\AA,0.5);
\draw[dotted,color=red] (0.7+\AA,0.4)--(0.9+\AA,0.5);
\draw[white,line width=3pt] (0.5+\AA,-0.1)--(0.5+\AA,0.1);    
\draw[dotted,color=red] (0.5+\AA,-0.1)--(0.5+\AA,0.1);
\node[scale=1] at (0.5+\AA,-0.5) {(a) triangle};
\draw[black, postaction={decorate},line width=2pt] (-0.5+\AAA,0) -- (1.5+\AAA,0);
\draw[white,postaction={sines2}](-0.5+\AAA,1) -- (1.5+\AAA,1);
\draw[white,postaction={sines,double}] (0+\AAA,0) -- (0+\AAA,1);
\draw[white,postaction={sines,double}] (1+\AAA,0) -- (1+\AAA,1);
\filldraw[color=black, fill=gray!70,  thin](0+\AAA,0) circle[radius=0.1] ;
\filldraw[color=black, fill=gray!70,  thin](1+\AAA,0) circle[radius=0.1] ;
\filldraw[color=black, fill=gray!70,  thin](1+\AAA,1) circle[radius=0.1] ;
\filldraw[color=black, fill=gray!70,  thin](0+\AAA,1) circle[radius=0.1] ;
\node[scale=1] at (-0.7+\AAA,0) {$1$};
\node[scale=1] at (-0.7+\AAA,1) {$2$};
\node[scale=1] at (1.7+\AAA,1) {$3$};
\node[scale=1] at (1.7+\AAA,0) {$4$};
\node[scale=1] at (-0.25+\AAA,0.3) {$\ell$};
\draw[-stealth,color=gray,line width=0.8pt] (-0.15+\AAA,0.4)--(-0.15+\AAA,0.1);
\draw[white,line width=3pt]  (-0.1+\AAA,0.5)--(0.1+\AAA,0.5);    
\draw[white,line width=3pt] (0.9+\AAA,0.5)--(1.1+\AAA,0.5);    
\draw[dotted,color=red] (-0.1+\AAA,0.5)--(0.1+\AAA,0.5);
\draw[dotted,color=red] (0.9+\AAA,0.5)--(1.1+\AAA,0.5);
\draw[white,line width=3pt] (0.5+\AAA,-0.1)--(0.5+\AAA,0.1);    
\draw[dotted,color=red] (0.5+\AAA,-0.1)--(0.5+\AAA,0.1);
\draw[white,line width=3pt] (0.5+\AAA,0.9)--(0.5+\AAA,1.1);    
\draw[dotted,color=red] (0.5+\AAA,0.9)--(0.5+\AAA,1.1);
\node[scale=1] at (0.5+\AAA,-0.5) {(b) box};
\draw[black, postaction={decorate},line width=2pt] (-0.5+\AAAA,0) -- (1.5+\AAAA,0);
\draw[white,postaction={sines2}](-0.5+\AAAA,1) -- (1.5+\AAAA,1);
\draw[white,postaction={sines3,double}] (0+\AAAA,0) -- (1+\AAAA,1);
\draw[white,postaction={sines3,double}] (1+\AAAA,0) -- (0+\AAAA,1);
\filldraw[color=black, fill=gray!70,  thin](0+\AAAA,0) circle[radius=0.1] ;
\filldraw[color=black, fill=gray!70,  thin](1+\AAAA,0) circle[radius=0.1] ;
\filldraw[color=black, fill=gray!70,  thin](1+\AAAA,1) circle[radius=0.1] ;
\filldraw[color=black, fill=gray!70,  thin](0+\AAAA,1) circle[radius=0.1] ;
\node[scale=1] at (-0.7+\AAAA,0) {$1$};
\node[scale=1] at (-0.7+\AAAA,1) {$2$};
\node[scale=1] at (1.7+\AAAA,1) {$3$};
\node[scale=1] at (1.7+\AAAA,0) {$4$};
\node[scale=1] at (0.1+\AAAA,0.45) {$\ell$};
\draw[-stealth,color=gray,line width=0.8pt] (0.2+\AAAA,0.4)--(0+\AAAA,0.2);
\draw[white,line width=3pt] (0.77+\AAAA,0.63)--(0.63+\AAAA,0.77);    
\draw[white,line width=3pt](0.63+\AAAA,0.23)--(0.77+\AAAA,0.37);    
\draw[dotted,color=red] (0.77+\AAAA,0.63)--(0.63+\AAAA,0.77);
\draw[dotted,color=red] (0.63+\AAAA,0.23)--(0.77+\AAAA,0.37);
\draw[white,line width=3pt] (0.5+\AAAA,-0.1)--(0.5+\AAAA,0.1);    
\draw[dotted,color=red](0.5+\AAAA,-0.1)--(0.5+\AAAA,0.1);
\draw[white,line width=3pt]  (0.5+\AAAA,0.9)--(0.5+\AAAA,1.1);    
\draw[dotted,color=red] (0.5+\AAAA,0.9)--(0.5+\AAAA,1.1);
\node[scale=1] at (0.5+\AAAA,-0.5) {(c) crossed box};
\end{tikzpicture}
\caption{The cut diagrams relevant to classical physics. The thick sold lines, wavy lines and double wavy lines represent massive spinning bodies, massless particles and gravitons, respectively. The cut propagators are labeled by the red dotted lines. $1$ and $2$ are ingoing, $3$ and $4$ are outgoing, and $\ell$ labels the loop momentum.   }\label{cutdiag}
\end{figure}
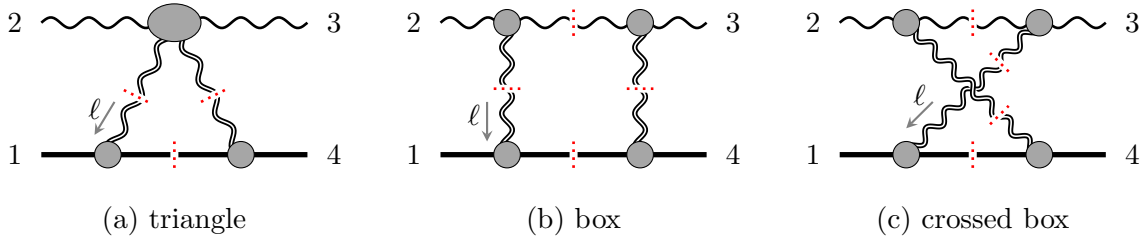

In this section, we compute the $\CO(G^2)$ eikonal phase for massless helicity states scattering of a general spinning body. Expanding LHS of \eqc{eq:eikphasedeforig} to $\CO(G^2)$ yields two terms.
\bl
\left. e^{i \delta} - 1 \right|_{\CO(G^2)} = i \delta^{(1)} - \frac{\left[ \delta^{(0)} \right]^2}{2} \,.
\el
The matching terms on the RHS of \eqc{eq:eikphasedeforig} is the one-loop ($\CO(G^2)$) amplitude, which can be expanded on a basis of master scalar integrals, whose coefficients can be computed via generalized unitarity methods~\cite{Bern:1994zx, Bern:1994cg, Bern:1997sc, Forde:2007mi,Kilgore:2007qr}. Only the scalar triangle integral with one massive propagator $\CI_\triangle$ and the box integrals $\CI_{\square}$ and $\CI_{\cSquare}$ are relevant for the classical limit~\cite{Holstein:2004dn}
\bl
\left. \widetilde{M}_4 \right|_{\CO(G^2)} = \left[c_\triangle\, \CI_{\triangle} +  c_{\square}\, \CI_{\square} + c_{\cSquare}\, \CI_{\cSquare} \right]\Big|_{\hbar\rightarrow 0} \,.
\el
The other triangle topology, which requires the Compton amplitude for evaluation, only contains massless propagators and become irrelevant in the classical limit. Thus, the kinematical setup allows us to compute higher spin orders without encountering any spurious singularities. The contributions from the box integrals, $c_{\square} \CI_{\square} + c_{\cSquare} \CI_{\cSquare}$, are known to reproduce the iteration terms of the eikonal phase, $- [ \delta^{(0)} ]^2 / 2$~\cite{Akhoury:2013yua,KoemansCollado:2019ggb}. Therefore for computing the one-loop eikonal phase $\delta^{(1)}$ it suffices to compute the triangle coefficient $c_\triangle$. However, to check consistency, we will also compute the box and cross box coefficients and study exponentiation. Interestingly, the exponentiation is manifest \emph{only if} we assume black hole values for the spin coupling.

We will give the leading $(\l / b)$ corrections from the triangle coefficient as an expansion in the spin $a/b$. This will capture the classical gravitational Faraday effect, and reproduce the linear in spin term  computed via null geodesic parallel transport computation in ref.\cite{Ishihara:1987dv}. We will give the subleading in $a/b$ corrections up to $a^{14}$. The $\CO(G^2)$ eikonal phase also provides a consistency test to the proposed universal spin-tensor structure conjectured in ref.\cite{Aoude:2022trd}  (and equivalently shift symmetry in ref.\cite{Bern:2022kto}) in the black hole limit for the scattering phase.

\subsection{Extraction of scalar integral coefficients}
It will be more convenient to use the triple cut of the loop integrals as the seed to compute the triangle, box and crossed box coefficients. The triple cut can be used to extract the triangle coefficient and, by imposing one more cut condition, can lead to quadruple-cuts to generate the box and crossed box coefficients. In our case, the triple-cut is formed by gluing a massless-massless-graviton-graviton amplitude to a product of two three-point graviton-massive-massive amplitudes together. 

Here we list the relevant tree amplitudes with the ingoing momenta of $1$ and $2$ and outgoing momenta $3$ and $4$. The explicit form of the massless-massless-graviton-graviton amplitude is 
\eqa
M_4^{EH}(1^{+h},2^{+2},3^{+2},4^{+h})&=&\left(-\frac{[ 24]}{[12]}\right)^{4-2h}\frac{[12]^4\langle 34\rangle^4}{s t u}\,,
\eqae
where $\pm h$ are the helicities of the massless particles. The product of two three-point graviton-massive-massive amplitudes can be found in ref.\cite{Chen:2021qkk},
\eqa
C_{4}(1^s,2^+,3^+,4^s)&\equiv& M_{3}(1^s,2^+,P)M_{3}(-P,3^+,4^s)=-\frac{[ 2|p_1|3\rangle^4}{t^2}\mathcal  S_{\{++\}}(1,2,3,4)\,,\\
C_{4}(1^s,2^-,3^+,4^s)&\equiv&M_{3}(1^s,2^-,P)M_{3}(-P,3^+,4^s)= - \frac{ \langle 23\rangle^4 }{t^2}  \mathcal S_{\{-+\}}(1,2,3,4)\,,
\eqae
where $P=p_1+k_2$ and 
\eqa
\label{spinf}\mathcal  S_{\{++\}}(1,2,3,4)&=& \oint \prod_{k=1,2} \frac{dz_k}{2\pi i z_k} \left( \sum_m C_{\text{S}^m} z_k^m \right)e^{  -i \left(\frac{v_1}{z_1}+\frac{v_2}{z_2} \right)  } \,,\\
\mathcal S_{\{-+\}}(1,2,3,4)&=&\oint \prod_{k=1,2} \frac{dz_k}{2\pi i z_k} \left( \sum_m C_{\text{S}^m} z_k^m \right)e^{- i\left(\frac{v_1}{z_1}-\frac{v_2^*}{z_2}+\frac{v_3}{z_1z_2}\right)}  \,,\\[2mm]
\notag v_1=-i (k_2\cdot S)\,,&&v_2=-\frac{i\left[(2m^2-t)(k_2\cdot S)+2 m^2(q\cdot S)+2 i(n\cdot S)\right]}{2 m^2}\,,\\
\notag &&
\hspace{1cm}v_3=\frac{-it (k_2\cdot S)+2 (n\cdot S)}{2m^2}\,,
\eqae
$v_2^*\,$ is the complex conjugate of $v_2$ and the irrelevant quantum contributions from $f(x)$ factor in ref.\cite{Chen:2021qkk} were omitted. The $\CO(S^n)$ contribution is obtained by expanding the exponential to $\CO(S^n)$, and then performing the auxiliary $z_k$ integrals.\par
According to the triple-cut diagram shown in fig.\ref{cutdiag}, the cut conditions are
\eqa
\label{triplec}\ell^2=\ell\cdot p_1=0\,,~~~2\ell\cdot q=-q^2\,.
\eqae
Also, the triple-cut involves two types of cuts, depending on whether the two cut gravitons have equal helicity. For external helicity $h<2$ only the opposite helicity cut is non-vanishing, while for gravitons, the equal helicity is non-vanishing for the helicity non-preserving case: 
\eqa \label{eq:cutintegrand}
\notag C_{\{++\}}^{\{h\}}&=&\left.\sum_{h'=\pm 2} C_4(1^{s_a},\ell^{+h'},-\ell_1^{+h'},4^{s_a})M_4^{EH}(2^{+h},\ell_1^{-h'},-\ell^{-h'},3^{+h})\right|_{\mathrm{eq.}\eqref{triplec}}\\
\notag &=&\left.\frac{
\left[-\mathrm{tr}_- \left(k_3 \ell p_1 \ell_1\right)\right]^{4-2h}}{t^3(2\ell_1\cdot k_3)(2\ell\cdot k_3)}
\left[\frac{\mathrm{tr}_-\left(k_2 \ell p_1 \ell_1 k_3 p_1\right)}{ [ 3|p_1|2\rangle}\right]^{2h}\mathcal S_{\{++\}} (1,\ell,-\ell_1,4)
\right.\\[-2mm]
\label{cpm}&&\hspace{8.6cm}
+\left(\ell\,\longleftrightarrow\,\ell_1\right)\bigg|_{\mathrm{eq.}\eqref{triplec}}
\,,~~~~~~\\[1mm]
\notag C_{\{-+\}}^{\{h= 2\}}&=&\sum_{h'=\pm 2}  C_4 (1^{s_a},\ell_{\,}^{-h'},-\ell_1^{+h'},4^{s_a})M_4^{EH}(2^{-2},\ell_1^{-h'},-\ell_{\,}^{+h'},3^{+2})\bigg|_{\mathrm{eq.}\eqref{triplec}}\\
&=&\frac{m^4 \langle 23\rangle^4 t }{(2\ell\cdot k_3)(2\ell_1\cdot k_3)}\mathcal S_{\{-+\}} (1,\ell,-\ell_1,4)+\left(\,\ell\,\longleftrightarrow\,\ell_1\,\right)\bigg|_{\mathrm{eq.}\eqref{triplec}}\,,\\
C_{\{-+\}}^{\{h\neq 2\}}&=&0\,,
\eqae
where $\ell_1=-\ell-q$, $\mathrm{tr}_-(\dots)=\frac{1}{2}\mathrm{tr}[(1-\gamma_5)\dots]$ and the subscripts denote the sign of $\{h_2, h_3\}$. As mentioned before, we will use these triple-cuts to compute all the relevant integral coefficients in the following subsections.

\subsubsection{Extraction of box and crossed box coefficients}
The box coefficient can be computed from the quadruple cut using unitarity methods~\cite{Bern:1994zx, Bern:1994cg, Bern:1997sc, Forde:2007mi,Kilgore:2007qr}. We can impose one more cut condition (see fig.\ref{cutdiag}) $\ell\cdot k_2=0$ to get the box cut or $\ell\cdot k_3=0$ to get the crossed box cut. For each quadruple cut, there are two solutions to loop momentum which solve the cut conditions. We use the loop momentum parametrisation of ref.\cite{Bern:2020buy} to compute the quadruple cut, with small modifications to account for masslessness of $k_2^\m$. 
\bl
\bld
\ell^\m &= \a p_1^\m + \b k_2^\m + \g q^\m + \delta \eta^\m \,,
\\ \eta^\m &= \bra{p_1^\flat} \s^\m \sket{k_2} \,,\, p_1^\flat = p_1 - \frac{m}{2\w} k_2 \,.
\eld
\el
$p_1^\flat$ is a null vector computed using Mandelstam parametrisations \eqc{eq:MandelstamDef}. The solutions to the cut conditions,
\bl
\ell^2 = p_1 \cdot \ell = k_2 \cdot \ell = 0 \,,\, q \cdot \ell = - \frac{q^2}{2} \,,
\el
are given as
\bl
\label{boxsol}\ell_{-}^\m = - \frac{q^2}{2 q \cdot \eta} \eta^\m \,,\, \ell_{+}^\m = \frac{- (2 q^2 \w) p_1^\m + 2 q^2 (m + \w) k_2^\m - 4 m \w^2 q^\m}{m q^2 + 2 q^2 \w + 4 m \w^2} + \frac{q^2}{2 q \cdot \eta} \eta^\m \,.
\el
The coefficient of the box integral can be computed as an average over the two loop momentum solutions $\ell_\pm$\,, 
\bl
\label{boxcoef} c_{\square,+\pm}^{\{h\}} &= 
\frac{1}{2} \sum_{\ell=\ell_\pm} \Big[ 
C_{\square}^{+\pm}(\ell,-\ell_1)\mathcal S_{\{+\pm\}} (1,\ell,-\ell_1,4)+C_{\square}^{+\pm}(\ell_1,-\ell)\mathcal S_{\{+\pm\}} (1,\ell_1,-\ell,4)\Big]_{\ell\cdot k_2=0} \,,
\el
where $C_{\square}^{\pm+}$ is the part of $[C_{\{\pm+\}}(-2\ell\cdot k_2)]_{\ell\cdot k_2=0}$ without the spin factor $\mathcal S_{\{\pm+\}}$. The (helicity factor stripped) crossed box coefficient can be obtained from that of the box by the substitution $k_2 \to - k_3$, $k_3 \to - k_2$, $h_2 \to - h_3$, and $h_3 \to - h_2$. For example,
\eqa
\label{crossr}\left.\frac{c_{\square,++}^{\{h\}}}{(\langle 3|p_1|2])^h}\right|_{\pm k_2\leftrightarrow \mp k_3}=\frac{c_{\cSquare,--}^{\{h\}}}{(\langle 2|p_1|3])^h}\,.
\eqae 
Similar relations hold for other helicity sectors such as $c_{\square,+-}^{\{h\}}$ and $c_{\cSquare,++}^{\{h\}}$.

\subsubsection{Extraction of triangle coefficients}
For the triple-cut as shown in fig.\ref{cutdiag},  the cut-satisfied loop momentum can be parametrized as 
\eqa
\label{lsol}\ell^\mu_\pm=-\left(\frac{t}{t-4m^2}\right) p_1^\mu+\left(\frac{2m^2}{t-4m^2}\right) q^\mu+ z P_\pm^\mu +\left[\frac{m^2 t}{4\gamma_\pm(t-4m^2)}\right]\frac{Q_\pm^\mu}{z}\,,
\eqae
where the $\pm$ labels two independent solutions for the cut conditions and 
\eqa
\gamma_\pm=\frac{1}{2}\left(-t\pm\sqrt{t(t-4m^2)}\right)\,.
\eqae
The explicit forms of $P$ and $Q$ are not important but they satisfy the following relations
\eqa
P_\pm^2=Q_\pm^2=P_\pm\cdot p_1=P_\pm\cdot q=Q_\pm\cdot p_1=Q_\pm\cdot q=0\,,~~~ P_\pm\cdot Q_\pm=2 \gamma_\pm\,,
\eqae
which imply a useful formula for the loop computation,
\eqa
\notag \hspace{-0.1cm}P_\pm^\mu Q_\pm^\nu&=&\gamma_\pm \eta^{\mu\nu}+ \frac{4\gamma_\pm}{t-4m^2} \left[p_1^\mu p_1^\nu+\frac{m^2q^\mu q^\nu}{t}+\frac{p_1^\mu q^\nu+q^\mu p_1^\nu}{2}-\frac{i (2m^2+\gamma_\pm)}{2\gamma_\pm}\epsilon^{\mu\nu\rho\sigma}(p_1)_\rho q_\sigma\right]\,.\\
\eqae
By averaging the residues of the triple-cut with loop momentum solutions eq.\eqref{lsol} at $z=\infty$, we can obtain the triangle coefficients $c_{\{+,\pm\}}^{\{h\}}$, formally as, 
\eqa
c_{\{+,\pm\}}^{\{h\}} =  \frac{1}{2}{\rm Res}_{z=\infty}\left[\sum_{\ell=\ell_\pm} C_{\{+\pm\}}^{\{h\}}(\ell)\right]\,.
\eqae
Note that for the helicity preserving case, we can rewrite eq.\eqref{cpm} as
\eqa
\notag C_{\{+,+\}}^{\{h\}}&=&-\left(\frac{[ 2 \ell]}{[ 3 \ell]}\right)^{2h}\frac{\mathrm{tr}\left[(1-\gamma_5)k_3 \ell p_1 \ell_1\right]^4 S_{\{++\}}(1,\ell,-\ell_1,4)}{t^3(2\ell\cdot k_3)(2\ell_1\cdot k_3)}\bigg|_{\mathrm{eq.}\eqref{triplec}}+(\ell\longleftrightarrow\ell_1) \,.\\
\eqae
In the leading order in $(q/\omega)$-expansion, $[ 2 \ell]\simeq[ 3 \ell]$, so $C_{\{+,+\}}^{\{h\}}$ is independent of $|h|$ and this can be understood as a manifestation of the equivalence principle. However, in the subleading order $C_{\{+,+\}}^{\{h\}}$ is no longer $|h|$ independent and the equivalence principle ceases to apply.

\subsection{Eikonal phase and its exponentiation}

The main contribution to tree order eikonal phase matrix comes from helicity preserving processes, therefore studying $c_{\square}^{++}$ is enough for checking exponentiation. The box coefficient for the helicity non-preserving case $c_{\square}^{+-}$ either vanishes ($|h|<2$) or is suppressed by $\CO((q/\w)^4)$ compared to $c_{\square}^{++}$ ($|h|=2$) which is beyond the regime where exponentiation holds.



An interesting behaviour of the quadruple cut \eqc{boxcoef} is that only one helicity configuration contributes to each loop momentum solution $\ell_+$ and $\ell_-$, thus the sum over all cut solutions and intermediate massless states becomes a sum over two terms 
\eqa
 c_{\square,++}^{\{h\}} &=& 
\notag\frac{1}{2}\Big[ 
C_{\square}^{++}(\ell_+,-\ell_{1+})\mathcal S_{\{++\}} (1,\ell_+,-\ell_{1+},4)\\
\label{eq:box_hel_cons}&&\hspace{3cm}+C_{\square}^{++}(\ell_{1-},-\ell_-)\mathcal S_{\{++\}} (1,\ell_{1-},-\ell_{-},4)\Big]_{\ell\cdot k_2=0} \,,~~~
\eqae
where the spin factors are
\bl
\CS_{\{++\}} (\ell_\pm) &= \oint \prod_{k = 1,2} \left[ \frac{dz_k}{2\pi i z_k} \left( \sum_j C_{\text{S}^j} z_k^j \right) \right] e^{i b_\pm^{\m\n} S_{\m\n}} \,,
\el
where $\CS_{\{++\}} (\ell_+)$ and $\CS_{\{++\}} (\ell_-)$ refer to the spin factors appearing in the first and second line of \eqref{eq:box_hel_cons} respectively. 
Note that the \emph{spin factors $\CS_{\{++\}} (\ell_\pm)$ are distinct}.
By using the loop momentum solutions \eqref{boxsol} and the explicit form of spin factor \eqref{spinf} subjected to the cut conditions 
 the exponent of the spin factor can be reorganised as
\bl
i b_\pm^{\m\n} S_{\m\n} &= - \left( \frac{1}{z_1} + \frac{1}{z_2} \right) \frac{i (n \cdot S)}{2 m^2 \w} \mp \left( \frac{1}{z_1} - \frac{1}{z_2} \right) \frac{(q \cdot S)}{2 m} + \CO(|\vec{q}|^2, \hbar^1) \,, \label{eq:box_spfactor}
\el
where the subleading classical terms of order $\CO(|\vec{q}|^2 \hbar^0)$ have been omitted, since they do not contribute to exponentiation. However, the distinct spin factors can be reorganised into a common factor that can be pulled out from the sum \eqc{eq:box_hel_cons}, due to the symmetry of the auxiliary integrals in $z_k$; $b_{+}^{\m\n}$ becomes $b_-^{\m\n}$ to the order considered in \eqc{eq:box_spfactor}, when exchanging the auxiliary variables $z_1 \leftrightarrow z_2$ for the cut solution $\ell_+$. The full box coefficient is now a product of the common spin factor and the scalar factor $c_{\square \,,\, (s=0)}^{++}$, which is the box integral coefficient for the massive scalar-massless scattering.
\bl
\bld
c_{\square,++}^{\{h\}} &= \oint \prod_{k = 1,2} \left[ \frac{dz_k}{2\pi i z_k} \left( \sum_j C_{\text{S}^j} z_k^j \right) \right] e^{i b^{\m\n} S_{\m\n}} \times c_{\square \,,\, (s=0)}^{++} \left[ 1 + \CO(|\vec{q}|^2, \hbar^1) \right] \,,
\\ c_{\square \,,\, (s=0)}^{++} &= \frac{C_{\square}^{++}(\ell_+,-\ell_{1+}) + C_{\square}^{++}(\ell_{1-},-\ell_-) }{2} \,,
\eld \label{eq:boxcoeff_spsc_factor}
\el
where we have omitted the subscript in $b_-^{\m\n} = b^{\m\n}$ for brevity. Up to the considered order, the spin factor $e^{i b^{\m\n} S_{\m\n}}$ is the same spin factor \eqc{eq:gravCompSFexpred} of the gravitational Compton amplitude \eqc{eq:gravComptonSF}.

Applying the rules \eqref{crossr} to \eqc{eq:boxcoeff_spsc_factor} yields the factorised form of the crossed box coefficient
\bl
\bld
c_{\cSquare}^{++} &= \oint \prod_{k = 1,2} \left[ \frac{dz_k}{2\pi i z_k} \left( \sum_j C_{\text{S}^j} z_k^j \right) \right] e^{i b^{\m\n} S_{\m\n}} \times c_{\cSquare \,,\, (s=0)}^{++} \left[ 1 + \CO(|\vec{q}|^2, \hbar^1) \right] \,.
\eld \label{eq:cboxcoeff_spsc_factor}
\el
The box coefficient and the crossed box coefficient for the massive scalar are equal in the classical limit, i.e.
\bl
c_{\cSquare \,,\, (s=0)}^{++} = c_{\square \,,\, (s=0)}^{++} \left[ 1 + \CO(\hbar^1) \right] \,,\, c_{\square \,,\, (s=0)}^{++} = 
16 (\k/2)^4 m^4 \w^4 \left[ 1 + \CO(|\vec{q}|^2, \hbar^1) \right] \,,
\el
which means the sum of \eqc{eq:boxcoeff_spsc_factor} and \eqc{eq:cboxcoeff_spsc_factor}, the full box topology contribution, also factorises into the spin factor and the scalar factor,
\bl
\bld
\left[ c_{\square}^{++} \CI_{\square} + c_{\cSquare}^{++} \CI_{\cSquare} \right] &= \oint \prod_{k = 1,2} \left[ \frac{dz_k}{2\pi i z_k} \left( \sum_j C_{\text{S}^j} z_k^j \right) \right] e^{i b^{\m\n} S_{\m\n}} \times c_{\square \,,\, (s=0)}^{++} \left[ \CI_{\square} + \CI_{\cSquare} \right] \,,
\eld \label{eq:BoxEikonal}
\el
up to subleading $\CO(|\vec{q}|^2, \hbar^1)$ terms, which are irrelevant for exponentiation. The validity of approximating box integral contributions by \eqc{eq:BoxEikonal} has been explicitly checked up to $\CO (S^4)$ order; the difference between the classical box and the classical crossed box coefficient for a specified spin sector appears at quantum subleading order
\bl
c_{\cSquare}^{++} &= c_{\square}^{++} \times \left[ 1 + \CO \left( \hbar^1 \right) \right] \,.
\el
The difference between $c_{\square}^{++}$ and $c_{\square}^{--}$, which signifies breaking of exponentiation as $\a_3 = 0$ at tree order, appears at $\CO((q/\w)^3)$ subleading order.
\bl
c_{\square}^{--} &= c_{\square}^{++} \times \left[ 1 + \CO \left( (q/\w)^3 , \hbar^1 \right) \right] \,.
\el
The difference term should be added as corrections to one-loop eikonal phase, corresponding to NLO for $\alpha$ (contributions to $\alpha^{(1,1)}$). Moreover, the $\CO(|\vec{q}|^2 \hbar^0)$ subleading classical terms of the spin factor \eqc{eq:box_spfactor} also results in terms not captured by exponentiation. These terms yield NLO corrections to $\bar{\delta}$ (contributions to $\bar{\delta}^{(1,1)}$).

In impact parameter space, \eqc{eq:BoxEikonal} scales as $\propto \w^2$ and provides the iteration term of tree eikonal phase~\cite{AccettulliHuber:2020oou}. In particular the non-spinning part can be matched to the square of the tree-eikonal phase:
\bl
c_{\square \,,\, (s=0)}^{++} \left[ \CI_{\square} + \CI_{\cSquare} \right] \Leftrightarrow - \frac{[\delta_{\text{S}^0}^{(0)} (b)]^2}{2} \,.
\el
The full spinning contribution in impact parameter space can be written as
\bl
\left[ c_{\square}^{++} \CI_{\square} + c_{\cSquare}^{++} \CI_{\cSquare} \right] &\Leftrightarrow \oint \prod_{k = 1,2} \left[ \frac{dz_k}{2\pi i z_k} \left( \sum_j C_{\text{S}^j} z_k^j \right) \right] e^{i b^{\m\n} S_{\m\n}} \times \left[ - \frac{[\delta_{\text{S}^0}^{(0)} (b)]^2}{2} \right] \,,
\el
where the spin factor exponent is a derivative operator
\bl
i b^{\m\n} S_{\m\n} &= \left( \frac{1}{z_1} + \frac{1}{z_2} \right) \frac{\vec{d} \cdot \nabla_b}{2} + \left( \frac{1}{z_1} - \frac{1}{z_2} \right) \frac{i ( \vec{a}_\perp \cdot \nabla_b)}{2} \,,
\el
acting on the ``square'' of non-spinning tree eikonal phase. Computation of $\CO(S^n)$ terms follow the same procedure of computing the box coefficients; the exponential is expanded to $\CO(S^n)$, the $z_k$ auxiliary integrals are performed, and then the derivative operators $\nabla_b$ are evaluated.

Note that for general Wilson coefficients the resulting expression will be different from ``squaring'' the tree eikonal phase \eqc{eq:treeeikphase}. For example, the finite part is different for the $\CO(S^2)$ term
\bl
\left[ c_{\square}^{++} \CI_{\square} + c_{\cSquare}^{++} \CI_{\cSquare} + \frac{[\delta^{(0)}(b)]^2}{2} \right]_{\CO(S^2)} &= \frac{4 G^2 m^2 \w^2 ((\vec{a}_{\perp}\cdot \vec{b})^2 - (\vec{d} \cdot \vec{b})^2)}{b^4} (C_{\text{S}^2} - 1) \,.
\el
where for the LHS we've taken the quadratic in spin part of the difference between the box contribution and the square of tree eikonal phase. 

Such mismatch similarly occurs for higher spin orders. In the Kerr limit $C_{\text{S}^j} = 1$ the spin factor simplifies to
\bl
\oint \prod_{k = 1,2} \left[ \frac{dz_k}{2\pi i z_k} \left( \sum_j C_{\text{S}^j} z_k^j \right) \right] e^{i b^{\m\n} S_{\m\n}} \stackrel{C_{\text{S}^j} \to 1}{\rightarrow} \exp \left( - \frac{i (n \cdot S)}{m^2 \w} \right) \Leftrightarrow \exp \left( \vec{d} \cdot \nabla_b \right) \,,
\el
which is a constant shift by $\vec{d} = \hat{k}_2 \times \vec{a}$ in impact parameter space. This is consistent with the fact that the tree eikonal phase for Kerr is obtained from the spinless case by a constant shift in impact parameter space~\cite{Bautista:2021wfy},
\bl
\delta^{(0)}_{\text{Kerr}} (\vec{b}) &= \delta^{(0)}_{\text{Scalar}} (\vec{b} + \vec{d}) \,,
\el
and exponentiation becomes manifest. The fact that the exponentiation of the eikonal phase correctly captures the eikonal limit of the amplitude only for black hole, or equivalently minimal coupling,  is reminiscent of similar statement for $g=2$ in exponentiation of electromagnetic scattering~\cite{Cheng:1971gf,Meng:1972xt,Weinberg:1971cdi,Czyz:1975bf}.

\subsection{Results}
We are interested in the helicity-preserving parts of the eikonal phase in the classical limit, which we decompose as
\bl
\bar{\delta} &:= \frac{\delta_{+,+} + \delta_{-,-}}{2} \,,\, \a := \frac{\delta_{-,-} - \delta_{+,+}}{2 |h|} \,.
\el
The superscripts and subscripts for $\delta(b)$ defined in \eqc{eq:eikphaseexp} will be used to denote the same perturbative expansions of $\bar{\delta}$ and $\a$. 
As explained in section~\ref{sec:NLOnAlpha}, $\a$ is identified as the polarisation plane rotation angle. For all spin sectors we find
\bl
\frac{\a}{\bar{\delta}} &\sim \CO \left( \frac{1}{\w b} \right) \,,
\el
confirming the expectation \eqc{eq:eikphaseheldepexp} based on geometric optics analysis.

The leading order terms of $\bar{\delta}$ and $\alpha$ have the dependence
\eq
{\rm LO}: 
\quad \bar{\delta}_{\mathrm{LO}} \sim \frac{G^2 m^2\omega}{b},\quad \alpha_{\mathrm{LO}} \sim \frac{G^2 m^2}{b^2}
\eqe
where each is multiplied by dimensionless combinations of spin vectors, impact parameter and momenta. The results are independent of helicity, which can be understood as the equivalence principle in the geometric optics limit. In the Kerr limit $C_{\text{S}^n} = 1$, we find that only particular spin structures appear, which can be connected to the shift symmetry \eqc{Shift1} considered by Bern et.al.~\cite{Bern:2022kto}. Furthermore, the leading terms 
$\bar{\delta}_{\mathrm{LO}}$ were checked to be consistent with eq.(5.38) of ref.\cite{Bautista:2021wfy}. We report triangle coefficients 
$c_{\mathrm{LO}}$ together with 
$\bar{\delta}_{\mathrm{LO}}$ and 
$\alpha_{\mathrm{LO}}$ to $\CO(S^6)$ in the main text, and report higher spin order terms in the ancillary file.

The NLO terms have the dependence
\eq
{\rm NLO}: \quad \bar{\delta}_{\mathrm{NLO}} \sim \frac{G^2 m^2}{b^2},\quad \alpha_{\mathrm{NLO}} \sim \frac{G^2 m^2}{b^3\omega}\,,
\eqe
which do not depend on $|h|$; the equivalence principle still holds. However, the NLO terms are purely imaginary in impact parameter space, and new spin structures that are inconsistent with the shift symmetry \eqc{Shift1} appear in the Kerr limit. Together with the triangle coefficients $c_{\bar\delta / \alpha}$, we also report residual contributions from box topology $c_\square^{\bar\delta / \alpha}$ that contribute to $\bar{\delta}_{\mathrm{NLO}}$ and $\alpha_{\mathrm{NLO}}$, up to $\CO(S^2)$ in the main text and to $\CO(S^4)$ in appendix \ref{app:eikNLO}.

\subsubsection{Leading order} \label{sec:2PMLOEik}
\begin{table}
\centering
\begin{tabular}{|c||c|}
            \hline
     non-vanishing coefficients in BH limit & vanishing coefficients in BH limit\\
     \hline
       $ \begin{array}{c}
       ~\\[-4mm]
        Z_{2, 1} = 1 + C_{\mathrm{S}^2}\\
        ~\\[-4mm]
        \end{array}$&$Z_{2, 2} = 1 - C_{\mathrm{S}^2}$\\
      \hline
       $ \begin{array}{c}
       ~\\[-4mm]
  Z_{3, 1} = 3 C_{\mathrm{S}^2} + C_{\mathrm{S}^3}\\
        ~\\[-4mm]
        \end{array}$&$Z_{3, 2} = C_{\mathrm{S}^2} - C_{\mathrm{S}^3}$\\
\hline
  $Z_{4, 1} = 3 C_{\mathrm{S}^2}^2 + 4 C_{\mathrm{S}^3} + C_{\mathrm{S}^4}$  &   $\begin{array}{c}
  ~\\[-4mm]
  Z_{4, 3} = C_{\mathrm{S}^2}^2 - C_{\mathrm{S}^4}
\\
  Z_{4, 4} = 3 C_{\mathrm{S}^2}^2 - 4 C_{\mathrm{S}^3} + C_{\mathrm{S}^4}\\
  ~\\[-4mm]
  \end{array}$
\\  
  \hline
 $Z_{5, 1} = 10 C_{\mathrm{S}^2} C_{\mathrm{S}^3} + 5 C_{\mathrm{S}^4} + C_{\mathrm{S}^5}$&$\begin{array}{c}
  ~\\[-4mm]
  Z_{5, 2} = 2 C_{\mathrm{S}^2} C_{\mathrm{S}^3} - C_{\mathrm{S}^4} - C_{\mathrm{S}^5} \\
   Z_{5, 3} = 2 C_{\mathrm{S}^2} C_{\mathrm{S}^3} - 3 C_{\mathrm{S}^4} + C_{\mathrm{S}^5}\\
   ~\\[-4mm]
  \end{array}$\\
  \hline
$Z_{6, 1} = 10 C_{\mathrm{S}^3}^2 + 15 C_{\mathrm{S}^2} C_{\mathrm{S}^4} + 6 C_{\mathrm{S}^5} + C_{\mathrm{S}^6}$
  &
  $\begin{array}{c}
   ~\\[-4mm]
  Z_{6, 2} = 2 C_{\mathrm{S}^3}^2 + C_{\mathrm{S}^2} C_{\mathrm{S}^4} - 2 C_{\mathrm{S}^5} - C_{\mathrm{S}^6}\\ 
  Z_{6, 3} = 10 C_{\mathrm{S}^3}^2 - 15 C_{\mathrm{S}^2} C_{\mathrm{S}^4} + 6 C_{\mathrm{S}^5} - C_{\mathrm{S}^6}\\
  Z_{6, 4} = 2 C_{\mathrm{S}^3}^2 - C_{\mathrm{S}^2} C_{\mathrm{S}^4} - 2 C_{\mathrm{S}^5} + C_{\mathrm{S}^6}\\ ~\\[-4mm]\end{array}$ \\
   \hline
\end{tabular}
\caption{The table provides the specific combination of Wilson coefficients used in the results of $\bar \delta$ and $\alpha$. Note that $Z_{4,2}$ appears in \cite{Bern:2022kto} which does not appear in our results.}
\end{table}

The triangle coefficients are normalised by $i M_4 = (\k/2)^4 c_\triangle \CI_\triangle + \cdots$. Our results can be summarized into the defined coefficient average $c_{\bar\delta} = \frac{c_{++} + c_{--}}{2}$ and coefficient difference $ c_\alpha = \frac{c_{--} - c_{++}}{2|h|}$. Both $c_{\bar \delta}$ and $c_{\alpha}$ are independent of $|h|$ at this order.

 To linear order in spin,
\bl
\bld
c_{\bar{\delta},\mathrm{LO}}^{\mathrm{S}^1}= 20 i m^2 \w (n \cdot S) \,,\, && 
c_{\alpha,\mathrm{LO}}^{\mathrm{S}^1}
= 5 m^3 q^2 (k_2 \cdot S) \,,
\\ 
\bar{\delta}^{\text{S}^1}_{\mathrm{LO}}
= - \frac{5 \pi G^2 m^2 \w ( \vec{d} \cdot \vec{b} )}{b^3} \,,\, && 
\a^{\text{S}^1}_{\mathrm{LO}}= \frac{5 \pi G^2 m^2 ( \hat{k}_2 \cdot \vec{a} )}{4 b^3} \,,
\eld
\el
where $\a$ matches with the null geodesic parallel transport computation of ref.\cite{Ishihara:1987dv}.

At quadratic order in spin
\bl
\notag 
c_{\bar{\delta},,\mathrm{LO}}^{\mathrm{S}^2}
&= \frac{5}{16} \Big\{Z_{2,1} \left[ 19 (n \cdot S)^2 + 7 m^2 q^2 (k_2 \cdot S)^2 \right] +12 Z_{2,2} m^2 \w^2 (q \cdot S)^2 \Big\}\,,
\\ 
\notag \bar{\delta}^{\text{S}^2}_{\mathrm{LO}} &= \frac{5 \pi G^2 m^2 \w}{64} \left\{Z_{2,1} \left[ 19 \left( \frac{3 (\vec{d} \cdot \vec{b})^2 - d^2 b^2}{b^5} \right) - 7 \frac{( \hat{k}_2 \cdot \vec{a})^2}{b^3} \right]\right.\\
\notag &\hspace{6cm}\left. + 12 \, Z_{2,2} \left[ \frac{3 (\vec{a} \cdot \vec{b})^2 - a_{\perp}^2 b^2}{b^5} \right] \right\}\,,
\\ 
 c_{\alpha,\mathrm{LO}}^{\mathrm{S}^2}&= - \frac{15 i Z_{2,1} m q^2 (n \cdot S) (k_2 \cdot S) }{8 \w}\,,~~
\a^{\text{S}^2}_{\mathrm{LO}} 
= - \frac{45 Z_{2,1} \pi G^2 m^2 (\hat{k}_2 \cdot \vec{a}) (\vec{d} \cdot \vec{b}) }{32 b^5} \,.
\el

At cubic order in spin
\bl
\notag c_{\bar{\delta},\mathrm{LO}}^{\mathrm{S}^3}
 &= - \frac{i (n \cdot S)}{8 m^2 \w} \left[ Z_{3,1} \left\{ 9 (n \cdot S)^2 + 7 m^2 q^2 (k_2 \cdot S)^2 \right\}+ 20 Z_{3,2} m^2 \w^2 (q \cdot S)^2 \right]\,,
\\ 
\notag \bar{\delta}^{\mathrm{S}^3}_{\mathrm{LO}} 
&= - \frac{3 \pi G^2 m^2 \w (\vec{d} \cdot \vec{b})}{32} \left\{ Z_{3,1} \left[ 9 \left( \frac{5 (\vec{d} \cdot \vec{b})^2 - 3 d^2 b^2}{b^7} \right) - 7 \left( \frac{(\hat{k}_2 \cdot \vec{a})^2}{b^5} \right) \right] \right. \nn
\\ \notag &\hspace{1cm} \phantom{=asdfasdf} \left. \phantom{asdfasdfasdf}+20 Z_{3,2} \left[ \frac{5 (\vec{a} \cdot \vec{b})^2 - a_\perp^2 b^2}{b^7} \right] \right\}\,,
\\ 
\notag c_{\alpha,\mathrm{LO}}^{\mathrm{S}^3}
&= - \frac{q^2 (k_2 \cdot S)}{96 m \w^2} \left[ Z_{3,1} \left\{ 39 (n \cdot S)^2 - 7 m^2 q^2 (k_2 \cdot S)^2 \right\}  
+ 60 Z_{3,2} m^2 \w^2 (q \cdot S)^2 \right\}\,,
\\ 
\bar{\a}^{\text{S}^3}_{\mathrm{LO}}
&= \frac{3 \pi G^2 m^2 (\hat{k}_2 \cdot \vec{a})}{128} \left\{Z_{3,1} \left[ 13 \left( \frac{5 (\vec{d} \cdot \vec{b})^2 - d^2 b^2}{b^7} \right) + 7 \frac{(\hat{k}_2 \cdot \vec{a})^2}{b^5} \right] \right.  \nn
\\ &\hspace{2cm}\phantom{=asdfasdfasdf} \left. \phantom{asdf} + 20 Z_{3,2} \left[ \frac{5 (\vec{a} \cdot \vec{b})^2 - a_\perp^2 b^2}{b^7} \right] \right\} \,.
\el

At quartic order in spin
\bl
\notag c_{\bar{\delta},\mathrm{LO}}^{\mathrm{S}^4}
&= \frac{-1}{1536 m^4 \w^2} \Big\{ Z_{4,1} \left[ 294 m^2 q^2 (n \cdot S)^2 (k_2 \cdot S)^2- 21 m^4 q^4 (k_2 \cdot S)^4 +239 (n \cdot S)^4 \right]  \\[1mm] 
\notag &~ + 60 Z_{4,3} m^2 \w^2 (q \cdot S)^2 \left[ 19 (n \cdot S)^2 + 7 m^2 q^2 (k_2 \cdot S)^2 \right] + 120 \,Z_{4,4} m^4 \w^4 (q \cdot S)^4 \Big\}\,,
\\ 
\bar{\delta}^{\text{S}^4}_{\mathrm{LO}}
 &= \frac{3 \pi G^2 m^2 \w}{256} \left\{ \frac{7 Z_{4,1}}{4} \left[ - 7 \, (\hat{k}_2 \cdot \vec{a})^2 \left( \frac{5 (\vec{d} \cdot \vec{b})^2 - d^2 b^2}{b^7} \right) \right. \right. \nn
\\ &\phantom{=} \left. \left. \phantom{asdfasdfasdf} - \frac{3 (\hat{k}_2 \cdot \vec{a})^4}{2 b^5} + \frac{239}{42} \left( \frac{35 (\vec{d} \cdot \vec{b})^4 - 30 (\vec{d} \cdot \vec{b})^2 d^2 b^2 + 3 d^4 b^4}{b^9} \right) \right] \right. \nn
\\ \notag &\phantom{=} \left. \phantom{asdf} + \frac{5 Z_{4,3}}{2} \left[ - 7 \, (\hat{k}_2 \cdot \vec{a})^2 \left( \frac{5 (\vec{a} \cdot \vec{b})^2 - a_\perp^2 b^2}{b^7} \right) \right. \right. \nn
\\\notag &\phantom{=} \left. \left. \phantom{asdfasd} + 19 \left( \frac{35 (\vec{a} \cdot \vec{b})^2 (\vec{d} \cdot \vec{b})^2 - 5 (\vec{a} \cdot \vec{b})^2 d^2 b^2 - 5 (\vec{d} \cdot \vec{b})^2 a_\perp^2 b^2 + d^2 a_\perp^2 b^4 }{b^9} \right) \right] \right. \nn
\\ &\phantom{=} \left. \left. \phantom{asdf} + 5  \,Z_{4,4} \left[ \frac{ 35 (\vec{a} \cdot \vec{b})^4 - 30 (\vec{a} \cdot \vec{b})^2 a_\perp^2 b^2 + 3 a_\perp^4 b^4 }{b^9} \right] \right. \right\}\,,
\\ 
c_{\alpha,\mathrm{LO}}^{\mathrm{S}^4}
\notag &= \frac{i q^2 (n \cdot S) (k_2 \cdot S)}{384 m^3 \w^3} \Big\{ Z_{4,1} \left[ 23 (n \cdot S)^2 - 7 m^2 q^2 (k_2 \cdot S)^2 \right]+ 90 Z_{4,3} m^2 \w^2 (q \cdot S)^2 \Big\}\,,
\\ 
\notag \a^{\text{S}^4}_{\mathrm{LO}}
&= - \frac{5 \pi G^2 m^2 (\hat{k}_2 \cdot \vec{a})(\vec{d} \cdot \vec{b})}{512} \left\{Z_{4,1} \left[ 23 \left( \frac{7 (\vec{d} \cdot \vec{b})^2 - 3 d^2 b^2}{b^9} \right) + 21 \frac{(\hat{k}_2 \cdot \vec{a})^2}{b^7} \right] \right.\\
&\left.\hspace{7.5cm}+ 90 Z_{4,3} \left[ \frac{7 (\vec{a} \cdot \vec{b})^2 - a_\perp^2 b^2}{b^9} \right] \right\} \,.
\el
At quintic order in spin

\bl
\notag c_{\bar\delta,\mathrm{LO}}^{\mathrm{S}^5}&=\frac{i  (n\cdot S)}{768 m^6 \omega ^3}\Big\{
Z_{5,1}
   \left[-3 m^4 q^4 (k_2\cdot S)^4+22 m^2 q^2 (k_2\cdot S)^2 (n\cdot S)^2+13 (n\cdot S)^4\right]\\
\notag &+12 Z_{5,2}m^2 \omega ^2  (q\cdot S)^2 \left[7 m^2 q^2 (k_2\cdot S)^2+9 (n\cdot S)^2\right]+40Z_{5,3} m^4 \omega ^4  (q\cdot S)^4\Big\}\,,\\[1mm]
\notag \bar{\delta}^{\text{S}^5}_{\mathrm{LO}}&=-\frac{5 \pi  G^2 m^2 \omega  (\vec d\cdot \vec b) }{1024 b^{11}}\left\{Z_{5,1}\left[-9\, b^4 (\vec k_2\cdot \vec a)^4+22 (\vec k_2\cdot \vec a)^2 \left(3 b^4 d^2-7b^2 (\vec d\cdot \vec b)^2\right)\right.\right.\\
\notag &\hspace{5cm}\left.+13 \left(-70 b^2 d^2 (\vec d\cdot \vec b)^2+63 (\vec d\cdot \vec b)^4+15 b^4 d^4\right)\right]
\\\notag & +12 Z_{5,2}\left[7 (\vec k_2\cdot \vec a)^2 \left(a_\perp^2 b^4-7 b^2 (\vec a\cdot \vec b)^2\right)-189 (\vec a\cdot \vec b)^2\left(b^2 d^2-3 (\vec d\cdot \vec b)^2\right)\right.\\
\notag &\left.\left.+9 a_\perp^2 b^2 \left(3 b^2 d^2-7 (\vec d\cdot \vec b)^2\right)\right]-120 Z_{5,3}\left(14 a_\perp^2 b^2 (\vec a\cdot \vec b)^2-21 (\vec a\cdot \vec b)^4-a_\perp^4 b^4\right)
   \right\}   \,,\\[1mm]
\notag c_{\alpha,\mathrm{LO}}^{\mathrm{S}^5}&=   \frac{ q^2 (k_2\cdot S)}{15360 m^5 \omega ^4}\Big\{
\notag Z_{5,1}
   \left[9 m^4 q^4 (k_2\cdot S)^4-50 m^2 q^2 (k_2\cdot S)^2 (n\cdot S)^2+105 (n\cdot S)^4\right]\\
\notag &+20 Z_{5,2}m^2 \omega ^2  (q\cdot S)^2 \left[39 (n\cdot S)^2-7 m^2 q^2 (k_2\cdot S)^2\right]+200 m^4 \omega ^4 Z_{5,3} (q\cdot S)^4\Big\}\,,\\[1mm]
   \notag \alpha^{\text{S}^5}_{\mathrm{LO}}&=\frac{15 \pi  G^2 m^2 (\vec k_2\cdot \vec a)}{4096 b^{11}} \left\{Z_{5,1}\left[9 b^4 (\vec k_2\cdot \vec a)^4-10 (\vec k_2\cdot \vec a)^2 \left(b^4 d^2-7 b^2 (\vec d\cdot \vec b)^2\right)\right.\right.\\
\notag   &\left. \hspace{5cm}+21
   \left(-14 b^2 d^2 (\vec d\cdot \vec b)^2+21 (\vec d\cdot \vec b)^4+b^4 d^4\right)\right]\\
\notag &  -4 Z_{5,2}\left[7 (\vec k_2\cdot \vec a)^2 \left(a_\perp^2 b^4-7 b^2 (\vec a\cdot \vec b)^2\right)+91 (\vec a\cdot \vec b)^2 \left(b^2 d^2-9
   (\vec d\cdot \vec b)^2\right)\right.\\
   &\left.\left.-13 a_\perp^2 b^2 \left(b^2 d^2-7 (\vec d\cdot \vec b)^2\right)\right]
   +40 Z_{5,3}\left(a_\perp^4 b^4
   -14 a_\perp^2 b^2(\vec a\cdot \vec b)^2+21 (\vec a\cdot \vec b)^4\right)\right\}\,.
       \el

At sextic order in spin
  \bl 
\notag c_{\bar\delta,\mathrm{LO}}^{\mathrm{S}^6}
&= \frac{1}{491520 m^8 \omega ^4}  \Big\{   
Z_{6,1} \left[33
   m^6 q^6 (k_2\cdot S)^6-345 m^4 q^4 (k_2\cdot S)^4 (n\cdot S)^2
   \right.\\
 \notag   &\left.    
  \hspace{5cm} +1615 m^2 q^2 (k_2\cdot S)^2 (n\cdot S)^4+745 (n\cdot S)^6\right]\\
\notag &-40  Z_{6,2}m^2 \omega ^2 (q\cdot S)^2 \left[21 m^4 q^4 (k_2\cdot S)^4-294 m^2 q^2 (k_2\cdot S)^2
   (n\cdot S)^2-239 (n\cdot S)^4\right]\\
\notag &+320 Z_{6,3}m^6 \omega ^6  (q\cdot S)^6+400 Z_{6,4}m^4 \omega ^4  (q\cdot S)^4 \left[7 m^2 q^2 (k_2\cdot S)^2+19 (n\cdot S)^2\right]\Big\}\,,\\[1mm]      
\notag \bar{\delta}^{\text{S}^6}_{\mathrm{LO}}&= -\frac{15 \pi  G^2 m^2 \omega }{131072 b^{13}} \left\{
Z_{6,1}\left[33 b^6 (\vec k_2\cdot \vec a)^6-69
   (\vec k_2\cdot \vec a)^4 \left(b^6 d^2-7 b^4 (\vec d\cdot \vec b)^2\right)\right.\right.\\
 \notag&  \hspace{3cm} +323 (\vec k_2\cdot \vec a)^2 \left(-14 b^4 d^2 (\vec d\cdot \vec b)^2+21 b^2 (\vec d\cdot \vec b)^4+b^6 d^4\right)\\
 \notag &\hspace{3cm} \left.+149 \left(-105 b^4 d^4 (\vec d\cdot \vec b)^2+315 b^2 d^2 (\vec d\cdot \vec b)^4-231 (\vec d\cdot \vec b)^6+5 b^6 d^6\right)\right]\\
\notag &+8 Z_{6,2}\left[98 (\vec k_2\cdot \vec a)^2 \left(a_\perp^2 b^4\left(b^2 d^2-7 (\vec d\cdot \vec b)^2\right)-7 (\vec a\cdot \vec b)^2 \left(b^4 d^2-9 b^2 (\vec d\cdot \vec b)^2\right)\right)\right.\\
\notag &
-21 (\vec k_2\cdot \vec a)^4 \left(a_\perp^2 b^6-7 b^4 (\vec a\cdot \vec b)^2\right)
+239 \left(a_\perp^2 b^2\left(-14 b^2 d^2 (\vec d\cdot \vec b)^2+21 (\vec d\cdot \vec b)^4+b^4 d^4\right)\right.\\
\notag &\left.\left.-7 (\vec a\cdot \vec b)^2 \left(-18 b^2 d^2 (\vec d\cdot \vec b)^2+33   (\vec d\cdot \vec b)^4+b^4 d^4\right)\right)\right]\\
\notag &+64 Z_{6,3}
   \left[-105 a_\perp^4 b^4 (\vec a\cdot \vec b)^2+315 a_\perp^2 b^2 (\vec a\cdot \vec b)^4-231 (\vec a\cdot \vec b)^6+5 a_\perp^6 b^6\right]\\
\notag &+80 Z_{6,4}\left[19 \left(-14 a_\perp^2 b^2 (\vec a\cdot \vec b)^2 \left(b^2 d^2-9
  (\vec d\cdot \vec b)^2\right)+21 (\vec a\cdot \vec b)^4 \left(b^2 d^2-11 (\vec d\cdot \vec b)^2\right)\right.\right.\\
  \notag &\left.\left.\left.+a_\perp^4 b^4 \left(b^2 d^2-7(\vec d\cdot \vec b)^2\right)\right)
+7 (\vec k_2\cdot \vec a)^2 \left(-14 a_\perp^2 b^4 (\vec a\cdot \vec b)^2+21 b^2 (\vec a\cdot \vec b)^4+a_\perp^4 b^6\right)\right]\right\}\,,\\[1mm]
\notag  c_{\alpha,\mathrm{LO}}^{\mathrm{S}^6}
  &=  -\frac{i  q^2 (k_2\cdot S) (n\cdot S)}{737280 m^7 \omega ^5}\Big\{3600 Z_{6,4} m^4 \omega ^4  (q\cdot S)^4\\
\notag &\hspace{2cm}+Z_{6,1}
   \left[81 m^4 q^4 (k_2\cdot S)^4-290 m^2 q^2 (k_2\cdot S)^2 (n\cdot S)^2+465 (n\cdot S)^4\right]\\
\notag &\hspace{2cm}+240 Z_{6,2} m^2 \omega ^2  (q\cdot S)^2 \left[23 (n\cdot S)^2-7 m^2 q^2 (k_2\cdot S)^2\right]\Big\}\,,\\[1mm]
\notag \alpha^{\text{S}^6}_{\mathrm{LO}}&= -\frac{105 \pi  G^2 m^2 (\vec k_2\cdot \vec a) (\vec d\cdot \vec b) }{65536 b^{13}}\left\{Z_{6,1}\left[27 b^4
   (\vec k_2\cdot \vec a)^4-58 (\vec k_2\cdot \vec a)^2 \left(b^4 d^2-3 b^2 (\vec d\cdot \vec b)^2\right)\right.\right.\\
\notag  &\hspace{6cm}\left.   +31 \left(-30 b^2 d^2 (\vec d\cdot \vec b)^2+33 (\vec d\cdot \vec b)^4+5 b^4
   d^4\right)\right]\\
 &+16 Z_{6,2}\left[23 \left((\vec a\cdot \vec b)^2 \left(33 (\vec d\cdot \vec b)^2-9 b^2
   d^2\right)+a_\perp^2 b^2 \left(b^2 d^2-3 (\vec d\cdot \vec b)^2\right)\right)\right.   \\
\notag   &\left.\left.-7 (\vec k_2\cdot \vec a)^2 \left(a_\perp^2 b^4-9 b^2
   (\vec a\cdot \vec b)^2\right)\right]
     +240 Z_{6,4}\left[a_\perp^4 b^4-18 a_\perp^2 b^2 (\vec a\cdot \vec b)^2+33 (\vec a\cdot \vec b)^4\right]\right\}\,.
 \el
\subsubsection{Subleading order}

The residuals from the box contribution not cancelled by iteration of tree contribute to subleading $\bar{\delta}$ and $\a$. The coefficients are normalised by $i M_4 = (\k/2)^4 c_\square [ \CI_\square + \CI_{\cSquare} ] + \cdots$ with coefficient average $c_\square^{\bar\delta} = \frac{c_{\square}^{ ++} + c_{\square}^{--}}{2}$ and coefficient difference $c_\square^{\bar\alpha} = \frac{c_{\square}^{ --} - c_{\square}^{ ++}}{2|h|}$. All integral coefficients ($c_{\bar\delta}$, $c_{\alpha}$, $c_\square^{\bar\delta}$, and $c_\square^{\alpha}$) are independent of $|h|$ at this order.

Note that $\bar{\delta}$ and $\a$ are purely imaginary at this order, which are written as a sum of two terms. The first part containing $\pi$ is from the triangle coefficients and the secont part without $\pi$ is from the box coefficients. The box coefficient contributions were obtained by removing $\CO(\w^2)$ terms corresponding to ``squaring'' of the tree eikonal phase.

At linear order in spin
\bl
&\notag c^{\,\text{S}^1}_{\bar\delta,\mathrm{NLO}} = 0 \,,~~~ \notag\bar{c}^{\,\text{S}^1,\,\bar\delta}_{\square,\mathrm{NLO}} = 4 i m^2 q^2 \omega  (n\cdot S) \,,~~~ \notag\bar{\delta}^{\,\text{S}^1}_{\mathrm{NLO}} = 0 + i \left[ \frac{8 G^2 m^2 (\vec{d}\cdot \vec{b} )}{b^4} \right] \,,
\\ &\notag c^{\,\text{S}^1}_{\alpha,\mathrm{NLO}}  = \frac{5 m^3 q^2 (q \cdot S) }{2} \,,~~~ \notag c^{\,\text{S}^1,\,\alpha}_{\square,\mathrm{NLO}} = 2 m^3 q^4 ( k_2 \cdot S ) \,,
\\& \a^{\,\text{S}^1}_{\mathrm{NLO}} = i \left[ \frac{15 \pi G^2 m^2 (\vec{a} \cdot \vec{b})}{8 b^5 \w} \right] - i \left[ \frac{8 G^2 m^2 (\hat{k_2}\cdot \vec{a})}{b^4 \omega } \right]\,.
\el

At quadratic order in spin
\bl
\notag c^{\,\text{S}^2}_{\bar\delta,\mathrm{NLO}}&= \frac{35 (1 + C_{\text{S}^2}) m^2 q^2 (q \cdot S) (k_2 \cdot S)}{16} \,,
\\\notag  \bar{c}^{\,\text{S}^2,\,\bar\delta}_{\square,\mathrm{NLO}} &= q^2 \left(C_{\text{S}^2}+1\right) \left(m^2 q^2 (k_2\cdot S )^2+2 (n\cdot S)^2\right) \,,
\\\notag  \frac{\bar{\delta}^{\,\text{S}^2}_{\mathrm{NLO}}}{i G^2 m^2} &= - \frac{105 (1 + C_{\text{S}^2}) \pi (\hat{k}_2 \cdot \vec{a}) (\vec{a} \cdot \vec{b})}{64 b^5} + \frac{4 (C_{\text{S}^2}+1) \left[b^2 a^2 - 4 (\vec{d}\cdot \vec{b})^2\right]}{b^6} \,,
\\\notag  c^{\,\text{S}^2}_{\alpha,\mathrm{NLO}}  &= - \frac{15 i (1 + C_{\text{S}^2}) m q^2 (q \cdot S) (n \cdot S) }{16 \w} \,,
\\\notag  c^{\,\text{S}^1,\,\alpha}_{\square,\mathrm{NLO}} &= -\frac{i m q^4 \left(C_{\text{S}^2}+1\right) (k_2\cdot S) (n\cdot S)}{\omega } \,,
\\ \frac{\w \,\a^{\,\text{S}^2}_{\mathrm{NLO}} }{iG^2m^2} &= - \frac{225 (1 + C_{\text{S}^2}) \pi (\vec{a} \cdot \vec{b}) (\vec{d} \cdot \vec{b})}{64 b^7 } + \frac{16  (C_{\text{S}^2}+1) (\vec{d}\cdot \vec{b}) (\hat{k_2}\cdot \vec{a})}{b^6 } \,.
\el
$a^2 = d^2+ (\hat{k_2}\cdot \vec{a} )^2 = a_\perp^2 + (\hat{k_2}\cdot \vec{a} )^2$ has been used to simplify $\bar{\delta}^{\,\text{S}^2}_{\mathrm{NLO}}$. 

\subsection{Discussions: equivalence principle, spin structures and massless limits}
In the geometric optics limit, the observables considered in this paper\textemdash the observables \eqc{eq:eik2obs1} derived from $\bar{\delta}$ and the rotation angle $\a$\textemdash has geometric interpretations in classical general relativity which do not depend on worldline degrees of freedom. This means the observables do not depend on the details of the particle, which can be understood as a manifestation of the equivalence principle. For the computations considered in this paper, the equivalence principle is realised as the independence of $\bar{\delta}$ and $\alpha$ from helicity $|h|$ of the massless particle. The term equivalence principle will be used to denote this helicity independence when referring to eikonal phase variables.

Since $\l/b$ is the classical expansion parameter that parametrises deviations from the geometric optics limit, we expect the equivalence principle to be obeyed by $\bar{\delta}$ and $\a$ at leading order in $\l/b$. Indeed this is the case. The LO $\bar{\delta}$ and $\a$ obeys the equivalence principle, at least up to one-loop order. The general expectation is that the equivalence principle for $\bar{\delta}$ and $\a$ is broken at sufficiently high orders in $\l/b$, since the corrections in $\l/b$ can be interpreted as wave property corrections to the point particle approximation; the equivalence principle only applies to the strict point particle limit where tidal effects can be neglected, while a characteristic property of waves is that they cannot be localised to a point. Interestingly, the equivalence principle is still obeyed at NLO for both $\bar{\delta}$ and $\a$.

\noindent \textbf{Spin structures} 

As mentioned in the introduction, the product of the two three point amplitudes transform nicely under the spin-shift eq.(\ref{Shift1}) in the black hole limit. In particular, this suggests that the 2PM classical Compton amplitude is invariant. Let us verify this explicitly. 

All LO triangle coefficients $c_{\bar\delta,\mathrm{LO}}$ and $c_{\alpha,\mathrm{LO}}$ have the special property that their dependence on $(q \cdot S)$ vanish in the Kerr limit $C_{\text{S}^j} \to 1$. That is, projecting the spin-vector on the basis of 
\eq
\{ (q \cdot S) \,,\, (k_2 \cdot S) \,,\, (n \cdot S) \}, 
\eqe
only the latter two appears. In the all massive setup relevant for the 2PM potential, similar phenomenon was observed, which lead to the conclusion that the result is invariant under the spin-shift. Indeed while $(n \cdot S)$ is manifestly invariant under the shift, the change in $(p_2 \cdot S_1)$ is a \emph{quantum} effect:  $(p_2 \cdot S_1) \sim \CO (\hbar^{-1})$ is shifted by $\frac{p_2 \cdot q}{q^2} \sim \CO (\hbar^0)$ where we've used momentum conservation $(p_j \cdot q) = \pm q^2/2$; $\frac{(p_j \cdot q)/q^2}{(p_j \cdot S_i)} = \CO(\hbar^1)$, thus the shift is a quantum effect and the classical result is invariant. In our massless-massive setup one naively encounters a violation of this symmetry. This is because the term $(k_2 \cdot S)\sim \CO (\hbar^{0})$ shifts as $\frac{k_2 \cdot q}{q^2} \sim \CO (\hbar^0)$, i.e. it is shifted by a classical effect. This observation poses a puzzle as the LO triangle coefficient is not shift invariant, but the integrand used to compute it was manifestly shift invariant.

The apparant paradox is resolved when NLO triangle coefficients are also included in the analysis. The NLO triangle coefficients all contain a factor of $(q \cdot S)$, which is manifestly not shift invariant. However, the terms arising from the shift $(q \cdot S) \to (q \cdot S) + \xi$ is \emph{no longer subleading} in the $\l/b$ expansion, and the shift-generated terms of both the LO and NLO triangle coefficients are of the same order in the $\l/b$ expansion. Explicit calculations show that the two terms cancel, and the shift symmetry is restored by the interplay of LO and NLO triangle coefficients.

As a concrete example, consider triangle coefficients at linear order in spin.
\bl
c_{\bar{\delta},\mathrm{LO}}^{\mathrm{S}^1} = 20 i m^2 \w (n \cdot S) \,,\,
c_{\alpha,\mathrm{LO}}^{\mathrm{S}^1} = 5 m^3 q^2 (k_2 \cdot S) \,,\, c^{\,\text{S}^1}_{\alpha,\mathrm{NLO}} = \frac{5 m^3 q^2 (q \cdot S) }{2} \,.
\el
Under the shift \eqc{Shift1} each coefficient transforms as
\bl
\bld
c_{\bar{\delta},\mathrm{LO}}^{\mathrm{S}^1} &\to c_{\bar{\delta},\mathrm{LO}}^{\mathrm{S}^1} \,,
\\ c_{\alpha,\mathrm{LO}}^{\mathrm{S}^1} &\to c_{\alpha,\mathrm{LO}}^{\mathrm{S}^1} - \frac{5 m^3 q^2}{2} \xi_1 \,,
\\ c^{\,\text{S}^1}_{\alpha,\mathrm{NLO}} &\to c^{\,\text{S}^1}_{\alpha,\mathrm{NLO}} + \frac{5 m^3 q^2}{2} \xi_1 \,,
\eld
\el
and shift-generated terms cancel due to the interference between LO and NLO contributions. Similar behaviour can be observed for terms of higher order in spin or $\l/b$ expansion.



\noindent \textbf{Leading order terms as the massless limit}

It is known that the leading order $\bar{\delta}_{\mathrm{LO}}$ can be obtained as the massless limit of the massive case.\footnote{JWK would like to thank Francesco Comberiati and Leonardo de la Cruz for pointing this out.} Denoting the massive eikonal phase as $\chi$, the explicit map is
\bl
\bar{\delta}_{\mathrm{LO}} &= \lim_{m_2 \to 0} \left. \chi_{S_1^n S_2^0} \right|_{m_2 \s = \w} \,.
\el
The subscript denotes that we substitute $\s = \w / m_2$ and then take the $m_2 \to 0$ limit. We also find that $\a_{\mathrm{LO}}$ can be obtained as the massless limit of the massive case. We first align the spin of the massive probe particle $m_2$ so that the probe particle has a well-defined helicity. In the rest frame of $m_1$, we can write the relevant four-vectors as
\bl
\bld
p_1^\m = (m_1 , \vec{0}) \,,\, p_2^\m = (\w , k \hat{z}) \,,\, S_2^\m = (hk/\w , h \hat{z}) \,,
\eld
\el
where $\w^2 = (k^2 + m_2^2)$ and $h$ is the helicity defined by $h = \frac{\vec{p}_2 \cdot \vec{S}_2 }{ |\vec{p}_2| }$. The aligned configuration sets $n \cdot S_2 = 0$, and ``factoring'' the helicity $h$ by setting $p_1 \cdot S_2 = - m_1 \s$ we find that $\a$ corresponds to the massless limit of $\chi_{S_2^1}$.\footnote{This scaling keeps Pauli-Lubanski vector $W_2^\m = - \half \e^{\m\a\b\g} k_{2\a} J_{\b\g} = m_2 S_2^\m$ finite in the massless limit.}
\bl
\a_{\mathrm{LO}} &= \lim_{m_2 \to 0} \left. \chi_{S_1^n S_2^1} \right|_{m_2 \s = \w \,, \, n \cdot S_2 = 0 \,,\, p_1 \cdot S_2 = - \frac{m_1 \w }{ m_2}} \,.
\el
The limit has been found by comparing to the results given in appendix B of ref.\cite{Chen:2021qkk}. It would be interesting to link the NLO imaginary terms of the eikonal phase variables to massive eikonal phase.

\noindent \textbf{Purely imaginary NLO terms}

Both $\bar{\delta}_{\mathrm{NLO}}$ and $\a_{\mathrm{NLO}}$ are purely imaginary in impact parameter space. Imaginary terms of the eikonal phase are usually interpreted as signs of non-conservative effects such as dissipation or radiation~\cite{DiVecchia:2020ymx,DiVecchia:2021ndb,DiVecchia:2021bdo,Bjerrum-Bohr:2021din,Brandhuber:2021eyq,Alessio:2022kwv}. However, the purely imaginary NLO terms $\bar{\delta}_{\mathrm{NLO}}$ and $\a_{\mathrm{NLO}}$ cannot be interpreted this way, as $2 \to 2$ scattering at one-loop order is purely elastic. On the other hand, the imaginary term is NLO in $\l/b$, which is the order the wave amplitude $a$ correction enters the geometric optics eikonal $\psi$ as can be seen in \eqc{eq:eikeqderiv}. Therefore a possible interpretation is that the imaginary terms encode (de)amplification of the waves due to spin of the massive body, which may have connections with polarising effects discussed in section 3.3 of ref.\cite{Bautista:2021wfy}.

\section{Conclusion}
In this paper, we derive the $\mathcal{O}(G^2)$ eikonal phase of graviton and photons scattering of general spinning objects. This corresponds to the classical limit of the one-loop amplitude, which in terms of scalar integral basis is given by the scalar triangle integral. We extract its coefficient by computing the triple cut, which is the product of massless four-point amplitude and two massive spinning three-point amplitudes.  From the result we extract the leading order gravitational Faraday effect, $\a:=-\alpha_3/|h|$ for polarized electromagnetic and gravitational waves, which is given as an expansion in the ratio of spin-length and impact parameter, $a/b$. While the linear term for Kerr black hole matches with result in~\cite{Ishihara:1987dv}, we provide the result for general spinning body up to 14th order in spin. 

In computing the 2PM eikonal phase, the contribution from the box and cross-box nicely reproduces the square of the 1 PM eikonal phase in the Kerr-limit, indicating that the exponentiation correctly captures the eikonal limit amplitude. This is no longer true for general Wilson coefficients. Such observation appears to be consistent with long known result that for electromagnetic scattering~\cite{Cheng:1971gf,Meng:1972xt,Weinberg:1971cdi,Czyz:1975bf}, the amplitude only exponentiates for $g=2$. However this appears to render the extraction of classical observables for neutron stars from eikonal amplitude challenging. We leave the resolution of this issue to future work.

In black hole limit, the $\mathcal{O}(G^2)$ eikonal phase is invariant under the shift transformation of the spin vector eq.(\ref{Shift1})~\cite{Aoude:2022trd, Bern:2022kto,Aoude:2022thd}, whose origin can be traced back to the exponentiated form of the product of spinning three-point amplitudes. As discussed in the introduction, for $\CO(G^2)$ calculation the spin factor gets multiplied by a factor of 
\eq
 \exp \left(- \xi  \frac{(\sum_{i{\in} {+}} k_i \cdot q){-}(\sum_{j{\in} {-}} k_j \cdot q)}{q^2}\right) \label{eq:shift_anom}
\eqe
which is in general non-unity. Beyond $\mathcal{O}(G^2)$, one would have to consider the product of $n$ three-point amplitudes in the cut, where $n \ge 3$. We expect the simple form of \eqc{eq:shift_anom} to be generalised for arbitrary multiple products of three-point amplitudes, when the calculations are reformulated in the framework such as HEFT~\cite{Brandhuber:2021kpo,Brandhuber:2021eyq}/HPET~\cite{Damgaard:2019lfh,Aoude:2020onz} where massive-massless coupling vertices are treated in a homogeneous manner. If the expectation is indeed the case, then the simplicity of the violation brings hope that one can write down the explicit form of the ``anomaly", by inserting this factor into the unitarity cuts. We leave this for future exploration.

Finally we find that at LO in $\l/b$ the eikonal phase is universal between photons and gravitons, while deviation occurs at NNLO. Given that the expansion in $\l/b$ is an expansion around the geometric optics approximation, the violation can be interpreted as the breakdown of eikonal approximation. Indeed we see that at this order, the overall box topology contribution no longer matches the square of the tree eikonal phase as required from exponentiation. It would be interesting nonetheless to explore whether sensible observables can be extracted from these terms.

\vskip 1cm 

\acknowledgments 

We are grateful for Zvi Bern, Andreas Brandhuber, Kays Haddad, Mich\`ele Levi, Andr\'es Luna, Donal O'Connell, Gabriele Travaglini, Radu Roiban 
and Justin Vines  for discussions and valuable feedback on the first draft of the manuscript.
JWK would like to thank Manuel Accettulli Huber and Stefano De Angelis for sharing computation details of ref.\cite{AccettulliHuber:2020oou} and Andr\'es Luna for sharing unlisted box coefficients of ref.\cite{Kosmopoulos:2021zoq} for comparison.
WMC is supported by in part by JSPS KAKENHI Grant Number 21F21317.
MZC and YTH is supported by MoST Grant No. 109-2112-M-002 -020 -MY3. 
JWK was supported by the Science and Technology Facilities Council (STFC) Consolidated Grant ST/T000686/1 \textit{``Amplitudes, Strings and Duality”}.
JWK would like to thank KITP for their hospitality during the stay at the programme \textit{``High-Precision Gravitational Waves''}, where part of this work was completed. This research was supported in part by the National Science Foundation under Grant No. NSF PHY-1748958.

\appendix
\section{Eikonal phase normalisation} \label{app:eiknorm}
We examine the different conventions and resulting differences in the definition of the eikonal phase. In the $2 \to 2$ convention for the $T$-matrix element $M_4$, where $S = \iden + i T$, the definitions for the eikonal phase used in the literature generically takes the form
\bl
\bld
e^{i \delta (b)} - 1 &:= \int \frac{d^D q}{(2 \pi)^D} \, (2 \pi) \delta [V \cdot q] \, (2 \pi) \delta [W \cdot q] \, e^{- i b \cdot q} \, i M_4
\\ &= \int \frac{d^{D-2} q}{(2 \pi)^{D-2}} \frac{i}{\sqrt{\left| V^2 W^2 - (V \cdot W)^2 \right|}} \left[ e^{i \vec{b} \cdot \vec{q}} M_4 (q) \right] \,,
\eld
\el
where $q^\mu = p_3^\mu - p_2^\mu$ and different definitions differ by the vectors $V^\m$ and $W^\m$ used to define the impact parameter space. For strict exponentiation of the classical terms at the diagram level the vectors are defined as~\cite{Brandhuber:2021eyq}
\bl
\bld
V = p_4 + p_1 
\,, \, W = p_3 + p_2 
\,,
\eld
\el
and the independent variable is average three-momentum $\vec{\bar{p}}$. The geometric interpretation of eikonal variables in Fig.1 of ref.\cite{Bern:2020gjj} becomes natural for this choice of $V^\m$ and $W^\m$. In centre of momentum (COM) frame the momenta are given as
\bl
\bld
p_1^\m = (\sqrt{m_1^2 + \vec{\bar{p}}^2 + \vec{q}^2 / 4} , - \vec{\bar{p}} + \vec{q} / 2) \,&,\, p_4^\m = (\sqrt{m_1^2 + \vec{\bar{p}}^2 + \vec{q}^2 / 4} , - \vec{\bar{p}} - \vec{q} / 2) \,,
\\ p_2^\m = (\sqrt{m_2^2 + \vec{\bar{p}}^2 + \vec{q}^2 / 4} , \vec{\bar{p}} - \vec{q} / 2) \,&,\, p_3^\m = (\sqrt{m_2^2 + \vec{\bar{p}}^2 + \vec{q}^2 / 4} , \vec{\bar{p}} + \vec{q} / 2) \,,
\eld
\el
with $\vec{\bar{p}} \cdot \vec{q} = 0$. The square of the Jacobian factor is
\bl
(V \cdot W)^2 - V^2 W^2 = 16 |\vec{\bar{p}}|^2 \left[ \sqrt{m_1^2 + \vec{\bar{p}}^2 + \vec{q}^2 / 4} + \sqrt{m_2^2 + \vec{\bar{p}}^2 + \vec{q}^2 / 4} \right]^2 \,.
\el
For both massive case ($m_1 \neq 0$ and $m_2 \neq 0$) we rescale $\vec{q} \to \hbar \vec{q}$. Only the $q$-dependent terms carry extra powers of $\hbar^2$ and the terms can be dropped in the classical limit.
\bl
(V \cdot W)^2 - V^2 W^2 = 16 |\vec{\bar{p}}|^2 \left[ \sqrt{m_1^2 + \vec{\bar{p}}^2} + \sqrt{m_2^2 + \vec{\bar{p}}^2} \right]^2 + \CO(\hbar^2)\,. \label{eq:avmomJF}
\el
For $m_2 = 0$ we rescale $\vec{\bar{p}} \to \hbar \vec{\bar{p}}$ and $\vec{q} \to \hbar \vec{q}$. The classical limit is given as
\bl
(V \cdot W)^2 - V^2 W^2 = 16 \hbar^2 m_1^2 |\vec{\bar{p}}|^2 \left[ 1 + \CO(\hbar^2) \right]\,. \label{eq:avmomNF}
\el
When comparing with the Mandelstam parametrisation using $\w$, we have the relation
\bl
\hbar m_1 \w = p_1 \cdot p_2 && \Rightarrow && \w = |\vec{\bar{p}}| \sqrt{1 + \frac{\vec{q}^2}{4 \vec{\bar{p}}^2}} \left[ 1 + \CO(\hbar^1) \right] \,. \label{eq:avmom2freq}
\el
Therefore to the first subleading order in $q/\w$ or $q / \bar{p}$, the two ratios can be considered equal $q / \w = q / \bar{p} \times [1 + \CO ((q/\bar{p})^2)]$.

We use a simpler prescription used in the literature~\cite{KoemansCollado:2019ggb,AccettulliHuber:2020oou}
\bl
\bld
V = 2 p_1 \,,\, W = 2 p_2 \,,\, (V \cdot W)^2 - V^2 W^2 = 16 m_1^2 m_2^2 \left( \s^2 - 1 \right) \,,
\eld
\el
where $\s := \frac{p_1 \cdot p_2}{m_1 m_2}$. A slightly different parametrisation for the COM frame momenta is more useful for this prescription:
\bl
\bld
p_1^\m = (\sqrt{m_1^2 + \vec{{p}}^2 } , - \vec{{p}} ) \,&,\, p_4^\m = (\sqrt{m_1^2 + \vec{{p}}^2 } , - \vec{{p}} - \vec{q} ) \,,
\\ p_2^\m = (\sqrt{m_2^2 + \vec{{p}}^2 } , \vec{{p}} ) \,&,\, p_3^\m = (\sqrt{m_2^2 + \vec{{p}}^2 } , \vec{{p}} + \vec{q} ) \,,
\eld
\el
which yields the squared Jacobian factor
\bl
(V \cdot W)^2 - V^2 W^2 = 16 m_1^2 m_2^2 \left( \s^2 - 1 \right) = 16 |\vec{{p}}|^2 \left[ \sqrt{m_1^2 + \vec{{p}}^2} + \sqrt{m_2^2 + \vec{{p}}^2} \right]^2 \,,
\el
which is exact. In the classical limit, this factor is equivalent to the factor \eqc{eq:avmomJF} up to the loop order considered in this paper if $\bar{p}$ is identified with $p$. For $m_2 = 0$ the factor simplifies to
\bl
(V \cdot W)^2 - V^2 W^2 = 16 \hbar^2 m_1^2 \w^2 \,, \label{eq:inmomNF}
\el
which gives the normalisation factor used in \eqc{eq:eikphasedeforig}. The difference of normalisation factors \eqc{eq:avmomNF} and \eqc{eq:inmomNF} starts at $\CO((q/\w)^2)$ order, therefore the two prescriptions are indistinguishable up to the first subleading order in $q/\w$.

\section{Linking eikonal phase to geometric optics} \label{app:eikphase2geo}
We present an argument why the eikonal phase $\delta(b)$ can be identified as the geometric optics eikonal $\psi$, which can be viewed as a simplified version of the arguments presented in ref.\cite{Cristofoli:2021jas}. A related approach is to identify the eikonal phase as the Hamilton's principal function for the massive case~\cite{Bern:2021dqo,Damgaard:2021ipf,Brandhuber:2021eyq,Kol:2021jjc}. We also comment that Hamiltonian mechanics was developed from the correspondence between the eikonal and Hamilton's principal function.

To build the connection, we identify the field component $f$ in \eqc{eq:fieldeikdecomp} as the following matrix element in a specified gauge, e.g. radiation gauge
\bl
a e^{i \psi} = f = \bra{\vec{p}'} A^\m (x) \ket{\vec{p}; \vec{k}, \ve} \,, \label{eq:eik2matelem}
\el
where $\vec{p}$ is the momentum of the incoming massive particle, $\vec{p}'$ is the momentum of the outgoing massive particle, and $\vec{k}$ is the momentum of the incoming photon with polarisation $\ve$. This matrix element can be converted to $2 \to 2$ scattering amplitude of a photon scattering off a massive particle by the LSZ reduction formula
\bl
M_4 (p;k,\ve;k',\ve';p') &= \int d^d x e^{ i k'_\m x^\m} [\ve'_\m]^\ast \partial^\n \partial_\n \bra{\vec{p}'} A^\m (x) \ket{\vec{p}; \vec{k}, \ve} \,,
\el
which can be ``inverted'' to obtain an asymptotic form of the matrix element in \eqc{eq:eik2matelem}
\bl
\bra{\vec{p}'} A_\m (x) \ket{\vec{p}; \vec{k}, \ve} \simeq \int \frac{d^d k'}{(2 \pi)^d} ~ \ve'_\m \,  M_4 (p;k,\ve;k',\ve';p') \, e^{- i k'_\m x^\m} \,, \label{eq:amp2matelem1}
\el
where $A_4$ is the scattering amplitude. The ``sum'' over $k'$ can be decomposed as a sum over transferred momentum $q = k' - k$ and the ``forward momentum'' $\bar{k} = (k + k')/2$, which we set as $\bar{k}^\m = (\w, \bar{k}_z \hat{z})$. Combining with the identification \eqc{eq:eik2matelem}, \eqc{eq:amp2matelem1} is then recast as
\bl
a e^{i \psi} \simeq \int \frac{d\w d \bar{k}_z}{(2\pi)^2} \int \frac{d^{d-2} q}{(2 \pi)^{d-2}} ~ J \, e^{- i ( \w t - \bar{k}_z z) } \left[ e^{i \vec{b} \cdot \vec{q}} M_4 (q) \right] \,. \label{eq:amp2matelem}
\el
where $\vec{b}$ is the impact parameter space coordinate\footnote{In this case, the coordinate values of $x^\m$ in \eqc{eq:amp2matelem1} except the time $t$ and $z$ component.} and $J$ denotes the unimportant terms; the on-shell condition deltas, the Jacobian, and the polarisation vector. Note that the sign of the Fourier factor $e^{i \vec{b} \cdot \vec{q}}$ originates from the positive frequency plane wave factor $e^{- i k'_\m x^\m}$ with outgoing momentum $k' = k + q$. Using \eqc{eq:eikphasedeforig}, \eqc{eq:amp2matelem} can be recast as
\bl
e^{i \psi + [\log a]} \simeq \int \frac{d\w d \bar{k}_z}{(2\pi)^2} ~ [\delta_{\text{o-s}}] \, e^{- i ( \w t - \bar{k}_z z ) + i \delta (b) + [ \log N' ] } \,, \label{eq:eik2eikphase}
\el
where $N' = J / N$ is a slowly varying factor and $\delta_{\text{o-s}}$ gives the on-shell conditions. In other words, we can approximate the eikonal $\psi$ of geometric optics as
\bl
\psi (t, z, \vec{b}) - i [\log a] &\simeq - \w t + \bar{k}_z z + \delta (\vec{b}) - i [\log N'] \,.
\el
where terms in the square brackets are ``small'' compared to other terms. 
The polarisation dependence is given by the ratio $\partial[\log a] / \partial \psi$, and estimating the amplitude variation length scale as $\partial [\log a] \sim b^{-1}$ we have the ratio $\partial[\log a] / \partial \psi \sim \l / b$.

\section{Simplifying the spin factor for long-distance physics} \label{app:KerrWilson}
The ansatz for the classical amplitude, \eqc{eq:tree4ptansatz}, is tailored to correctly capture the $t$-channel residue. While the $t$-channel is the only physical channel for $h \le 1$, this is no longer true for $h > 1$ and $s$- and $u$-channels will also become physical. For example, the gravitational Compton amplitude eq.(B.12) of ref.\cite{Chen:2021qkk}, reproduced below with $2 \to 2$ continuation from $4 \to 0$, has a different spin factor from \eqc{eq:tree4ptansatz}.
\bl
\bgd
M_4^{s} = M_4^{s=0} \oint \prod_{k=1,2} \left[ \frac{dz_k}{2\pi i z_k} \left( \sum_j C_{\text{S}^j} z_k^j \right) \right] \exp \left( -i \bar{K}^\m L^\n J_{\m\n} \right) \,,
\\ \bar{K}^\m = \frac{k_2^\m}{z_1} - \frac{k_3^\m}{z_2} \,,\, L^\m = \frac{\bra{3} \s^\m \sket{2}}{\bra{3} p_1 \sket{2}} \,.
\egd \label{eq:gravComptonSF}
\el
The scalar factor $A_4^{s=0}$ is the $h=2$ case of \eqc{eq:tree4ptscalar}, and the integration contours of the spin factor encircle the origins $z_{1,2} = 0$ counter-clockwise. While the spin factor is motivated to capture the $s$- and $u$-channel poles, it also correctly captures the $t$-channel pole. This is because in the $\ket{2} \propto \ket{3}$ limit ($\sket{2} \propto \sket{3}$ limit) the vector $L^\m$ becomes proportional to $k_2^\m$ ($k_3^\m$), so the $z_1$ ($z_2$) dependence of $\bar{K}^\m$ drops out and the spin factor of \eqc{eq:gravComptonSF} effectively reduces to that of \eqc{eq:tree4ptansatz}.

The gravitational Compton amplitude \eqc{eq:gravComptonSF} was motivated by the fact that all physical channel poles are correctly captured, therefore it will only differ from the ``real'' Compton amplitude by unknown polynomial terms, at least up to $\CO(J^4)$ since there are no spurious poles. Since polynomial terms become (derivatives of) Dirac delta interactions in impact parameter space and thus cannot affect long-distance physics, it is concluded that the amplitude \eqc{eq:gravComptonSF} correctly captures the long-distance physics of graviton scattering from a spinning compact object. The conclusion poses a mystery; while \eqc{eq:tree4ptansatz} is \emph{linear} in Wilson coefficients, \eqc{eq:gravComptonSF} is \emph{quadratic}. How can their long-distance behaviours be equivalent?



In fact, there is a nontrivial cancellation for \eqc{eq:gravComptonSF} and the long-distance behaviour reduces to that of \eqc{eq:tree4ptansatz}. To simplify the analysis, we recast the spin factor exponent of \eqc{eq:gravComptonSF} in a form similar to \eqc{eq:treeansatzexpred}. Substituting the Lorentz generator $J^{\m\n}$ by the spin tensor $S^{\m\n}$ and using the definition $q^\m = k_3^\m - k_2^\m$,
\bl
- i \bar{K}^\m L^\n S_{\m\n} = - \left( \frac{1}{z_1} + \frac{1}{z_2} \right) \frac{i (n \cdot S)}{2 m^2 \w} + \left( \frac{1}{z_1} - \frac{1}{z_2} \right) \frac{(q \cdot S)}{2 m} + \CO(|\vec{q}|^2, \hbar^1) \,. \label{eq:gravCompSFexpred}
\el
Due to the antisymmetry of $z_1$ and $z_2$, any terms with odd powers of $(q \cdot S)$ become irrelevant and only even powers of $(q \cdot S)$ need to be considered. The even powers of $(q \cdot S)$ can be traded for even powers of $(n \cdot S)$
\bl
\left[ \frac{(q \cdot S)}{m} \right]^2 = \left[ - \frac{i (n \cdot S)}{m^2 \w} \right]^2 + \CO (q^2) \,, \label{eq:qSnSequiv}
\el
which is equivalent to the relation $(\vec{d} \cdot \del_{b})^2 = a_\perp^2 \del_{b}^2 - (\vec{a}_\perp \cdot \del_{b})^2$ in impact parameter space. The Laplacian cannot contribute to the long-distance behaviour at tree level, therefore it can be ignored in the analysis.

We now focus on the (effective) coefficient of $(n \cdot S)^N$ term when the exponentials in \eqc{eq:tree4ptansatz} and \eqc{eq:gravComptonSF} are expanded to $\CO(S^N)$ order. We will ignore the irrelevant $(N!)^{-1}$ factor from the exponential function in both cases. For \eqc{eq:tree4ptansatz} we get
\bl
\oint \frac{dz}{2\pi i z} \left( \sum_j C_{\text{S}^j} z^j \right) \left( - \frac{i (n \cdot S)}{z m^2 \w} \right)^N = C_{\text{S}^N} \left( - \frac{i (n \cdot S)}{m^2 \w} \right)^N \,. \label{eq:treeansatzWilson}
\el
For \eqc{eq:gravComptonSF} the expansion is rather involved, as all even power contributions from $(q \cdot S)$ must be included. Using the binomial theorem, collecting even powers of $(q \cdot S)$, and using \eqc{eq:qSnSequiv} to reorganise the expression as $(n \cdot S)^N$, the relevant $\CO(S^N)$ term from \eqc{eq:gravComptonSF} can be written as
\bl
\bgd
\oint \prod_{k=1,2} \left[ \frac{dz_k}{2\pi i z_k} \left( \sum_j C_{\text{S}^j} z_k^j \right) \right] g_N (z_1,z_2) \left( - \frac{i (n \cdot S)}{m^2 \w} \right)^N \,,
\\ g_N (z_1,z_2) := \frac{1}{2^{N}} \sum_{M=0}^{\lfloor N/2 \rfloor} {N \choose 2M} \left( \frac{1}{z_1} + \frac{1}{z_2} \right)^{N-2M} \left( \frac{1}{z_1} - \frac{1}{z_2} \right)^{2M} \,,
\egd \label{eq:gravCompWilson}
\el
where $\lfloor x \rfloor$ is the floor function. The sum in $g_N(z_1,z_2)$ can be explicitly evaluated.
\bl
\bld
g_N (z_1,z_2) &= \frac{1}{2^{N}} \sum_{J=0}^{N} \frac{1 + (-1)^J}{2} {N \choose J} \left( \frac{1}{z_1} + \frac{1}{z_2} \right)^{N-J} \left( \frac{1}{z_1} - \frac{1}{z_2} \right)^{J}
\\ &= \frac{1}{2^{N+1}} \left[ \sum_{J=0}^{N} {N \choose J} \left( \frac{1}{z_1} + \frac{1}{z_2} \right)^{N-J} \left( \frac{1}{z_1} - \frac{1}{z_2} \right)^{J} \right.
\\ &\phantom{=asdfasdfasdf} \left. + \sum_{J=0}^{N} {N \choose J} \left( \frac{1}{z_1} + \frac{1}{z_2} \right)^{N-J} \left( \frac{1}{z_2} - \frac{1}{z_1} \right)^{J} \right]
\\ &= \frac{1}{2} \left( \frac{1}{z_1^N} + \frac{1}{z_2^N} \right) \,.
\eld
\el
Therefore \eqc{eq:gravCompWilson} becomes
\bl
\oint \prod_{k=1,2} \left[ \frac{dz_k}{2\pi i z_k} \left( \sum_j C_{\text{S}^j} z_k^j \right) \right] \left( \frac{1}{2 z_1^N} + \frac{1}{2 z_2^N} \right) \left( - \frac{i (n \cdot S)}{m^2 \w} \right)^N = C_{\text{S}^N} \left( - \frac{i (n \cdot S)}{m^2 \w} \right)^N \,,
\el
which is the same as \eqc{eq:treeansatzWilson}. In conclusion, the ansatz \eqc{eq:tree4ptansatz} may be used in place of the ``correct'' gravitational Compton amplitude \eqc{eq:gravComptonSF} for evaluating $\CO(G^1)$ eikonal phase.

\section{Eikonal phase at NLO to $\CO(S^4)$} \label{app:eikNLO}
The relation $a^2 = d^2+ (\hat{k_2}\cdot \vec{a} )^2 = a_\perp^2 + (\hat{k_2}\cdot \vec{a} )^2$ was used to simplify some expressions.
\subsection{Cubic order in spin}
\bl
c^{\,\text{S}^3}_{\bar\delta,\mathrm{NLO}} &= - \frac{7 i q^2 (3 C_{\text{S}^2} + C_{\text{S}^3}) (q \cdot S) (k_2 \cdot S) (n \cdot S)}{8 \w} \,,
\\ \bar{c}^{\,\text{S}^3,\,\bar\delta}_{\square,\mathrm{NLO}} &= - \frac{i q^2 (n\cdot S) }{2 m^2 \omega } \left[ (C_{\text{S}^3}+3 C_{\text{S}^2}) \{ m^2 q^2 (k_2 \cdot S)^2 + (n\cdot S)^2 \} \right. \nn
\\ &\phantom{=asdfasdfasdf} \left. + (C_{\text{S}^2}-C_{\text{S}^3}) m^2 \omega ^2 (q\cdot S)^2 \right] \,,
\\ \frac{\bar{\delta}^{\,\text{S}^3}_{\mathrm{NLO}}}{i G^2 m^2} &= \frac{105 \pi (3 C_{\text{S}^2} + C_{\text{S}^3}) (\hat{k}_2 \cdot \vec{a}) (\vec{d} \cdot \vec{b}) (\vec{a} \cdot \vec{b})}{32 b^7} \nn
\\ &\phantom{=as} + \frac{4 (\vec{d}\cdot \vec{b})}{b^8} \left[ (C_{\text{S}^3}+3 C_{\text{S}^2} ) \left\{ 6 (\vec{d}\cdot \vec{b})^2 - b^2 ( d^2 + 2 a^2 )\right\} \right. \nn
\\ &\phantom{=asdfasdfasdf} \left. - (C_{\text{S}^2}-C_{\text{S}^3}) \left\{ b^2 a_{\perp}^2-6 (\vec{a}\cdot \vec{b})^2\right\} \right] \,,
\\ c^{\,\text{S}^3}_{\a,\mathrm{NLO}} &= - \frac{q^2 (q \cdot S)}{64 m \w^2} \left[ (3 C_{\text{S}^2} + C_{\text{S}^3}) \left\{ 13 (n \cdot S)^2 - 7 m^2 q^2 (k_2 \cdot S)^2 \right\}  \right. \nn
\\ & \phantom{=} \left. \phantom{asdfasdfasdf} - 20 \, (C_{\text{S}^3} - C_{\text{S}^2}) m^2 \w^2 (q \cdot S)^2 \right] \,,
\\ \bar{c}^{\,\text{S}^3,\,\a}_{\square,\mathrm{NLO}} &= -\frac{q^4 (k_2\cdot S) \left( (C_{\text{S}^2}-C_{\text{S}^3}) m^2 \omega ^2 (q\cdot S)^2 + (C_{\text{S}^3}+3 C_{\text{S}^2}) (n\cdot S)^2\right)}{4 m \omega ^2} \,,
\\ \frac{\w \,\a^{\,\text{S}^3}_{\mathrm{NLO}} }{i G^2 m^2} &= - \frac{75 \pi (\vec{a} \cdot \vec{b})}{64 b^9} \left[ (C_{\text{S}^3} - C_{\text{S}^2}) \, \{ 7 (\vec{a} \cdot \vec{b})^2 - 3 a_\perp^2 b^2 \} \right. \nn 
\\ &\phantom{=} \left. \phantom{asdf} - \frac{(3 C_{\text{S}^2} + C_{\text{S}^3})}{20} \left\{ 21 (\hat{k}_2 \cdot \vec{a})^2 b^2 + 13 \left( 7 (\vec{d} \cdot \vec{b})^2 - d^2 b^2 \right) \right\} \right] \nn
\\ &\phantom{=} + \frac{4 (\hat{k_2}\cdot \vec{a})}{b^8} \left[ (C_{\text{S}^3}+3 C_{\text{S}^2}) \left\{ b^2 d^2-6 (\vec{d}\cdot \vec{b})^2\right\} \right. \nn
\\ &\phantom{=asdfasdfasdfasdf} \left. + (C_{\text{S}^2}-C_{\text{S}^3}) \left\{ b^2 a_{\perp}^2-6 (\vec{a}\cdot \vec{b})^2\right\} \right] \,.
\el

\subsection{Quartic order in spin}
\bl
c^{\,\text{S}^4}_{\bar\delta,\mathrm{NLO}} &= - \frac{7 q^2 (q\cdot S)(k_2 \cdot S)}{256 m^2 \w^2} \left[ 10 (C_{\text{S}^2}^2 - C_{\text{S}^4}) m^2 \w^2 (q \cdot S)^2 \right. \nn
\\ &\phantom{=asdf}\left. + (3 C_{\text{S}^2}^2 + 4 C_{\text{S}^3} + C_{\text{S}^4}) \left\{ 7 (n \cdot S)^2 - m^2 q^2 (k_2 \cdot S)^2 \right\} \right] \,,
\\ \bar{c}^{\,\text{S}^4,\,\bar\delta}_{\square,\mathrm{NLO}} &= - \frac{q^2}{24 m^4 \omega ^2} \left[ (C_{\text{S}^4}+4 C_{\text{S}^3}+3 C_{\text{S}^2}^2) (n\cdot S)^2 \left\{ 2 (n\cdot S)^2 + 3 m^2 q^2 (k_2\cdot S)^2 \right\} \right. \nn
\\ &\phantom{=asdf} \left. \phantom{asdf} + 3 (C_{\text{S}^2}^2-C_{\text{S}^4}) m^2 \w^2 (q\cdot S)^2 \left\{ 2 (n \cdot S)^2 + m^2 q^2 (k_2\cdot S)^2 \right\} \right] \,,
\\ \frac{\bar{\delta}^{\,\text{S}^4}_{\mathrm{NLO}}}{i G^2 m^2} &= - \frac{105 \pi (\hat{k}_2 \cdot \vec{a}) (\vec{a} \cdot \vec{b})}{512 b^9} \left[ 5  \, (C_{\text{S}^2}^2 - C_{\text{S}^4}) \, \{ 7 (\vec{a} \cdot \vec{b})^2 - 3 a_\perp^2 b^2 \}  \right. \nn
\\ &\phantom{=} \left. \phantom{asdf} + \frac{(3 C_{\text{S}^2}^2 + 4 C_{\text{S}^3} + C_{\text{S}^4})}{2} \left\{ 3 (\hat{k}_2 \cdot \vec{a})^2 b^2 + 7 \left( 7 (\vec{d} \cdot \vec{b})^2 - d^2 b^2 \right) \right\} \right]  \nn
\\ &\phantom{=} - \frac{2}{b^{10}} \left[ (C_{\text{S}^2}^2-C_{\text{S}^4}) \left\{ 6 (\vec{a}\cdot \vec{b})^2 (8 (\vec{d}\cdot \vec{b})^2 - b^2 a^2 ) + b^2 a_{\perp}^2 (b^2 a^2 - 6 (\vec{d}\cdot \vec{b})^2) \right\} \right. \nn
\\ &\phantom{=as} \left. + (C_{\text{S}^4}+4 C_{\text{S}^3}+3 C_{\text{S}^2}^2) \left\{ b^4 d^2 a^2 - 6 b^2 (\vec{d}\cdot \vec{b})^2 (d^2+a^2) + 16 (\vec{d}\cdot \vec{b})^4 \right\} \right]\,,
\\ c^{\,\text{S}^1}_{\a,\mathrm{NLO}} &= \frac{i q^2 (q \cdot S) (n \cdot S)}{768 m^3 \w^3} \left[ (3 C_{\text{S}^2}^2 + 4 C_{\text{S}^3} + C_{\text{S}^4}) \left\{ 23 (n \cdot S)^2 - 21 m^2 q^2 (k_2 \cdot S)^2 \right\} \right. \nn
\\ & \phantom{=} \left. \phantom{asdfasdfasdfasdf} + 90 (C_{\text{S}^2}^2 - C_{\text{S}^4}) m^2 \w^2 (q \cdot S)^2 \right] \,,
\\ \bar{c}^{\,\text{S}^4,\,\a}_{\square,\mathrm{NLO}} &= \frac{i q^4 (k_2\cdot S) (n\cdot S)}{24 m^3 \omega ^3} \left[(C_{\text{S}^4}+4 C_{\text{S}^3}+3 C_{\text{S}^2}^2) (n\cdot S)^2 \right. \nn
\\ & \phantom{=asdf} \left. \phantom{asdfasdfasdfasdf} + 3 m^2 \omega ^2 (C_{\text{S}^2}^2-C_{\text{S}^4}) (q\cdot S)^2\right] \,,
\\ \frac{\w \,\a^{\,\text{S}^4}_{\mathrm{NLO}} }{i G^2 m^2} &= - \frac{105 \pi (\vec{d} \cdot \vec{b}) (\vec{a} \cdot \vec{b})}{512 b^{11}} \left[ 45  \, (C_{\text{S}^2}^2 - C_{\text{S}^4}) \, \{ 3 (\vec{a} \cdot \vec{b})^2 - a_\perp^2 b^2 \} \right. \nn \nn
\\ &\phantom{=} \left. \phantom{asdf} + \frac{(3 C_{\text{S}^2}^2 + 4 C_{\text{S}^3} + C_{\text{S}^4})}{2} \left\{ 21 (\hat{k}_2 \cdot \vec{a})^2 b^2 + 23 \left( 3 (\vec{d} \cdot \vec{b})^2 - d^2 b^2 \right) \right\} \right]  \nn
\\ &\phantom{=} + \frac{4(\hat{k_2}\cdot \vec{a}) (\vec{d}\cdot \vec{b})}{b^{10}} \left[ (C_{\text{S}^4}+4 C_{\text{S}^3}+3 C_{\text{S}^2}^2) \left(8 (\vec{d}\cdot \vec{b})^2-3 b^2 d^2 \right) \right. \nn
\\ &\phantom{=asdf} \left. \phantom{asdf}  - 3 (C_{\text{S}^2}^2-C_{\text{S}^4}) \left(b^2 a_{\perp}^2-8 (\vec{a}\cdot \vec{b})^2\right) \right] \,.
\el

\newpage

\bibliography{mybib}{}

\providecommand{\href}[2]{#2}\begingroup\raggedright\begin{thebibliography}{100}

\bibitem{Ishihara:1987dv}
H.~Ishihara, M.~Takahashi and A.~Tomimatsu, \emph{{GRAVITATIONAL FARADAY
  ROTATION INDUCED BY KERR BLACK HOLE}},
  \href{https://doi.org/10.1103/PhysRevD.38.472}{\emph{Phys. Rev. D} {\bfseries
  38} (1988) 472}.

\bibitem{Bern:2020buy}
Z.~Bern, A.~Luna, R.~Roiban, C.-H.~Shen and M.~Zeng, \emph{{Spinning black hole
  binary dynamics, scattering amplitudes, and effective field theory}},
  \href{https://doi.org/10.1103/PhysRevD.104.065014}{\emph{Phys. Rev. D}
  {\bfseries 104} (2021) 065014}
  [\href{https://arxiv.org/abs/2005.03071}{{\ttfamily 2005.03071}}].

\bibitem{Aoude:2020ygw}
R.~Aoude, K.~Haddad and A.~Helset, \emph{{Tidal effects for spinning
  particles}}, \href{https://doi.org/10.1007/JHEP03(2021)097}{\emph{JHEP}
  {\bfseries 03} (2021) 097}
  [\href{https://arxiv.org/abs/2012.05256}{{\ttfamily 2012.05256}}].

\bibitem{Kosmopoulos:2021zoq}
D.~Kosmopoulos and A.~Luna, \emph{{Quadratic-in-spin Hamiltonian at $
  \mathcal{O} $(G$^{2}$) from scattering amplitudes}},
  \href{https://doi.org/10.1007/JHEP07(2021)037}{\emph{JHEP} {\bfseries 07}
  (2021) 037} [\href{https://arxiv.org/abs/2102.10137}{{\ttfamily
  2102.10137}}].

\bibitem{Chen:2021qkk}
W.-M.~Chen, M.-Z.~Chung, Y.-t.~Huang and J.-W.~Kim, \emph{{The 2PM Hamiltonian
  for binary Kerr to quartic in spin}},
  \href{https://arxiv.org/abs/2111.13639}{{\ttfamily 2111.13639}}.

\bibitem{Aoude:2022trd}
R.~Aoude, K.~Haddad and A.~Helset, \emph{{Searching for Kerr in the 2PM
  amplitude}},  \href{https://arxiv.org/abs/2203.06197}{{\ttfamily
  2203.06197}}.

\bibitem{Bern:2022kto}
Z.~Bern, D.~Kosmopoulos, A.~Luna, R.~Roiban and F.~Teng, \emph{{Binary Dynamics
  Through the Fifth Power of Spin at $\mathcal{O}(G^2)$}},
  \href{https://arxiv.org/abs/2203.06202}{{\ttfamily 2203.06202}}.

\bibitem{Aoude:2022thd}
R.~Aoude, K.~Haddad and A.~Helset, \emph{{Classical gravitational
  spinning-spinless scattering at $\mathcal{O}(G^{2} S^{\infty})$}},
  \href{https://arxiv.org/abs/2205.02809}{{\ttfamily 2205.02809}}.

\bibitem{LIGOScientific:2016aoc}
{\scshape LIGO Scientific, Virgo} collaboration, \emph{{Observation of
  Gravitational Waves from a Binary Black Hole Merger}},
  \href{https://doi.org/10.1103/PhysRevLett.116.061102}{\emph{Phys. Rev. Lett.}
  {\bfseries 116} (2016) 061102}
  [\href{https://arxiv.org/abs/1602.03837}{{\ttfamily 1602.03837}}].

\bibitem{LIGOScientific:2017vwq}
{\scshape LIGO Scientific, Virgo} collaboration, \emph{{GW170817: Observation
  of Gravitational Waves from a Binary Neutron Star Inspiral}},
  \href{https://doi.org/10.1103/PhysRevLett.119.161101}{\emph{Phys. Rev. Lett.}
  {\bfseries 119} (2017) 161101}
  [\href{https://arxiv.org/abs/1710.05832}{{\ttfamily 1710.05832}}].

\bibitem{Duff:1973zz}
M.J.~Duff, \emph{{Quantum Tree Graphs and the Schwarzschild Solution}},
  \href{https://doi.org/10.1103/PhysRevD.7.2317}{\emph{Phys. Rev. D} {\bfseries
  7} (1973) 2317}.

\bibitem{Holstein:2004dn}
B.R.~Holstein and J.F.~Donoghue, \emph{{Classical physics and quantum loops}},
  \href{https://doi.org/10.1103/PhysRevLett.93.201602}{\emph{Phys. Rev. Lett.}
  {\bfseries 93} (2004) 201602}
  [\href{https://arxiv.org/abs/hep-th/0405239}{{\ttfamily hep-th/0405239}}].

\bibitem{Donoghue:2001qc}
J.F.~Donoghue, B.R.~Holstein, B.~Garbrecht and T.~Konstandin, \emph{{Quantum
  corrections to the Reissner-Nordstr\"om and Kerr-Newman metrics}},
  \href{https://doi.org/10.1016/S0370-2693(02)01246-7}{\emph{Phys. Lett. B}
  {\bfseries 529} (2002) 132}
  [\href{https://arxiv.org/abs/hep-th/0112237}{{\ttfamily hep-th/0112237}}].

\bibitem{Bertotti:1956pxu}
B.~Bertotti, \emph{{On gravitational motion}},
  \href{https://doi.org/10.1007/bf02746175}{\emph{Nuovo Cim.} {\bfseries 4}
  (1956) 898}.

\bibitem{Kerr:1959zlt}
R.P.~Kerr, \emph{{The Lorentz-covariant approximation method in general
  relativity I}}, \href{https://doi.org/10.1007/bf02732767}{\emph{Nuovo Cim.}
  {\bfseries 13} (1959) 469}.

\bibitem{Bertotti:1960wuq}
B.~Bertotti and J.~Plebanski, \emph{{Theory of gravitational perturbations in
  the fast motion approximation}},
  \href{https://doi.org/10.1016/0003-4916(60)90132-9}{\emph{Annals Phys.}
  {\bfseries 11} (1960) 169}.

\bibitem{Portilla:1979xx}
M.~Portilla, \emph{{MOMENTUM AND ANGULAR MOMENTUM OF TWO GRAVITATING
  PARTICLES}}, \href{https://doi.org/10.1088/0305-4470/12/7/025}{\emph{J. Phys.
  A} {\bfseries 12} (1979) 1075}.

\bibitem{Westpfahl:1979gu}
K.~Westpfahl and M.~Goller, \emph{{GRAVITATIONAL SCATTERING OF TWO RELATIVISTIC
  PARTICLES IN POSTLINEAR APPROXIMATION}},
  \href{https://doi.org/10.1007/BF02817047}{\emph{Lett. Nuovo Cim.} {\bfseries
  26} (1979) 573}.

\bibitem{Portilla:1980uz}
M.~Portilla, \emph{{SCATTERING OF TWO GRAVITATING PARTICLES: CLASSICAL
  APPROACH}}, \href{https://doi.org/10.1088/0305-4470/13/12/017}{\emph{J. Phys.
  A} {\bfseries 13} (1980) 3677}.

\bibitem{Bel:1981be}
L.~Bel, T.~Damour, N.~Deruelle, J.~Ibanez and J.~Martin,
  \emph{{Poincar\'e-invariant gravitational field and equations of motion of
  two pointlike objects: The postlinear approximation of general relativity}},
  \href{https://doi.org/10.1007/BF00756073}{\emph{Gen. Rel. Grav.} {\bfseries
  13} (1981) 963}.

\bibitem{Westpfahl:1985tsl}
K.~Westpfahl, \emph{{High-Speed Scattering of Charged and Uncharged Particles
  in General Relativity}},
  \href{https://doi.org/10.1002/prop.2190330802}{\emph{Fortsch. Phys.}
  {\bfseries 33} (1985) 417}.

\bibitem{Ledvinka:2008tk}
T.~Ledvinka, G.~Schaefer and J.~Bicak, \emph{{Relativistic Closed-Form
  Hamiltonian for Many-Body Gravitating Systems in the Post-Minkowskian
  Approximation}},
  \href{https://doi.org/10.1103/PhysRevLett.100.251101}{\emph{Phys. Rev. Lett.}
  {\bfseries 100} (2008) 251101}
  [\href{https://arxiv.org/abs/0807.0214}{{\ttfamily 0807.0214}}].

\bibitem{Damour:2017zjx}
T.~Damour, \emph{{High-energy gravitational scattering and the general
  relativistic two-body problem}},
  \href{https://doi.org/10.1103/PhysRevD.97.044038}{\emph{Phys. Rev. D}
  {\bfseries 97} (2018) 044038}
  [\href{https://arxiv.org/abs/1710.10599}{{\ttfamily 1710.10599}}].

\bibitem{Donoghue:1994dn}
J.F.~Donoghue, \emph{{General relativity as an effective field theory: The
  leading quantum corrections}},
  \href{https://doi.org/10.1103/PhysRevD.50.3874}{\emph{Phys. Rev. D}
  {\bfseries 50} (1994) 3874}
  [\href{https://arxiv.org/abs/gr-qc/9405057}{{\ttfamily gr-qc/9405057}}].

\bibitem{Bjerrum-Bohr:2002gqz}
N.E.J.~Bjerrum-Bohr, J.F.~Donoghue and B.R.~Holstein, \emph{{Quantum
  gravitational corrections to the nonrelativistic scattering potential of two
  masses}}, \href{https://doi.org/10.1103/PhysRevD.71.069903}{\emph{Phys. Rev.
  D} {\bfseries 67} (2003) 084033}
  [\href{https://arxiv.org/abs/hep-th/0211072}{{\ttfamily hep-th/0211072}}].

\bibitem{Bjerrum-Bohr:2013bxa}
N.E.J.~Bjerrum-Bohr, J.F.~Donoghue and P.~Vanhove, \emph{{On-shell Techniques
  and Universal Results in Quantum Gravity}},
  \href{https://doi.org/10.1007/JHEP02(2014)111}{\emph{JHEP} {\bfseries 02}
  (2014) 111} [\href{https://arxiv.org/abs/1309.0804}{{\ttfamily 1309.0804}}].

\bibitem{Neill:2013wsa}
D.~Neill and I.Z.~Rothstein, \emph{{Classical Space-Times from the S Matrix}},
  \href{https://doi.org/10.1016/j.nuclphysb.2013.09.007}{\emph{Nucl. Phys. B}
  {\bfseries 877} (2013) 177}
  [\href{https://arxiv.org/abs/1304.7263}{{\ttfamily 1304.7263}}].

\bibitem{Cheung:2018wkq}
C.~Cheung, I.Z.~Rothstein and M.P.~Solon, \emph{{From Scattering Amplitudes to
  Classical Potentials in the Post-Minkowskian Expansion}},
  \href{https://doi.org/10.1103/PhysRevLett.121.251101}{\emph{Phys. Rev. Lett.}
  {\bfseries 121} (2018) 251101}
  [\href{https://arxiv.org/abs/1808.02489}{{\ttfamily 1808.02489}}].

\bibitem{Bern:2019nnu}
Z.~Bern, C.~Cheung, R.~Roiban, C.-H.~Shen, M.P.~Solon and M.~Zeng,
  \emph{{Scattering Amplitudes and the Conservative Hamiltonian for Binary
  Systems at Third Post-Minkowskian Order}},
  \href{https://doi.org/10.1103/PhysRevLett.122.201603}{\emph{Phys. Rev. Lett.}
  {\bfseries 122} (2019) 201603}
  [\href{https://arxiv.org/abs/1901.04424}{{\ttfamily 1901.04424}}].

\bibitem{Bern:2019crd}
Z.~Bern, C.~Cheung, R.~Roiban, C.-H.~Shen, M.P.~Solon and M.~Zeng, \emph{{Black
  Hole Binary Dynamics from the Double Copy and Effective Theory}},
  \href{https://doi.org/10.1007/JHEP10(2019)206}{\emph{JHEP} {\bfseries 10}
  (2019) 206} [\href{https://arxiv.org/abs/1908.01493}{{\ttfamily
  1908.01493}}].

\bibitem{Bern:2021dqo}
Z.~Bern, J.~Parra-Martinez, R.~Roiban, M.S.~Ruf, C.-H.~Shen, M.P.~Solon et~al.,
  \emph{{Scattering Amplitudes and Conservative Binary Dynamics at ${\cal
  O}(G^4)$}}, \href{https://doi.org/10.1103/PhysRevLett.126.171601}{\emph{Phys.
  Rev. Lett.} {\bfseries 126} (2021) 171601}
  [\href{https://arxiv.org/abs/2101.07254}{{\ttfamily 2101.07254}}].

\bibitem{Einstein:1938yz}
A.~Einstein, L.~Infeld and B.~Hoffmann, \emph{{The Gravitational equations and
  the problem of motion}}, \href{https://doi.org/10.2307/1968714}{\emph{Annals
  Math.} {\bfseries 39} (1938) 65}.

\bibitem{Ohta:1973je}
T.~Ohta, H.~Okamura, T.~Kimura and K.~Hiida, \emph{{Physically acceptable
  solution of einstein's equation for many-body system}},
  \href{https://doi.org/10.1143/PTP.50.492}{\emph{Prog. Theor. Phys.}
  {\bfseries 50} (1973) 492}.

\bibitem{Jaranowski:1997ky}
P.~Jaranowski and G.~Schaefer, \emph{{Third postNewtonian higher order ADM
  Hamilton dynamics for two-body point mass systems}},
  \href{https://doi.org/10.1103/PhysRevD.57.7274}{\emph{Phys. Rev. D}
  {\bfseries 57} (1998) 7274}
  [\href{https://arxiv.org/abs/gr-qc/9712075}{{\ttfamily gr-qc/9712075}}].

\bibitem{Damour:1999cr}
T.~Damour, P.~Jaranowski and G.~Schaefer, \emph{{Dynamical invariants for
  general relativistic two-body systems at the third postNewtonian
  approximation}},
  \href{https://doi.org/10.1103/PhysRevD.62.044024}{\emph{Phys. Rev. D}
  {\bfseries 62} (2000) 044024}
  [\href{https://arxiv.org/abs/gr-qc/9912092}{{\ttfamily gr-qc/9912092}}].

\bibitem{Blanchet:2000nv}
L.~Blanchet and G.~Faye, \emph{{Equations of motion of point particle binaries
  at the third postNewtonian order}},
  \href{https://doi.org/10.1016/S0375-9601(00)00360-1}{\emph{Phys. Lett. A}
  {\bfseries 271} (2000) 58}
  [\href{https://arxiv.org/abs/gr-qc/0004009}{{\ttfamily gr-qc/0004009}}].

\bibitem{Damour:2001bu}
T.~Damour, P.~Jaranowski and G.~Schaefer, \emph{{Dimensional regularization of
  the gravitational interaction of point masses}},
  \href{https://doi.org/10.1016/S0370-2693(01)00642-6}{\emph{Phys. Lett. B}
  {\bfseries 513} (2001) 147}
  [\href{https://arxiv.org/abs/gr-qc/0105038}{{\ttfamily gr-qc/0105038}}].

\bibitem{Damour:2014jta}
T.~Damour, P.~Jaranowski and G.~Sch\"afer, \emph{{Nonlocal-in-time action for
  the fourth post-Newtonian conservative dynamics of two-body systems}},
  \href{https://doi.org/10.1103/PhysRevD.89.064058}{\emph{Phys. Rev. D}
  {\bfseries 89} (2014) 064058}
  [\href{https://arxiv.org/abs/1401.4548}{{\ttfamily 1401.4548}}].

\bibitem{Jaranowski:2015lha}
P.~Jaranowski and G.~Sch\"afer, \emph{{Derivation of local-in-time fourth
  post-Newtonian ADM Hamiltonian for spinless compact binaries}},
  \href{https://doi.org/10.1103/PhysRevD.92.124043}{\emph{Phys. Rev. D}
  {\bfseries 92} (2015) 124043}
  [\href{https://arxiv.org/abs/1508.01016}{{\ttfamily 1508.01016}}].

\bibitem{Holstein:2008sx}
B.R.~Holstein and A.~Ross, \emph{{Spin Effects in Long Range Gravitational
  Scattering}},  \href{https://arxiv.org/abs/0802.0716}{{\ttfamily 0802.0716}}.

\bibitem{Vaidya:2014kza}
V.~Vaidya, \emph{{Gravitational spin Hamiltonians from the S matrix}},
  \href{https://doi.org/10.1103/PhysRevD.91.024017}{\emph{Phys. Rev. D}
  {\bfseries 91} (2015) 024017}
  [\href{https://arxiv.org/abs/1410.5348}{{\ttfamily 1410.5348}}].

\bibitem{Arkani-Hamed:2017jhn}
N.~Arkani-Hamed, T.-C.~Huang and Y.-t.~Huang, \emph{{Scattering amplitudes for
  all masses and spins}},
  \href{https://doi.org/10.1007/JHEP11(2021)070}{\emph{JHEP} {\bfseries 11}
  (2021) 070} [\href{https://arxiv.org/abs/1709.04891}{{\ttfamily
  1709.04891}}].

\bibitem{Maybee:2019jus}
B.~Maybee, D.~O'Connell and J.~Vines, \emph{{Observables and amplitudes for
  spinning particles and black holes}},
  \href{https://doi.org/10.1007/JHEP12(2019)156}{\emph{JHEP} {\bfseries 12}
  (2019) 156} [\href{https://arxiv.org/abs/1906.09260}{{\ttfamily
  1906.09260}}].

\bibitem{Guevara:2018wpp}
A.~Guevara, A.~Ochirov and J.~Vines, \emph{{Scattering of Spinning Black Holes
  from Exponentiated Soft Factors}},
  \href{https://doi.org/10.1007/JHEP09(2019)056}{\emph{JHEP} {\bfseries 09}
  (2019) 056} [\href{https://arxiv.org/abs/1812.06895}{{\ttfamily
  1812.06895}}].

\bibitem{Chung:2018kqs}
M.-Z.~Chung, Y.-T.~Huang, J.-W.~Kim and S.~Lee, \emph{{The simplest massive
  S-matrix: from minimal coupling to Black Holes}},
  \href{https://doi.org/10.1007/JHEP04(2019)156}{\emph{JHEP} {\bfseries 04}
  (2019) 156} [\href{https://arxiv.org/abs/1812.08752}{{\ttfamily
  1812.08752}}].

\bibitem{Chung:2019duq}
M.-Z.~Chung, Y.-T.~Huang and J.-W.~Kim, \emph{{Classical potential for general
  spinning bodies}}, \href{https://doi.org/10.1007/JHEP09(2020)074}{\emph{JHEP}
  {\bfseries 09} (2020) 074}
  [\href{https://arxiv.org/abs/1908.08463}{{\ttfamily 1908.08463}}].

\bibitem{Guevara:2019fsj}
A.~Guevara, A.~Ochirov and J.~Vines, \emph{{Black-hole scattering with general
  spin directions from minimal-coupling amplitudes}},
  \href{https://doi.org/10.1103/PhysRevD.100.104024}{\emph{Phys. Rev. D}
  {\bfseries 100} (2019) 104024}
  [\href{https://arxiv.org/abs/1906.10071}{{\ttfamily 1906.10071}}].

\bibitem{Arkani-Hamed:2019ymq}
N.~Arkani-Hamed, Y.-t.~Huang and D.~O'Connell, \emph{{Kerr black holes as
  elementary particles}},
  \href{https://doi.org/10.1007/JHEP01(2020)046}{\emph{JHEP} {\bfseries 01}
  (2020) 046} [\href{https://arxiv.org/abs/1906.10100}{{\ttfamily
  1906.10100}}].

\bibitem{Aoude:2020onz}
R.~Aoude, K.~Haddad and A.~Helset, \emph{{On-shell heavy particle effective
  theories}}, \href{https://doi.org/10.1007/JHEP05(2020)051}{\emph{JHEP}
  {\bfseries 05} (2020) 051}
  [\href{https://arxiv.org/abs/2001.09164}{{\ttfamily 2001.09164}}].

\bibitem{Aoude:2021oqj}
R.~Aoude and A.~Ochirov, \emph{{Classical observables from coherent-spin
  amplitudes}}, \href{https://doi.org/10.1007/JHEP10(2021)008}{\emph{JHEP}
  {\bfseries 10} (2021) 008}
  [\href{https://arxiv.org/abs/2108.01649}{{\ttfamily 2108.01649}}].

\bibitem{Porto:2005ac}
R.A.~Porto, \emph{{Post-Newtonian corrections to the motion of spinning bodies
  in NRGR}}, \href{https://doi.org/10.1103/PhysRevD.73.104031}{\emph{Phys. Rev.
  D} {\bfseries 73} (2006) 104031}
  [\href{https://arxiv.org/abs/gr-qc/0511061}{{\ttfamily gr-qc/0511061}}].

\bibitem{Porto:2008jj}
R.A.~Porto and I.Z.~Rothstein, \emph{{Next to Leading Order Spin(1)Spin(1)
  Effects in the Motion of Inspiralling Compact Binaries}},
  \href{https://doi.org/10.1103/PhysRevD.78.044013}{\emph{Phys. Rev. D}
  {\bfseries 78} (2008) 044013}
  [\href{https://arxiv.org/abs/0804.0260}{{\ttfamily 0804.0260}}].

\bibitem{Levi:2015msa}
M.~Levi and J.~Steinhoff, \emph{{Spinning gravitating objects in the effective
  field theory in the post-Newtonian scheme}},
  \href{https://doi.org/10.1007/JHEP09(2015)219}{\emph{JHEP} {\bfseries 09}
  (2015) 219} [\href{https://arxiv.org/abs/1501.04956}{{\ttfamily
  1501.04956}}].

\bibitem{Levi:2019kgk}
M.~Levi, S.~Mougiakakos and M.~Vieira, \emph{{Gravitational cubic-in-spin
  interaction at the next-to-leading post-Newtonian order}},
  \href{https://doi.org/10.1007/JHEP01(2021)036}{\emph{JHEP} {\bfseries 01}
  (2021) 036} [\href{https://arxiv.org/abs/1912.06276}{{\ttfamily
  1912.06276}}].

\bibitem{Levi:2020uwu}
M.~Levi, A.J.~Mcleod and M.~Von~Hippel, \emph{{N$^{3}$LO gravitational
  quadratic-in-spin interactions at G$^{4}$}},
  \href{https://doi.org/10.1007/JHEP07(2021)116}{\emph{JHEP} {\bfseries 07}
  (2021) 116} [\href{https://arxiv.org/abs/2003.07890}{{\ttfamily
  2003.07890}}].

\bibitem{Levi:2020lfn}
M.~Levi and F.~Teng, \emph{{NLO gravitational quartic-in-spin interaction}},
  \href{https://doi.org/10.1007/JHEP01(2021)066}{\emph{JHEP} {\bfseries 01}
  (2021) 066} [\href{https://arxiv.org/abs/2008.12280}{{\ttfamily
  2008.12280}}].

\bibitem{Kim:2021rfj}
J.-W.~Kim, M.~Levi and Z.~Yin, \emph{{Quadratic-in-spin interactions at fifth
  post-Newtonian order probe new physics}},
  \href{https://arxiv.org/abs/2112.01509}{{\ttfamily 2112.01509}}.

\bibitem{Liu:2021zxr}
Z.~Liu, R.A.~Porto and Z.~Yang, \emph{{Spin Effects in the Effective Field
  Theory Approach to Post-Minkowskian Conservative Dynamics}},
  \href{https://doi.org/10.1007/JHEP06(2021)012}{\emph{JHEP} {\bfseries 06}
  (2021) 012} [\href{https://arxiv.org/abs/2102.10059}{{\ttfamily
  2102.10059}}].

\bibitem{Jakobsen:2021lvp}
G.U.~Jakobsen, G.~Mogull, J.~Plefka and J.~Steinhoff, \emph{{Gravitational
  Bremsstrahlung and Hidden Supersymmetry of Spinning Bodies}},
  \href{https://doi.org/10.1103/PhysRevLett.128.011101}{\emph{Phys. Rev. Lett.}
  {\bfseries 128} (2022) 011101}
  [\href{https://arxiv.org/abs/2106.10256}{{\ttfamily 2106.10256}}].

\bibitem{Jakobsen:2021zvh}
G.U.~Jakobsen, G.~Mogull, J.~Plefka and J.~Steinhoff, \emph{{SUSY in the sky
  with gravitons}}, \href{https://doi.org/10.1007/JHEP01(2022)027}{\emph{JHEP}
  {\bfseries 01} (2022) 027}
  [\href{https://arxiv.org/abs/2109.04465}{{\ttfamily 2109.04465}}].

\bibitem{Jakobsen:2022fcj}
G.U.~Jakobsen and G.~Mogull, \emph{{Conservative and Radiative Dynamics of
  Spinning Bodies at Third Post-Minkowskian Order Using Worldline Quantum Field
  Theory}}, \href{https://doi.org/10.1103/PhysRevLett.128.141102}{\emph{Phys.
  Rev. Lett.} {\bfseries 128} (2022) 141102}
  [\href{https://arxiv.org/abs/2201.07778}{{\ttfamily 2201.07778}}].

\bibitem{Chung:2020rrz}
M.-Z.~Chung, Y.-t.~Huang, J.-W.~Kim and S.~Lee, \emph{{Complete Hamiltonian for
  spinning binary systems at first post-Minkowskian order}},
  \href{https://doi.org/10.1007/JHEP05(2020)105}{\emph{JHEP} {\bfseries 05}
  (2020) 105} [\href{https://arxiv.org/abs/2003.06600}{{\ttfamily
  2003.06600}}].

\bibitem{Chiodaroli:2021eug}
M.~Chiodaroli, H.~Johansson and P.~Pichini, \emph{{Compton Black-Hole
  Scattering for $s \leq 5/2$}},
  \href{https://arxiv.org/abs/2107.14779}{{\ttfamily 2107.14779}}.

\bibitem{Falkowski:2020aso}
A.~Falkowski and C.S.~Machado, \emph{{Soft Matters, or the Recursions with
  Massive Spinors}}, \href{https://doi.org/10.1007/JHEP05(2021)238}{\emph{JHEP}
  {\bfseries 05} (2021) 238}
  [\href{https://arxiv.org/abs/2005.08981}{{\ttfamily 2005.08981}}].

\bibitem{Aoude:2020mlg}
R.~Aoude, M.-Z.~Chung, Y.-t.~Huang, C.S.~Machado and M.-K.~Tam, \emph{{Silence
  of Binary Kerr Black Holes}},
  \href{https://doi.org/10.1103/PhysRevLett.125.181602}{\emph{Phys. Rev. Lett.}
  {\bfseries 125} (2020) 181602}
  [\href{https://arxiv.org/abs/2007.09486}{{\ttfamily 2007.09486}}].

\bibitem{Chen:2021huj}
B.-T.~Chen, M.-Z.~Chung, Y.-t.~Huang and M.K.~Tam, \emph{{Minimal spin
  deflection of Kerr-Newman and supersymmetric black hole}},
  \href{https://doi.org/10.1007/JHEP10(2021)011}{\emph{JHEP} {\bfseries 10}
  (2021) 011} [\href{https://arxiv.org/abs/2106.12518}{{\ttfamily
  2106.12518}}].

\bibitem{Guevara:2017csg}
A.~Guevara, \emph{{Holomorphic Classical Limit for Spin Effects in
  Gravitational and Electromagnetic Scattering}},
  \href{https://doi.org/10.1007/JHEP04(2019)033}{\emph{JHEP} {\bfseries 04}
  (2019) 033} [\href{https://arxiv.org/abs/1706.02314}{{\ttfamily
  1706.02314}}].

\bibitem{Damgaard:2019lfh}
P.H.~Damgaard, K.~Haddad and A.~Helset, \emph{{Heavy Black Hole Effective
  Theory}}, \href{https://doi.org/10.1007/JHEP11(2019)070}{\emph{JHEP}
  {\bfseries 11} (2019) 070}
  [\href{https://arxiv.org/abs/1908.10308}{{\ttfamily 1908.10308}}].

\bibitem{Bautista:2019tdr}
Y.F.~Bautista and A.~Guevara, \emph{{From Scattering Amplitudes to Classical
  Physics: Universality, Double Copy and Soft Theorems}},
  \href{https://arxiv.org/abs/1903.12419}{{\ttfamily 1903.12419}}.

\bibitem{JustinNew}
Y.F.~Bautista, A.~Guevara, C.~Kavanagh and J.~Vines, \emph{{to be published}},
  .

\bibitem{Bjerrum-Bohr:2014zsa}
N.E.J.~Bjerrum-Bohr, J.F.~Donoghue, B.R.~Holstein, L.~Plant\'e and P.~Vanhove,
  \emph{{Bending of Light in Quantum Gravity}},
  \href{https://doi.org/10.1103/PhysRevLett.114.061301}{\emph{Phys. Rev. Lett.}
  {\bfseries 114} (2015) 061301}
  [\href{https://arxiv.org/abs/1410.7590}{{\ttfamily 1410.7590}}].

\bibitem{Bjerrum-Bohr:2016hpa}
N.E.J.~Bjerrum-Bohr, J.F.~Donoghue, B.R.~Holstein, L.~Plante and P.~Vanhove,
  \emph{{Light-like Scattering in Quantum Gravity}},
  \href{https://doi.org/10.1007/JHEP11(2016)117}{\emph{JHEP} {\bfseries 11}
  (2016) 117} [\href{https://arxiv.org/abs/1609.07477}{{\ttfamily
  1609.07477}}].

\bibitem{Bai:2016ivl}
D.~Bai and Y.~Huang, \emph{{More on the Bending of Light in Quantum Gravity}},
  \href{https://doi.org/10.1103/PhysRevD.95.064045}{\emph{Phys. Rev. D}
  {\bfseries 95} (2017) 064045}
  [\href{https://arxiv.org/abs/1612.07629}{{\ttfamily 1612.07629}}].

\bibitem{Chi:2019owc}
H.-H.~Chi, \emph{{Graviton Bending in Quantum Gravity from One-Loop
  Amplitudes}}, \href{https://doi.org/10.1103/PhysRevD.99.126008}{\emph{Phys.
  Rev. D} {\bfseries 99} (2019) 126008}
  [\href{https://arxiv.org/abs/1903.07944}{{\ttfamily 1903.07944}}].

\bibitem{Camanho:2014apa}
X.O.~Camanho, J.D.~Edelstein, J.~Maldacena and A.~Zhiboedov, \emph{{Causality
  Constraints on Corrections to the Graviton Three-Point Coupling}},
  \href{https://doi.org/10.1007/JHEP02(2016)020}{\emph{JHEP} {\bfseries 02}
  (2016) 020} [\href{https://arxiv.org/abs/1407.5597}{{\ttfamily 1407.5597}}].

\bibitem{AccettulliHuber:2020oou}
M.~Accettulli~Huber, A.~Brandhuber, S.~De~Angelis and G.~Travaglini,
  \emph{{Eikonal phase matrix, deflection angle and time delay in effective
  field theories of gravity}},
  \href{https://doi.org/10.1103/PhysRevD.102.046014}{\emph{Phys. Rev. D}
  {\bfseries 102} (2020) 046014}
  [\href{https://arxiv.org/abs/2006.02375}{{\ttfamily 2006.02375}}].

\bibitem{Cheng:1969eh}
H.~Cheng and T.T.~Wu, \emph{{High-energy elastic scattering in quantum
  electrodynamics}},
  \href{https://doi.org/10.1103/PhysRevLett.22.666}{\emph{Phys. Rev. Lett.}
  {\bfseries 22} (1969) 666}.

\bibitem{Abarbanel:1969ek}
H.D.I.~Abarbanel and C.~Itzykson, \emph{{Relativistic eikonal expansion}},
  \href{https://doi.org/10.1103/PhysRevLett.23.53}{\emph{Phys. Rev. Lett.}
  {\bfseries 23} (1969) 53}.

\bibitem{Levy:1969cr}
M.~Levy and J.~Sucher, \emph{{Eikonal approximation in quantum field theory}},
  \href{https://doi.org/10.1103/PhysRev.186.1656}{\emph{Phys. Rev.} {\bfseries
  186} (1969) 1656}.

\bibitem{Amati:1987wq}
D.~Amati, M.~Ciafaloni and G.~Veneziano, \emph{{Superstring Collisions at
  Planckian Energies}},
  \href{https://doi.org/10.1016/0370-2693(87)90346-7}{\emph{Phys. Lett. B}
  {\bfseries 197} (1987) 81}.

\bibitem{Amati:1987uf}
D.~Amati, M.~Ciafaloni and G.~Veneziano, \emph{{Classical and Quantum Gravity
  Effects from Planckian Energy Superstring Collisions}},
  \href{https://doi.org/10.1142/S0217751X88000710}{\emph{Int. J. Mod. Phys. A}
  {\bfseries 3} (1988) 1615}.

\bibitem{Kabat:1992tb}
D.N.~Kabat and M.~Ortiz, \emph{{Eikonal quantum gravity and Planckian
  scattering}}, \href{https://doi.org/10.1016/0550-3213(92)90627-N}{\emph{Nucl.
  Phys. B} {\bfseries 388} (1992) 570}
  [\href{https://arxiv.org/abs/hep-th/9203082}{{\ttfamily hep-th/9203082}}].

\bibitem{Akhoury:2013yua}
R.~Akhoury, R.~Saotome and G.~Sterman, \emph{{High Energy Scattering in
  Perturbative Quantum Gravity at Next to Leading Power}},
  \href{https://doi.org/10.1103/PhysRevD.103.064036}{\emph{Phys. Rev. D}
  {\bfseries 103} (2021) 064036}
  [\href{https://arxiv.org/abs/1308.5204}{{\ttfamily 1308.5204}}].

\bibitem{KoemansCollado:2019ggb}
A.~Koemans~Collado, P.~Di~Vecchia and R.~Russo, \emph{{Revisiting the second
  post-Minkowskian eikonal and the dynamics of binary black holes}},
  \href{https://doi.org/10.1103/PhysRevD.100.066028}{\emph{Phys. Rev. D}
  {\bfseries 100} (2019) 066028}
  [\href{https://arxiv.org/abs/1904.02667}{{\ttfamily 1904.02667}}].

\bibitem{Nouri-Zonoz:1999jls}
M.~Nouri-Zonoz, \emph{{Gravoelectromagnetic approach to the gravitational
  Faraday rotation in stationary space-times}},
  \href{https://doi.org/10.1103/PhysRevD.60.024013}{\emph{Phys. Rev. D}
  {\bfseries 60} (1999) 024013}
  [\href{https://arxiv.org/abs/gr-qc/9901011}{{\ttfamily gr-qc/9901011}}].

\bibitem{Sereno:2004jx}
M.~Sereno, \emph{{Gravitational Faraday rotation in a weak gravitational
  field}}, \href{https://doi.org/10.1103/PhysRevD.69.087501}{\emph{Phys. Rev.
  D} {\bfseries 69} (2004) 087501}
  [\href{https://arxiv.org/abs/astro-ph/0401295}{{\ttfamily
  astro-ph/0401295}}].

\bibitem{Brodutch:2011eh}
A.~Brodutch and D.R.~Terno, \emph{{Polarization rotation, reference frames and
  Mach's principle}},
  \href{https://doi.org/10.1103/PhysRevD.84.121501}{\emph{Phys. Rev. D}
  {\bfseries 84} (2011) 121501}
  [\href{https://arxiv.org/abs/1107.1274}{{\ttfamily 1107.1274}}].

\bibitem{Farooqui:2013rga}
A.~Farooqui, N.~Kamran and P.~Panangaden, \emph{{An Exact Expression for Photon
  Polarization in Kerr Geometry}},
  \href{https://doi.org/10.4310/ATMP.2014.v18.n3.a3}{\emph{Adv. Theor. Math.
  Phys.} {\bfseries 18} (2014) 659}
  [\href{https://arxiv.org/abs/1306.6292}{{\ttfamily 1306.6292}}].

\bibitem{Shoom:2020zhr}
A.A.~Shoom, \emph{{Gravitational Faraday and spin-Hall effects of light}},
  \href{https://doi.org/10.1103/PhysRevD.104.084007}{\emph{Phys. Rev. D}
  {\bfseries 104} (2021) 084007}
  [\href{https://arxiv.org/abs/2006.10077}{{\ttfamily 2006.10077}}].

\bibitem{Deriglazov:2021gwa}
A.A.~Deriglazov, \emph{{Massless polarized particle and Faraday rotation of
  light in the Schwarzschild spacetime}},
  \href{https://doi.org/10.1103/PhysRevD.104.025006}{\emph{Phys. Rev. D}
  {\bfseries 104} (2021) 025006}
  [\href{https://arxiv.org/abs/2103.07794}{{\ttfamily 2103.07794}}].

\bibitem{Chakraborty:2021bsb}
C.~Chakraborty, \emph{{Gravitational analogue of Faraday rotation in the
  magnetized Kerr and Reissner-Nordstr\"om spacetimes}},
  \href{https://arxiv.org/abs/2106.03520}{{\ttfamily 2106.03520}}.

\bibitem{Li:2022izh}
Z.~Li, W.~Zhao and X.~Er, \emph{{Gravitational Faraday rotation of
  gravitational waves by a Kerr black hole}},
  \href{https://arxiv.org/abs/2204.10512}{{\ttfamily 2204.10512}}.

\bibitem{Cristofoli:2021vyo}
A.~Cristofoli, R.~Gonzo, D.A.~Kosower and D.~O'Connell, \emph{{Waveforms from
  Amplitudes}},  \href{https://arxiv.org/abs/2107.10193}{{\ttfamily
  2107.10193}}.

\bibitem{Brandhuber:2021eyq}
A.~Brandhuber, G.~Chen, G.~Travaglini and C.~Wen, \emph{{Classical
  gravitational scattering from a gauge-invariant double copy}},
  \href{https://doi.org/10.1007/JHEP10(2021)118}{\emph{JHEP} {\bfseries 10}
  (2021) 118} [\href{https://arxiv.org/abs/2108.04216}{{\ttfamily
  2108.04216}}].

\bibitem{Asada:2000vn}
H.~Asada and M.~Kasai, \emph{{Can we see a rotating gravitational lens?}},
  \href{https://doi.org/10.1143/PTP.104.95}{\emph{Prog. Theor. Phys.}
  {\bfseries 104} (2000) 95}
  [\href{https://arxiv.org/abs/astro-ph/0006157}{{\ttfamily
  astro-ph/0006157}}].

\bibitem{landau1975classical}
L.~Landau, E.~Lifshitz and M.~Hamermesh, \emph{The Classical Theory of Fields:
  Volume 2}, Course of theoretical physics, Elsevier Science (1975).

\bibitem{Isaacson:1968hbi}
R.A.~Isaacson, \emph{{Gravitational Radiation in the Limit of High Frequency.
  I. The Linear Approximation and Geometrical Optics}},
  \href{https://doi.org/10.1103/PhysRev.166.1263}{\emph{Phys. Rev.} {\bfseries
  166} (1968) 1263}.

\bibitem{Bautista:2021wfy}
Y.F.~Bautista, A.~Guevara, C.~Kavanagh and J.~Vines, \emph{{From Scattering in
  Black Hole Backgrounds to Higher-Spin Amplitudes: Part I}},
  \href{https://arxiv.org/abs/2107.10179}{{\ttfamily 2107.10179}}.

\bibitem{Kim:2020cvf}
J.-W.~Kim and M.~Shim, \emph{{Gravitational Dyonic Amplitude at One-Loop and
  its Inconsistency with the Classical Impulse}},
  \href{https://doi.org/10.1007/JHEP02(2021)217}{\emph{JHEP} {\bfseries 02}
  (2021) 217} [\href{https://arxiv.org/abs/2010.14347}{{\ttfamily
  2010.14347}}].

\bibitem{Bern:1994zx}
Z.~Bern, L.J.~Dixon, D.C.~Dunbar and D.A.~Kosower, \emph{{One loop n point
  gauge theory amplitudes, unitarity and collinear limits}},
  \href{https://doi.org/10.1016/0550-3213(94)90179-1}{\emph{Nucl. Phys. B}
  {\bfseries 425} (1994) 217}
  [\href{https://arxiv.org/abs/hep-ph/9403226}{{\ttfamily hep-ph/9403226}}].

\bibitem{Bern:1994cg}
Z.~Bern, L.J.~Dixon, D.C.~Dunbar and D.A.~Kosower, \emph{{Fusing gauge theory
  tree amplitudes into loop amplitudes}},
  \href{https://doi.org/10.1016/0550-3213(94)00488-Z}{\emph{Nucl. Phys. B}
  {\bfseries 435} (1995) 59}
  [\href{https://arxiv.org/abs/hep-ph/9409265}{{\ttfamily hep-ph/9409265}}].

\bibitem{Bern:1997sc}
Z.~Bern, L.J.~Dixon and D.A.~Kosower, \emph{{One loop amplitudes for e+ e- to
  four partons}},
  \href{https://doi.org/10.1016/S0550-3213(97)00703-7}{\emph{Nucl. Phys. B}
  {\bfseries 513} (1998) 3}
  [\href{https://arxiv.org/abs/hep-ph/9708239}{{\ttfamily hep-ph/9708239}}].

\bibitem{Forde:2007mi}
D.~Forde, \emph{{Direct extraction of one-loop integral coefficients}},
  \href{https://doi.org/10.1103/PhysRevD.75.125019}{\emph{Phys. Rev. D}
  {\bfseries 75} (2007) 125019}
  [\href{https://arxiv.org/abs/0704.1835}{{\ttfamily 0704.1835}}].

\bibitem{Kilgore:2007qr}
W.B.~Kilgore, \emph{{One-loop Integral Coefficients from Generalized
  Unitarity}},  \href{https://arxiv.org/abs/0711.5015}{{\ttfamily 0711.5015}}.

\bibitem{Cheng:1971gf}
H.~Cheng and T.T.~Wu, \emph{{High-energy scattering of a fermion with anomalous
  magnetic moment - nonexponentiation}},
  \href{https://doi.org/10.1103/PhysRevD.3.2394}{\emph{Phys. Rev. D} {\bfseries
  3} (1971) 2394}.

\bibitem{Meng:1972xt}
T.-C.~Meng, \emph{{High-energy scattering of a charged vector meson in a static
  field - simple exponentiation and s-channel helicity conservation}},
  \href{https://doi.org/10.1103/PhysRevD.6.1169}{\emph{Phys. Rev. D} {\bfseries
  6} (1972) 1169}.

\bibitem{Weinberg:1971cdi}
S.~Weinberg, \emph{{Exponentiation and sum rules}},
  \href{https://doi.org/10.1016/0370-2693(71)90354-6}{\emph{Phys. Lett. B}
  {\bfseries 37} (1971) 494}.

\bibitem{Czyz:1975bf}
W.~Czyz and P.K.~Kabir, \emph{{High-Energy Scattering of Spinning Particles by
  External Fields and 'Exponentiation'}},
  \href{https://doi.org/10.1103/PhysRevD.11.2219}{\emph{Phys. Rev. D}
  {\bfseries 11} (1975) 2219}.

\bibitem{DiVecchia:2020ymx}
P.~Di~Vecchia, C.~Heissenberg, R.~Russo and G.~Veneziano, \emph{{Universality
  of ultra-relativistic gravitational scattering}},
  \href{https://doi.org/10.1016/j.physletb.2020.135924}{\emph{Phys. Lett. B}
  {\bfseries 811} (2020) 135924}
  [\href{https://arxiv.org/abs/2008.12743}{{\ttfamily 2008.12743}}].

\bibitem{DiVecchia:2021ndb}
P.~Di~Vecchia, C.~Heissenberg, R.~Russo and G.~Veneziano, \emph{{Radiation
  Reaction from Soft Theorems}},
  \href{https://doi.org/10.1016/j.physletb.2021.136379}{\emph{Phys. Lett. B}
  {\bfseries 818} (2021) 136379}
  [\href{https://arxiv.org/abs/2101.05772}{{\ttfamily 2101.05772}}].

\bibitem{DiVecchia:2021bdo}
P.~Di~Vecchia, C.~Heissenberg, R.~Russo and G.~Veneziano, \emph{{The eikonal
  approach to gravitational scattering and radiation at $ \mathcal{O}
  $(G$^{3}$)}}, \href{https://doi.org/10.1007/JHEP07(2021)169}{\emph{JHEP}
  {\bfseries 07} (2021) 169}
  [\href{https://arxiv.org/abs/2104.03256}{{\ttfamily 2104.03256}}].

\bibitem{Bjerrum-Bohr:2021din}
N.E.J.~Bjerrum-Bohr, P.H.~Damgaard, L.~Plant\'e and P.~Vanhove, \emph{{The
  amplitude for classical gravitational scattering at third Post-Minkowskian
  order}}, \href{https://doi.org/10.1007/JHEP08(2021)172}{\emph{JHEP}
  {\bfseries 08} (2021) 172}
  [\href{https://arxiv.org/abs/2105.05218}{{\ttfamily 2105.05218}}].

\bibitem{Alessio:2022kwv}
F.~Alessio and P.~Di~Vecchia, \emph{{Radiation reaction for spinning black-hole
  scattering}},  \href{https://arxiv.org/abs/2203.13272}{{\ttfamily
  2203.13272}}.

\bibitem{Brandhuber:2021kpo}
A.~Brandhuber, G.~Chen, G.~Travaglini and C.~Wen, \emph{{A new gauge-invariant
  double copy for heavy-mass effective theory}},
  \href{https://doi.org/10.1007/JHEP07(2021)047}{\emph{JHEP} {\bfseries 07}
  (2021) 047} [\href{https://arxiv.org/abs/2104.11206}{{\ttfamily
  2104.11206}}].

\bibitem{Bern:2020gjj}
Z.~Bern, H.~Ita, J.~Parra-Martinez and M.S.~Ruf, \emph{{Universality in the
  classical limit of massless gravitational scattering}},
  \href{https://doi.org/10.1103/PhysRevLett.125.031601}{\emph{Phys. Rev. Lett.}
  {\bfseries 125} (2020) 031601}
  [\href{https://arxiv.org/abs/2002.02459}{{\ttfamily 2002.02459}}].

\bibitem{Cristofoli:2021jas}
A.~Cristofoli, R.~Gonzo, N.~Moynihan, D.~O'Connell, A.~Ross, M.~Sergola et~al.,
  \emph{{The Uncertainty Principle and Classical Amplitudes}},
  \href{https://arxiv.org/abs/2112.07556}{{\ttfamily 2112.07556}}.

\bibitem{Damgaard:2021ipf}
P.H.~Damgaard, L.~Plante and P.~Vanhove, \emph{{On an Exponential
  Representation of the Gravitational S-Matrix}},
  \href{https://arxiv.org/abs/2107.12891}{{\ttfamily 2107.12891}}.

\bibitem{Kol:2021jjc}
U.~Kol, D.~O'connell and O.~Telem, \emph{{The Radial Action from Probe
  Amplitudes to All Orders}},
  \href{https://arxiv.org/abs/2109.12092}{{\ttfamily 2109.12092}}.

\end{thebibliography}\endgroup
\bibliographystyle{JHEP}
\end{document}